\documentclass[twocolumn,epjc3]{svjour3}

\usepackage[numbers,sort&compress]{natbib}
\RequirePackage[T1]{fontenc}
\smartqed
\RequirePackage{graphicx}
\RequirePackage[colorlinks,citecolor=blue,urlcolor=blue,linkcolor=blue]{hyperref}
\graphicspath{{figures/}} 

\usepackage{epstopdf}
\usepackage{amsmath}
\usepackage{amsfonts}
\usepackage{amssymb}
\usepackage{textcomp}
\usepackage[utf8]{inputenc}
\usepackage{hyperref}
\usepackage{slashed}
\usepackage{feynmp-auto}
\usepackage{bm}
\usepackage{subfigure}

\usepackage{ulem}
\usepackage{cancel}

\journalname{Eur. Phys. J. A}

\begin{document}

\title{Exploring Baryon Resonances with Transition Generalized Parton Distributions: Status and Perspectives}

\author{S.~Diehl\thanksref{addr1,addr1a,e5,e4}
		\and K.~Joo\thanksref{addr1a,e5}
  		\and K.~Semenov-Tian-Shansky\thanksref{addr10,addr10a,e5} 
        \and C.~Weiss\thanksref{addr5,e5} 
		\and V.~Braun\thanksref{addr2}
	    \and W.C.~Chang\thanksref{addr4}
	      \and P.~Chatagnon\thanksref{addr5}
		\and M.~Constantinou\thanksref{addr6}
		\and Y.~Guo\thanksref{addr7}
		\and P.~T.~P.~Hutauruk\thanksref{addr8}
		\and H.S.~Jo\thanksref{addr10}
		\and A.~Kim\thanksref{addr1a}
		\and J.-Y.~Kim\thanksref{addr5}
		\and P.~Kroll\thanksref{addr14}
		\and S.~Kumano\thanksref{addr15}
		\and C.-H.~Lee\thanksref{addr16}
		\and S.~Liuti\thanksref{addr17}
		\and R.~McNulty\thanksref{addr18}
		\and H.-D.~Son\thanksref{addr22}
		\and P.~Sznajder\thanksref{addr23}
		\and A.~Usman\thanksref{addr24}
		\and C.~ Van~Hulse\thanksref{addr25}
		\and M.~Vanderhaeghen\thanksref{addr26}
		\and M.~Winn\thanksref{addr28}
}

\thankstext{e5}{Editors}
\thankstext{e4}{Corresponding author e-mail: stefan.diehl@exp2.physik.uni-giessen.de, sdiehl@jlab.org}

\institute{Justus Liebig Universit\"at Gie{\ss}en, 35390 Gie{\ss}en, Germany \label{addr1}
          \and
          University of Connecticut, Storrs, CT 06269, USA \label{addr1a}
          \and 
          Kyungpook National University, Daegu 41566, Korea \label{addr10}
		   \and
         NRC ``Kurchatov Institute'' - PNPI, Gatchina 188300, Russia
          \label{addr10a}
          \and
          Thomas Jefferson National Accelerator Facility, Newport News, VA 23606, USA\label{addr5}
          \and
          Institut f\"ur Theoretische Physik, Universit\"at Regensburg, 93040 Regensburg, Germany \label{addr2}
					\and
          Institute of Physics, Academia Sinica, Taipei 11529, Taiwan \label{addr4}
					\and
          Physics Department, Temple University, Philadelphia, PA 19122-1801, USA \label{addr6}
          \and
          Nuclear Science Division, Lawrence Berkeley National Laboratory, Berkeley, CA 94720, USA \label{addr7}
					\and
          Department of Physics, Pukyong National University (PKNU), Busan 48513, Korea \label{addr8}
					\and
          University of Wuppertal, D-42097 Wuppertal, Germany \label{addr14}
					\and
          Department of Mathematics, Physics, and Computer Science, Faculty of Science, Japan Women's University, Tokyo 112-8681, Japan and Theory Center, Institute of Particle and Nuclear Studies, KEK, Tsukuba, Ibaraki, 305-0801, Japan \label{addr15}
					\and
          Department of Physics, Pusan National University, Busan 46241, Korea and Asia Pacific Center for Theoretical Physics, POSTECH, Pohang 37673, Korea \label{addr16}
					\and
          Physics Department, University of Virginia, Charlottesville, VA 22904-4714, USA \label{addr17}
					\and
          University College Dublin, School of Physics, Science Centre Belfield Dublin 4, Ireland  \label{addr18}
					\and
          Department of Physics, Inha University, Incheon 22212, Korea \label{addr22}
					\and
          National Centre for Nuclear Research (NCBJ), 02-093 Warsaw, Poland \label{addr23}
					\and
          University of Regina, Regina SK  S4S~0A2 Canada \label{addr24}
					\and
          Universidad de Alcal\'{a}, 28801  Alcal\'{a} de Henares, Spain \label{addr25}
					\and
          Institut f\"ur Kernphysik and $\text{PRISMA}^+$ Cluster of Excellence, Johannes Gutenberg Universit\"at, 55099 Mainz, Germany \label{addr26}
					\and
          DPhN/Irfu, CEA Saclay, 91191 Gif sur Yvette, France \label{addr28}
          }

\date{\today}

\maketitle

\sloppy 


\begin{abstract}
QCD gives rise to a rich spectrum of excited baryon states.
Understanding their internal structure is important for many areas of nuclear physics,
such as	nuclear	forces,	dense matter, and neutrino-nucleus interactions.
Generalized parton distributions (GPDs) are an established tool for characterizing the QCD
structure of the ground-state nucleon. They are	used to	create 3D tomographic images of
the quark/gluon structure and quantify the mechanical properties such as the distribution
of mass, angular momentum, and forces in the system.
Transition GPDs	extend these concepts to $N \rightarrow N^\ast$ transitions and can be
used to characterize the 3D structure and mechanical properties of baryon resonances.
They can be probed in high-momentum-transfer exclusive electroproduction processes	with
resonance transitions $e + N \rightarrow e' + M + N^\ast$, such as deeply-virtual
Compton scattering ($M = \gamma$) or meson production ($M = \pi, K$, {\it etc.}), and in
related	photon/hadron-induced processes.

This White Paper describes a research program aiming to explore baryon resonance structure with transition GPDs.
This includes the properties and interpretation of the transition GPDs, theoretical methods	for
structures and processes, first experimental results from JLab 12 GeV, future measurements
with existing and planned facilities (JLab detector and energy upgrades, COMPASS/AMBER,
EIC, EicC, J-PARC, LHC ultraperipheral collisions), and the theoretical and experimental developments needed to realize this program.

\end{abstract}


\newpage
\setcounter{tocdepth}{2}
\tableofcontents

\section{Introduction}

Hadrons are emergent phenomena of Quantum Chromodynamics (QCD), see 
{\it e.g.}
\cite{Gross:2022hyw}. The elementary dynamics are expressed
in both the spectrum and the structure of the hadronic states. Exploring both expressions together
is essential for a complete understanding of strong interactions. There are many examples of
connections between the internal motion and the excitation spectrum in nonrelativistic quantum
systems (atoms, nuclei). Similar connections are expected in the more complex relativistic
quantum systems presented by the hadrons in QCD.

A rich spectrum of baryon resonances is known to emerge from QCD.  They are observed in scattering
experiments with electron, photon, and hadron beams at energies of a few GeV. These are excited
states of the nucleon (proton or neutron), the basic building block of nuclei, and visible matter
in the universe. The spectrum of these excited states was essential for establishing the quark model
as a precursor to QCD. The baryon resonances play an important role in the theory of nuclear forces
({\it e.g.}\ the $\Delta$ isobar), in the behavior of matter at high densities and temperatures (early
universe, stellar structure), and in the existence of hypothetical strange matter (neutron stars) \cite{PhysRevD.102.063008,Marquez:2022gmu}. 
They are also needed for describing the interactions of neutrinos with nuclei at energies of a few GeV (neutrino detection in oscillation experiments) \cite{Amaro:2019zos}.  Understanding the internal
structure of the baryon resonances is thus of fundamental interest and practical importance.

Some information on the structure of baryon resonances is available from the electromagnetic
transition form factors, {see {\it e.g.} \cite{Ramalho:2023hqd} for a review}, measured in electroproduction experiments. They describe the spatial
distribution of charge and current in the dynamical system. Further information from other probes is needed to characterize the excited baryon states.

The study of ground-state nucleon structure has significantly improved in the last two decades.
The concept of generalized parton distributions (GPDs)
\cite{Muller:1994ses,Radyushkin:1997ki,Ji:1996nm}
has provided a rigorous formulation of the
spatial distributions of quarks and gluons in the nucleon and offered new opportunities for
characterizing nucleon structure, see {\it e.g.} \cite{Diehl:2023nmm} for an overview. The GPDs unify the concepts of the elastic nucleon form factors
and the quark/gluon particle densities. They enable the construction of 3D tomographic images of the
distribution of quarks and gluons, which have great potential for the visualization of nucleon
structure and the discussion of internal motion. They also quantify the distribution of energy,
momentum, angular momentum, and forces in the nucleon, which allows one to discuss the mechanical
properties of the quantum system in analogy with classical systems.  The GPDs are extracted using a
combination of theoretical methods and experimental data from various scattering processes,
particularly exclusive processes at energy and momentum transfers $\gg 1$ GeV. Programs of GPD
measurements are underway at Jefferson Lab (JLab) with 12 GeV electron beam \cite{Burkert:2020akg, PAC42hallC}, CERN COMPASS \cite{COMPASS:2018pup}, Large Hadron Collider (LHC)
\cite{Klein:2020fmr}, and Japan Proton Accelerator Research Complex (J-PARC)
\cite{Aoki:2021cqa}, and planned with the
future Electron-Ion Collider (EIC) \cite{AbdulKhalek:2021gbh,Burkert:2022hjz} and  Electron-ion collider in China (EicC)~\cite{Anderle:2021wcy}. 

Recent advances have made it possible to extend the framework of GPDs to excited baryon states and
use it to characterize baryon resonance structures. Theoretical studies have formulated the concept
of $N \rightarrow \Delta, \,N^{*}$ transition GPDs and extended the method of quark/gluon tomography to baryon
resonances. Experimental efforts have produced the first measurements of exclusive processes with
$N \rightarrow \Delta$ transitions and demonstrated their feasibility.  The transition GPDs allow
one to ask and answer new questions about baryon resonance structure:
\begin{itemize}
\item What is the spatial distribution of quarks in excited baryon states, and how does it differ from
the ground state? Can we construct tomographic images of the baryon resonances?

\item What are the distributions of energy, momentum, and angular momentum carried by quarks and gluons
in baryon resonances? Can we quantify the mechanical properties of the baryon resonances?

\item What is the distribution of quark tensor charge in baryon resonances?  What is the gluonic
structure of resonances? Can we excite baryon resonances using transition operators with quantum
numbers other than the vector current operator used in traditional photo/electroexcitation?
\end{itemize}
These questions greatly expand the scope of baryon resonance structure studies and raise it to the
same level as current ground-state nucleon structure studies.

A workshop ``Exploring resonance structure with transition GPDs'' was held at the European Center for
Theoretical Studies in Nuclear Physics and Related Areas (ECT*) in Trento, Italy, on August 21-25,
2023, supported jointly by ECT* and the Asia Pacific Center for Theoretical Physics (APCTP)
~[\href{https://indico.ectstar.eu/event/176/}{https://indico.ectstar.eu/event/176/}]. Twenty-seven presentations were given over five
days, followed by extensive discussions. This white-paper summarizes the status of theoretical and
experimental studies of transition GPDs and baryon resonance structure, the prospects for future experimental programs using electromagnetic and hadronic probes, and the developments needed for their realization.

The article is organized as follows:
Section \ref{sec:overview} presents a general introduction to the concepts and an overview of the scientific program. Section \ref{sec:theory} provides a theoretical introduction to transition GPDs, while section \ref{sec:physics} shows their connection to higher level observables of the baryon resonance structure. In section \ref{sec:methods} an overview of the methods for the theoretical treatment of transition GPDs is provided.
Section \ref{sec:observables} introduces the description of lepto-production processes sensitive to transition GPDs and section \ref{sec:current_results} presents the first experimental results from 12 GeV JLab on deeply virtual $N  \to  N ^{*}$ meson production and $N  \to  N ^{*}$ Compton scattering.
Sections \ref{sec:transGPDhadrons} and \ref{sec:transGPDdiffractive} introduce the description and measurement of transition GPDs in hadron scattering and diffractive scattering processes.
Section \ref{sec:perspectives} provides an overview of future perspectives at JLab, but also at hadron facilities like J-PARC and other places.
Finally, the strategy for the extraction of transition GPDs from experimental observables is discussed in section \ref{sec:extraction}
and an overview of the future transition GPD program is provided in section \ref{sec:future}.


\section{Overview}
\label{sec:overview}

Hadron structure in QCD is expressed in the matrix elements of QCD operators between hadronic states.
These gauge-invariant operators are composed of the fundamental quark and gluon fields and measure
certain features of their distribution in the hadronic state.
Two basic types of QCD operators are accessible in scattering experiments.
One type is the vector and axial-vector currents of the quark fields, which are probed in
elastic lepton-nucleon or neutrino-nucleon scattering mediated by electromagnetic or weak interactions.
These operators are local and measure the distribution of electric and axial charge and
current in the hadronic state (form factors).
Another type is QCD operators arising from the factorization of deep-inelastic
scattering processes at energy and momentum transfer much larger than the hadronic scale ($\gg$ 1 GeV). 
These operators are non-local in space-time (light-like separation of fields) and measure the
momentum distributions of quarks and gluons in the hadronic state (parton distributions).
Inclusive deep-inelastic lepton-nucleon scattering and related processes have provided
extensive information on the quark and gluon parton densities (PDFs).

\begin{figure}[t]
\centering
\includegraphics[width = 0.77\columnwidth]{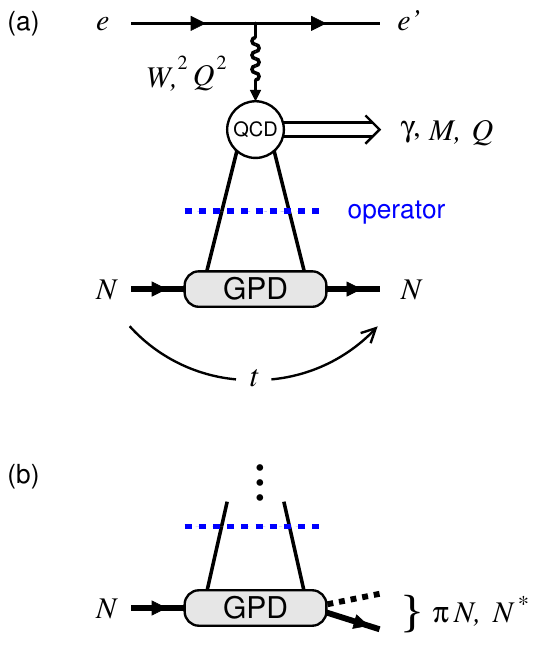}
\caption{(a)~QCD factorization of hard exclusive processes 
$e + N \rightarrow e' + \gamma N, M N, \mathcal{Q}N$.
{
The production process happens in the scattering from a quark or gluon,
whose emission/absorption by the target is described by the 
matrix elements of a QCD light-cone operator.} The $N \rightarrow N$ matrix elements
of the operator are parametrized by the nucleon GPDs. 
(b)~Exclusive processes with transition $N \rightarrow \pi N, \Delta, N^\ast$.
The $N \rightarrow \pi N,  \Delta, N^\ast$ matrix elements of the QCD 
light-cone operator are parametrized by the transition GPDs.}
\label{fig:hardexcl}
\end{figure}

Exclusive processes in lepton-nucleon scattering at energy and momentum transfer $W^2, Q^2 \gg$ 1 GeV$^2$ are another class of processes to which the factorization method can be applied. Examples include the electroproduction of real photons (deeply-virtual Compton scattering, 
or DVCS)
$e + N \rightarrow e' + \gamma + N'$, electroproduction of light mesons
$e + N \rightarrow e' + M + N'$, or photo/electroproduction of heavy quarkonia, $e + N \rightarrow e' + \mathcal{Q} + N'$.
{
In the asymptotic regime 
where $Q^2, W^2 \gg 1 \, \textrm{GeV}^2$, with the ratio $Q^2/W^2$
fixed and described by a scaling variable, and with the momentum transfer between the
initial and final nucleon state $|t| \lesssim 1\, \textrm{GeV}^2$, the amplitude of
the production process can be separated into two parts: a quark/gluon scattering amplitude involving
perturbative QCD interactions at momenta $\sim Q$, and the amplitude for the nucleon to emit/absorb
the quarks/gluons in the relevant configurations at the hadronic scale 
(see Fig.~\ref{fig:hardexcl}a). 
The latter is described by the nucleon matrix elements of certain QCD operators with light-like separation,
parameterized by the GPDs; see
Refs.~\cite{Goeke:2001tz,Diehl:2003ny,Belitsky:2005qn} for a review.}
These are non-forward matrix elements, involving a longitudinal and transverse momentum transfer of the order of the hadronic scale between the initial and
final nucleon, resulting in a rich structure. The GPDs 
thus unify the concepts of elastic form factors and
parton densities and contain both as limiting cases.
%
%

\begin{figure*}[t]
\centering
\includegraphics[width = 0.6\textwidth]{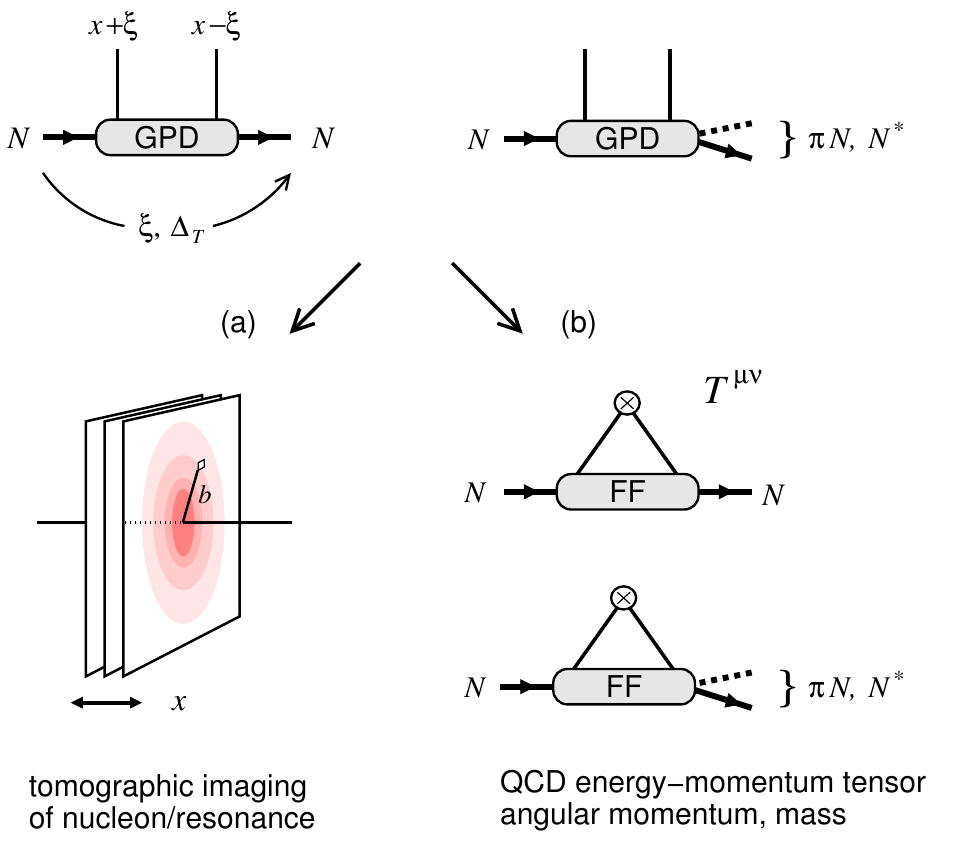}
\caption{Applications of GPDs to nucleon structure: (a) Transverse spatial distributions of partons
(tomographic imaging). (b) Form factors of QCD energy-momentum tensor describing distributions of
angular momentum, mass, and forces (mechanical properties). The concepts and structures can be
extended to $N \rightarrow \pi N, N^\ast$ transitions.}
\label{fig:gpd_interpretation}
\end{figure*}

GPDs provide new information on nucleon structure beyond what is available from conventional
probes (form factors, PDFs), see Fig.~\ref{fig:gpd_interpretation}).
(i) The transverse Fourier transform of the GPDs (transverse momentum
transfer $\Delta_T \rightarrow$ transverse coordinate $b$) describes the transverse spatial distribution
of quarks and gluons with given longitudinal momentum fraction $x$
\cite{Burkardt:2000za,Burkardt:2002hr}. 
It provides a 2 + 1 dimensional
``tomographic image'' of the nucleon. This spatial representation is appropriate for the nucleon
as a relativistic quantum system and allows one to visualize it as an extended object in 
{the transverse space \cite{Dupre:2016mai}.}
The spatial structure can be connected with the internal motion of the quarks and gluons and their
polarization and provides a new framework for the discussion of structure and dynamics.
``3D imaging'' of the nucleon based on GPDs, which requires measurements with high precision, is the object of experimental programs at JLab and EIC.
(ii) The moments of the GPDs (weighted integrals over the momentum fraction $x$) represent matrix elements
of local QCD operators of spin $n > 1$ (so-called generalized form factors). This makes it possible to
probe nucleon structure with local operators beyond the spin-1 operators accessible with the electroweak currents.
The spin-2 operators obtained from GPDs contain the QCD energy-momentum tensor, the matrix elements of which describe
the distribution of momentum, angular momentum, forces, and mass and are of fundamental interest for
nucleon structure. Characterizing the ``mechanical properties'' of hadrons based on the QCD energy-momentum tensor
has emerged as a field of study in its own right, with many theoretical and experimental results; see Refs.~\cite{Polyakov:2018zvc,Lorce:2018egm} for a review.

QCD factorization can be applied not only to exclusive processes with $N \rightarrow N$ transitions
on the target side, but also to processes in which the nucleon undergoes a transition to a
low-mass hadronic state, $N \rightarrow \pi N, \pi\pi N, N^\ast, ...$ (see Fig.~\ref{fig:hardexcl}b) \cite{Goeke:2001tz}.
Examples are DVCS with $N \rightarrow \pi N$ transitions,
$e + N \rightarrow e' + \gamma + \pi N (\textrm{low-mass})$; or exclusive pion production
with $N \rightarrow \pi N$ transitions, $e + N \rightarrow e' + \pi + \pi N (\textrm{low-mass})$.
The QCD operators representing the high-energy processes are the same as in the case of $N \rightarrow N$
transitions. The matrix elements are now taken between the initial $N$ and the final $\pi N$ state,
$\langle \pi N | \mathcal{O}_{\rm QCD}| N\rangle$, and are parameterized by so-called transition GPDs.
This extends the concepts of GPDs to transitions between low-mass hadronic states. The final $\pi N$ state here
can be non-resonant or resonant \cite{Thiel:2022xtb}. Matrix elements for transitions to a baryon resonance $N \rightarrow N^\ast$
can be defined rigorously through analytic continuation in invariant mass of the $\pi N$ system
(resonance pole). This allows one to establish the concept of GPDs for baryon resonance transitions.

The nucleon is known to possess a rich spectrum of excited states \cite{ParticleDataGroup:2022pth}. 
They are observed in production and decay
processes induced by electromagnetic and hadronic probes. The spectrum has been explained using models based
on effective degrees of freedom (quark model, chiral soliton) and is increasingly being confirmed by lattice QCD (LQCD)
calculations. Exploring the structure of the excited states in terms of QCD degrees of freedom is the
next step in their study. In the hadronic picture, the excited states are as fundamental as the ground state,
and the QCD structure of all the states is needed for a complete understanding of strong interaction dynamics.
Some information on $N^\ast$ structure is available from the transition form factors of the vector current
measured in electroexcitation processes. Much more information on $N^\ast$ can be obtained from the
transition GPDs measured in $N \rightarrow N^\ast$ exclusive processes and the concepts derived from them
(see Fig.~\ref{fig:gpd_interpretation}).
The transition GPDs allow one to construct tomographic images of the $N^\ast$ at the same level as
the $N$ and discuss the QCD structure of resonances in these terms. They provide access to the transition matrix elements of the QCD EMT and allow one to discuss the mechanical properties of the resonances.
The factorization of exclusive processes provides QCD operators with quantum numbers that are
not easily accessible otherwise, such as chiral-odd quark operators (pion production) and gluonic
operators (heavy-quarkonium production), and these operators can be used for resonance excitation and
structure studies. While extracting the structures from experiments poses considerable challenges,
the concepts derived from transition GPDs can also be explored using dynamical models and LQCD results
and enrich $N^\ast$ structure studies in this way. Altogether, transition GPDs have the potential
to greatly expand the range of baryon structure studies in QCD.

After first studies of associated electroproduction of real photons, $e + p \rightarrow e'+ \gamma+ \pi N$, in the $\Delta(1232)$-resonance region with HERMES \cite{Duren:2014ola}, recent experiments at JLab have shown the feasibility of measuring exclusive processes with
$N \rightarrow N^\ast$ transitions with significantly increased precision and background separation capabilities. Measurements of exclusive electroproduction $e + p \rightarrow e' + \pi^- + \Delta^{++}$ have been performed at JLab CLAS12 and are being interpreted in terms of transition GPDs \cite{CLAS:2023akb}. Similar measurements of $e + p \rightarrow e' + \pi^+ + \Delta^{0}$ have been performed at JLab Hall C and are being analyzed.
Measurements of DVCS with $N \rightarrow \Delta$ and other $N \rightarrow N^\ast$ transitions
will be possible with the forthcoming CLAS12 data.

A program for exploring resonance structure using transition GPDs is emerging.
To realize it, it is necessary to develop theoretical and experimental methods,
define the scientific objectives, simulate the proposed measurements, and optimize the
analysis and extraction procedure.

Theoretical methods needed for the transition GPD program include the structural decomposition
of the $N \rightarrow N^\ast$ matrix elements of the QCD operators and their parametrization
in terms of transition GPDs. The structure of the $N \rightarrow \Delta$ matrix elements of
the chiral-even QCD operators was discussed in Refs.~\cite{Goeke:2001tz,Belitsky:2005qn}. Recent work revisited the
definition of the chiral-even $N \rightarrow \Delta$ transition GPDs and extended the analysis
to the chiral-odd sector \cite{Kroll:2022roq}. The $N \rightarrow \Delta$ transition matrix elements of the QCD
energy-momentum tensor were studied in Ref.~\cite{Ozdem:2019pkg,Polyakov:2020rzq,Kim:2022bwn}.

Also needed is the physical interpretation of the transition matrix elements, especially
the $N \rightarrow N^\ast$ transition matrix elements of the QCD energy-momentum tensor
and the mechanical properties derived from it. Recent work has extended the concept of
QCD angular momentum to $N \rightarrow \Delta$ transitions, using a formulation
in terms of light-front densities appropriate for transitions between baryon states
with different masses and quantum numbers \cite{Kim:2023xvw}.

Theoretical efforts also focus on making quantitative predictions of the transition of GPDs
using methods of nonperturbative QCD. Particularly useful are methods that can connect the
$N \rightarrow N$ and $N \rightarrow \pi N, \Delta$ {\it etc.} \ matrix elements in a systematic fashion.
The $1/N_c$ expansion of QCD is a general method for analyzing matrix elements of QCD operators
in meson and baryon states. It is based on the dynamical spin-flavor symmetry of QCD in the large-$N_c$ limit,
where $N$ and $\Delta$ states are in the same multiplet, and transitions between them are
connected by symmetry. The method can classify and predict the spin-flavor components of
$N \rightarrow \Delta$ transition GPDs and express them in terms of the $N \rightarrow N$ GPDs.
The $1/N_c$ expansion was applied to 
{vector and axial-vector}
transition GPDs \cite{Goeke:2001tz}, and was recently extended
to study 
{tensor transition}
GPDs \cite{Schweitzer:2016jmd,Kroll:2022roq}. Methods based on chiral dynamics (soft-pion theorems, chiral perturbation theory)
can connect $N \rightarrow N$ and $N \rightarrow \pi N$ transition matrix elements near threshold and predict
transition GPDs in this regime \cite{Pobylitsa:2001cz,Chen:2003jm,Guichon:2003ah,Birse:2005hh}. They can also be extended to processes with $N \rightarrow \Delta$ transitions \cite{Alharazin:2023zzc,Alharazin:2023uhr}. Models based on the holographic representation of QCD (gauge-string duality)
posit a close connection between hadron structure and the spectrum of excited states and are
useful for the study of GPDs and transition GPDs~\cite{Mamo:2022jhp}.

LQCD methods for computing nucleon matrix elements of local QCD operators are well-developed.
They have recently been extended to enable the computation of $x$-dependent partonic structure,
using matrix elements of nonlocal operators with spacelike separation in fast-moving hadronic states
(quasi/pseudo distributions). Applications to GPDs are in progress \cite{Alexandrou:2020zbe,Bhattacharya:2022aob,Bhattacharya:2023jsc}.
At the same time, LQCD methods have been adapted to extract excited hadronic states,
using arrays of Euclidean correlation functions of operators projecting on ground and excited states
(generalized eigenvalue problem, distillation) \cite{Edwards:2011jj,xQCD:2019jke}.
While the combination of both techniques is challenging and requires major development,
there is the long-term prospect of LQCD calculations of transition GPDs.

Theory developments currently focus on reaction theory 
and observables for DVCS with $N \rightarrow \Delta$ and $N \rightarrow N^\ast$ transitions \cite{Semenov-Tian-Shansky:2023bsy} as well as
$\pi$ electroproduction with $N \rightarrow \Delta$ transitions \cite{Kroll:2022roq}.

Experimental developments at JLab focus on improving the measurement of pion production with
$N \rightarrow \Delta$ transitions (CLAS12, Hall C), and preparing measurements of DVCS
with $N \rightarrow \Delta$ and other $N \rightarrow N^\ast$ transitions (CLAS12). It is planned to extend the studies to the strangeness sector.
The luminosity upgrade of CLAS12, which is currently in preparation, will help to increase the statistics for these studies. Further detector modifications to increase the detection efficiency of the processes with CLAS12 will be investigated. A potential energy upgrade of JLab
\cite{Accardi:2023chb}
would help to increase the accessible $Q^{2}$ range (factorization of the processes) and the phase space of the invariant masses above the background rejection cuts (detection efficiency).

The future EIC will enable an extended program of GPD measurements
on the proton at high energies ($10^{-3} \lesssim x_B \lesssim 0.1$). The final-state proton or
neutron moves forward in the ion beam direction and will be detected with the far-forward
a detection system (magnetic spectrometer with several detector subsystems for proton,
zero-degree calorimeter for neutron) \cite{AbdulKhalek:2021gbh}. The possibility of $\pi N$ and other transition
GPD measurements with this setup are being explored and present interesting challenges for 
far-forward detection and event reconstruction.

Ultraperipheral proton-nucleus collisions and central exclusive production in proton-proton
collisions at LHC enable measurements of photon-proton scattering
at energies $W \sim 10^3$ GeV, the highest energies available for electromagnetic scattering
(an order of magnitude larger than at the electron-proton HERA collider) \cite{Baltz:2007kq}. The setup is used to
measure gluon GPDs in the photoproduction of heavy quarkonia, $\gamma + p \rightarrow \bar Q Q + p$,
reaching momentum fractions $x \sim 10^{-6}$ \cite{ALICE:2014eof,ALICE:2018oyo}. This exclusive scattering process is diffractive
(vacuum exchange) and can be discussed using concepts of diffractive scattering. The scattering
either leaves the proton intact, $p \rightarrow p$, or causes it to dissociate into a
low-mass hadronic state, $p \rightarrow X(\textrm{low-mass})$. The latter process measures the
gluon GPD for $p \rightarrow X$ transitions, which can be connected with the quantum fluctuations
of the gluon density in the proton \cite{Frankfurt:2008vi}.

Hadron-induced scattering processes can also be used to access GPDs under certain conditions,
providing a complementary method to electromagnetic processes. Exclusive dilepton production
(Drell-Yan pair production) \cite{Berger:2001zn,Sawada:2016mao} as well as 2 $\to$ 3 processes \cite{Kumano:2009he}, have been proposed as a way of measuring GPDs and transition GPDs. These hadron-induced processes have a natural connection to transition GPDs.


\section{Transition GPDs}
\label{sec:theory}

\subsection{GPDs of the ground state nucleon}
\label{sec:GPDdef}
{
GPDs describe the hadronic matrix elements of partonic QCD 
operators -- nonlocal operators with a lightlike separation between the quark
or gluon fields \cite{Muller:1994ses,Ji:1996ek,Radyushkin:1996nd}. 
The partonic quark operators are defined as
\begin{align}
\bar q (-\tau n /2) \, [-\tau n/2, \tau n/2] \, \Gamma \, q(\tau n /2),
\label{partonic_operator}
\end{align}
where $q$ and $\bar q$ are the QCD quark fields of flavor $q = u, d, ...$;
$n$ is a light-like 4-vector, $n^2 = 0$, and $\tau$ is a parameter specifying the
distance along the light-like direction. $[..., ...]$ denotes the gauge link (Wilson line)
connecting the fields along the light-like path; it can be set to unity by choosing an
appropriate light-like gauge and is omitted in the following. A basis of light-like
4-vectors is constructed by complementing $n$ with a second independent light-like 
vector $\tilde n$ with $\tilde n^2 = 0$ and $n \cdot\tilde n = 1$, and the 
light-cone components of a 4-vector are 
defined as\footnote{The definition of light-cone components in Eq.~(\ref{lightcone}) 
in terms of the vectors $n$ and $\tilde n$ is covariant and does not rely on a 
specific coordinate system. In the standard coordinate system where 
$n^\mu = (n^0, n^1, n^2, n^3) = (1/\sqrt{2}, 0, 0, -1/\sqrt{2})$
and 
$\tilde n^\mu = (1/\sqrt{2}, 0, 0, 1/\sqrt{2})$, 
the light-cone component take the conventional form
$v^\pm = (v^0 \pm v^3)/\sqrt{2}, \bm{v}_\perp = (v^1, v^2)$.}
\begin{align}
v^+ \equiv n\cdot v, \hspace{1em} v^- \equiv \tilde{n} \cdot v,
\hspace{1em}
v = v^+ \tilde n + v^- n + v_\perp
\label{lightcone}
\end{align}
In Eq.~(\ref{partonic_operator}) $\Gamma$ denotes a spinor matrix defining the
spin projection of the quark fields. In the class of so-called twist-2 operators,
three different spin projections are considered:
\begin{alignat}{3}
\Gamma = \gamma^+ & \hspace{1em} \text{vector} && \text{unpolarized} &&
\label{operator_vector}
\\
\gamma^+\gamma^5 & \hspace{1em} \text{axial vector} \hspace{2em} && \text{helicity-polarized} &&
\label{operator_axial}
\\
\sigma^{+T} & \hspace{1em} \text{tensor} && \text{transversity-polarized} &&
\label{operator_tensor}
\end{alignat}
The vector, axial-vector, and tensor operator measure the unpolarized (spin-averaged), 
helicity-polarized (spin difference in the longitudinal direction), 
and transversity-polarized (spin difference in transverse space) 
distribution of quarks. The operators can be expanded in local operators of
twist-2 and have known renormalization properties (scale dependence, evolution).
Similar partonic operators are constructed from gluon fields.

The transition matrix elements of the operators Eq.~(\ref{partonic_operator}) between nucleon states $(N \rightarrow N)$
are parametrized by the GPDs. For the twist-2 quark operators, Eqs.~(\ref{operator_vector})--(\ref{operator_tensor}),
for each quark flavor $q$, there are 8 independent structures characterized by the quantum numbers of the quark operator (or quark spin projection), 
the nucleon spin variables, and the momentum transfer between the nucleon
states, see {\it e.g.}, \cite{Diehl:2003ny, Belitsky:2005qn}.
The matrix element of the vector operator Eq.~(\ref{operator_vector})
is parametrized by 2 vector 
(or unpolarized) GPDs,
\begin{align}
 \int & \frac{d z^-}{4\pi}e^{i \tau (n \cdot P) x} \nonumber \\ &
 \langle N(p', \lambda_N') \left|  \bar{q}(-\tau n /2) \gamma^+ q(\tau n /2)  \right| N(p, \lambda_N) \rangle 
\nonumber \\
 = & \frac{1}{2  (n \cdot P)  } \, \bar{u} (p',\lambda'_N) 
 \left[ H^q (x,\xi,t) \gamma^+  \right. \nonumber \\  & \left.
   + E^q (x,\xi,t)  \frac{i \sigma^{+ \alpha} \Delta_\alpha}{2 \, M_N}
 \right ] u (p, \lambda_N);
\label{eqn:gpd-n}
\end{align}
the matrix element of the axial vector operator Eq.~(\ref{operator_axial}) is parametrized by 2 axial vectors (or helicity) GPDs,
\begin{align}
 \int & \frac{d \tau}{4\pi}e^{i \tau (n \cdot P) x} \nonumber \\ &
 \left< N(p', \lambda_N') \left|  \bar{q}(-\tau n /2) \gamma^+ \gamma_5 q(\tau n/2)  \right| N(p, \lambda_N) \right> 
\nonumber \\
 = & \frac{1}{2  (n \cdot P)} \, \bar{u} (p',\lambda'_N) 
 \left [ \tilde{H}^q (x,\xi,t) \gamma^+ \gamma_5 \nonumber \right. \\ & \left.
   + \tilde{E}^q (x,\xi,t)
   \frac{\gamma_5 \Delta^+}{2 \, M_N}
 \right ] u (p,\lambda_N);
\label{eqn:gpd-p}
\end{align}
and the matrix element of the tensor operator Eq.~(\ref{operator_tensor}) is parametrized by 4 
tensor (or transversity GPDs) \cite{Diehl:2001pm},
\begin{align}
&  \int \frac{d \tau}{4 \pi} e^{i \tau (n \cdot P) x} \nonumber \\ &
\left\langle N(p^{\prime}, \lambda_N^{\prime})\right| \bar{q}(-\tau n /2) i \sigma^{+i} q (\tau n /2) \left|N(p, \lambda_N)\right\rangle 
\nonumber
\\
& =\frac{1}{2 (n \cdot P)} \bar{u}\left(p^{\prime}, \lambda_N^{\prime}\right)\left[H_T^q(x,\xi,t)
i \sigma^{+i} \phantom{\frac{0}{0}}
\right.
\nonumber \\
&+\tilde{H}_T^q(x,\xi,t) \frac{P^{+} \Delta^i-\Delta^{+} P^i}{M_N^2}
\nonumber \\
& +E_T^q(x,\xi,t) \frac{\gamma^{+} \Delta^i-\Delta^{+} \gamma^i}{2 M_N}
\nonumber \\
& \left. +\tilde{E}_T^q(x,\xi,t) \frac{\gamma^{+} P^i-P^{+} \gamma^i}{M_N}\right] u\left(p, \lambda_N\right).
\label{eqn:gpd-t}
\end{align} 
Here $ \left| N(p, \lambda_N) \right>$  ($ \left| N(p', \lambda'_N) \right>$) denotes the proton state with momentum $p$ ($p'$), polarization $\lambda_N$ ($\lambda'_N$) and mass $M_N$, $u(p,\lambda_N)$ stand for the Dirac spinor of the nucleon;  $P \equiv (p+p')/2$ is the average momentum; $\Delta \equiv p'-p$ is the momentum transfer.

In the matrix elements Eqs.~(\ref{eqn:gpd-n})--(\ref{eqn:gpd-t}) the dependence of the operator
on the light-like distance $\tau$ is Fourier-transformed and converted to 
a dependence on the conjugate momentum variable $(n\cdot P)x$. The partonic variable $x$ can be interpreted as the ratio of the light-cone ``plus'' momentum of the quark field modes in the operator to the average plus momentum of the initial and final nucleon. The GPDs also depend on two variables characterizing the momentum transfer between the nucleon states: the ``skewness'' variable
\begin{align}
\xi \equiv - \frac{(n \cdot \Delta)}{2 (n \cdot P)}, \hspace{2em} 0 < \xi < 1,
\end{align}
describing the fractional plus momentum transfer to the nucleon and the invariant momentum transfer
\begin{align}
t \equiv \Delta^2 < 0.
\end{align}

The GPDs also depend on the renormalization scale of the partonic operator $\mu^2$, 
which is not shown explicitly in Eqs.~(\ref{eqn:gpd-n})--(\ref{eqn:gpd-t}) and most
of the following. This dependence is governed by the QCD evolution equations for
the partonic operator.

\begin{figure}[t]
\centering
\includegraphics[width=0.48\textwidth]{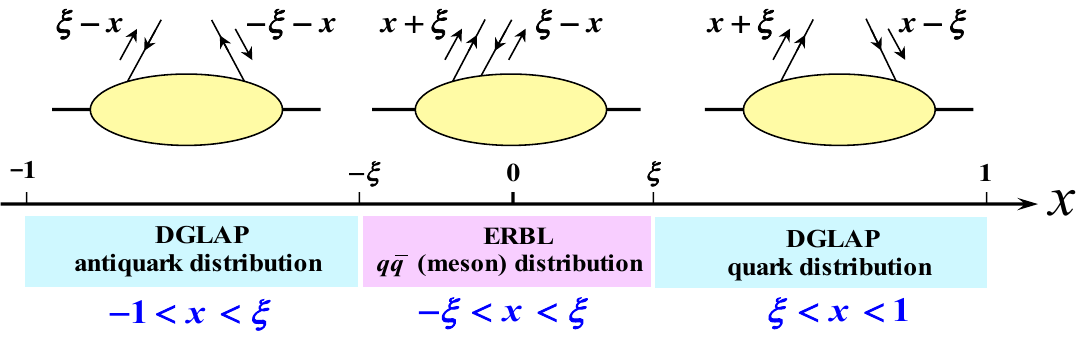}
\caption{ERBL and DGLAP regions of the GPDs.}
\label{fig:ERBL-DGLAP}
\end{figure}

In the dependence of the GPDs on the partonic variable $x$
one distinguishes three regions, determined by the position of $x$ relative to the
skewness variable $\xi$, see Fig.\,\ref{fig:ERBL-DGLAP}.
In the regions $\xi<x<1$ ($-1<x<-\xi$) the GPDs describe the emission of a
quark (antiquark) with plus momentum fraction $x+\xi$ ($\xi-x$) and 
subsequent absorption of a quark (antiquark) with fraction $x-\xi$ ($-\xi-x$).
In these two regions, the GPDs have the characteristics of quark or anti-quark
density in the nucleon (so-called DGLAP, or Dokshitzer-Gribov-Lipatov-Altarelli-Parisi regions).
In the region $-\xi<x<\xi$ the GPDs describe the emission of a quark-antiquark pair
with momentum fractions $x+\xi$ and $\xi-x$ by the nucleon. In this region, the GPD
has the characteristics of a meson distribution amplitude
(so-called ERBL, or Efremov-Radyushkin-Brodsky-Lepage region).

The GPDs unify the concepts of the nucleon parton densities and elastic form	factors	and
contain	both as	limiting cases.	In the limit of zero momentum transfer $\xi = 0, t = 0$ (forward limit),
the vector GPD $H^q$ coincides with the quark/antiquark parton density (PDF) in the nucleon,
\begin{align}
H^q (x, \xi = 0, t = 0) \; = \; 
\left\{ \begin{array}{rr}
q(x) & \hspace{2em} (x > 0), 
\\
-\bar q(-x) & (x < 0).
\end{array}
\right.
\end{align}
The vector GPD $E^q$ in the forward limit cannot be connected with a known PDF
because the structure is non-diagonal in the nucleon helicity.	The integral of the vector GPDs
$H^q$ and $E^q$ over $x$ (so-called first moment) reduces the vector partonic operator to the local vector current operator, whose matrix element is parametrized by	the nucleon's Dirac Pauli form factors of quark flavor $q$,
and therefore
\begin{align}
\int_{-1}^1 dx \; H^q (x, \xi, t) &= F_1^q(t),
\\
\int_{-1}^1 dx \; E^q (x, \xi, t) &= F_2^q(t).
\end{align}
In a similar way the axial-vector GPD $\tilde H^q$ is related to the quark helicity-polarized PDFs $\Delta q(x)$ and $\Delta \bar q(x)$, and the first moments of $\tilde H$ and $\tilde E$	give the nucleon's axial vector and induced pseudoscalar form factors. The tensor GPD $H_T^q$ is related to the quark transversity-polarized PDFs $\delta q(x)$ and $\delta \bar q(x)$, and the first moments of the tensor GPDs $H_T^q, E_T^q, \tilde H_T^q$ give the form factors of the local tensor operator; the first moment of $\tilde E_T^q$ is zero because of time reversal invariance (see further discussion in Sec.~\ref{subsec:tensor_charge}).

The partonic QCD operators Eqs.~(\ref{operator_vector})--(\ref{operator_tensor}) and their GPDs
are also distinguished by their chiral properties. The vector and axial vector operators,
Eqs.~(\ref{operator_vector}) and (\ref{operator_axial}), are chirally even and conserve the
helicity of the quark/antiquark emitted and absorbed by the nucleon (for the massless quarks
the chirality is the same as the helicity). The tensor operator Eq.~(\ref{operator_tensor}) is
chirally odd and flips the helicity of the quark/antiquark. 
Correspondingly, the GPDs $H^q, E^q, \tilde{H}^q, \tilde{E}^q$ are referred to as the chiral-even GPDs,
while $H_T^q, E_T^q, \tilde{H}_T^q, \tilde{E}_T^q$ are referred to as the chiral-odd GPDs
\cite{Collins:1996fb,Diehl:2003ny,Diehl:2001pm}.
The chiral properties of the operators determine the type of exclusive process in which 
the GPDs can be sampled (see Sec.~\ref{sec:observables}).
}

\subsection{$N \rightarrow \Delta$ transition GPDs}
\label{sec:transGPDdef}

Transition GPDs describe transitions from a ground-state nucleon to a resonating meson-nucleon system
and occur in a description of several classes of hard-exclusive reactions within the collinear factorization framework. 
The first studies are focused on the nucleon transitions to $\Delta$ and nucleon resonances with the lowest masses.
A possible more general description in terms of $N \to \pi N$ transition GPDs
\cite{Polyakov:1998sz,Polyakov:2006dd}
was sketched in \cite{Goeke:2001tz}.
In the threshold region, the $N \to \pi N$ GPDs 
are related to the nucleon GPDs through the soft-pion theorems, allowing the parameter-free predictions for the $N \to \pi N$ DVCS at the pion production threshold \cite{Guichon:2003ah}.

For the special case of the  
$ N \to$~$\Delta$ transition, a total of 16
twist-2
transition GPDs \cite{Belitsky:2005qn,Kroll:2022roq} can be defined.
Several equivalent definitions are available in the literature.  
Here, for the cases of vector and axial-vector operators,
Eqs.~(\ref{operator_vector}) and (\ref{operator_axial}),
we present the definitions of the set of 
the proton-to-$\Delta^{++}$
transition
GPDs of Ref.~\cite{Belitsky:2005qn} employed in the recent analysis 
\cite{Kroll:2022roq}. 
There are $4$ 
{vector} transition GPDs,
\begingroup\makeatletter\def\f@size{9}\check@mathfonts
 \begin{eqnarray}
&&      \int  \frac{d \tau}{2\pi} e^{i \tau (n \cdot P) x} \nonumber \\ &&
\langle \Delta^{++}(p_R,\lambda_R)|\bar{u}(-\tau n/2)\gamma^+ d(\tau n/2)|N^p(p,\lambda_N)\rangle
       \nonumber \\ &&
    = 
    \frac1{(n \cdot P)}\,
    \bar{ \cal U}_\delta(p_R,\lambda_R)\left\{ \frac{\Delta^\delta n^\mu - \Delta^\mu n^\delta}{M_N} \Big(\gamma_\mu G_1(x,\xi,t)   \right.\nonumber\\
     && \left. + \frac{P_\mu}{M_N} G_2(x,\xi,t) + \frac{\Delta_\mu}{M_N} G_3(x,\xi,t)\Big)  \right.
     \nonumber\\
     && 
     \left. + \frac{\Delta^+\Delta^\delta}{M_N^2} G_4(x,\xi,t)\right\} \gamma_5 u(p,\lambda_N);
   \label{eq:even-pD-GPDs}
 \end{eqnarray}
  \endgroup
 and  
 {another 4 axial-vector}
 ones,
  \begingroup\makeatletter\def\f@size{9}\check@mathfonts
  \begin{eqnarray}
  &&    \int  \frac{d \tau}{2\pi} e^{i \tau (n \cdot P) x  } \nonumber \\ &&
    \langle \Delta^{++}(p_R,\lambda_R)|\bar{u}(-\tau n/2)\gamma^+\gamma^5 d(\tau n /2)|N^p(p,\lambda_N)\rangle
      \nonumber\\
      && =
    \frac1{(n \cdot P)}\,
    \bar{\cal U}_\delta(p_R,\lambda_R)\left\{ \frac{\Delta^\delta n^\mu - \Delta^\mu n^\delta}{M_N} \Big(\gamma_\mu \tilde{G}_1(x,\xi,t)   \right.\nonumber\\
    && + \frac{P_\mu}{M_N} \tilde{G}_2(x,\xi,t)\Big)  + n^\delta \tilde{G}_3(x,\xi,t) 
    \nonumber\\ &&
   \left. 
    + \frac{\Delta^+\Delta^\delta}{M_N^2} \tilde{G}_4(x,\xi,t)\right\} u(p,\lambda_N).
 \label{eq:odd-pD-GPDs}
 \end{eqnarray}
 \endgroup
Here, 
$P$ is the average hadron momentum $P\equiv \frac{p+p_R}{2}$ 
{of the proton $|N^p(p,\lambda_N)\rangle$ and $\Delta^{++}$ 
$|\Delta^{++}(p_R,\lambda_R) \rangle$
states;}
and $\Delta \equiv  p_R-p$  is the momentum transfer;
$p (\lambda_N,M_N)$ and $p_R(\lambda_R,M_\Delta)$ denote the proton and the $\Delta(1232)$ momenta (helicities and masses),
  respectively. The quantity ${ \cal U}_\delta(p_R,\lambda_R)$ stands for the Rarita-Schwinger spinor describing the $\Delta$ particle, and $u(p,\lambda_N)$ is the nucleon's Dirac spinor.
The skewness parameter $\xi$ is defined by the ratio of light-cone plus components of $\Delta$ and
$P$:
$
  \xi = - \frac{\Delta^+}{2P^+}
$.
The Mandelstam variable $t$ is given by
\begin{equation}
  t = \Delta^2 = t_0 - \frac{\mathbf{\Delta}_\perp^2}{1-\xi^2},
\end{equation}
where $t_0$ is the minimal value of $-t$ implied by the positivity of the transverse squared momentum transfer $\mathbf{\Delta}_\perp^2$
\begin{equation}
  t_0 = - \frac{2\xi}{1-\xi^2}\Big[ (1+\xi)(M^2_\Delta-M^2_N) + 2\xi M_N^2\Big]\,.
\label{eq:t0}
\end{equation}
So far the $\Delta(1232)$ is considered to be a stable particle. Therefore, one can apply the usual time-reversal invariance arguments
to show that the $p-\Delta$ GPDs turn to be real-valued functions of the momentum fraction $x$, the skewness, and $t$
\cite{Belitsky:2005qn}.
The GPDs also depend on the factorization scale. For convenience, this dependence is suppressed here.

An alternative parametrization \cite{Pascalutsa:2006up} relying on a different set of spin-tensor structures involves $4$ 
{vector}
GPDs 
$h_{M,\,E,\,C,\,4}(x,\xi,t)$ 
and 
$4$  
{axial-vector}
GPDs 
$C_{1,\,2,\,3,\,4}(x,\xi,t)$; 
and is employed in the analysis of 
Ref.~\cite{Semenov-Tian-Shansky:2023bsy}.
The relation of this transition GPD set to that of the heritage parametrization of Ref.~\cite{Goeke:2001tz} 
is specified in 
Ref.~\cite{Pascalutsa:2006up}.

The set of 
$8$
{tensor proton-to-$\Delta^{++}$ transition}
GPDs
$G_{T1, \ldots, T8}(x,\xi,t)$
has been defined in \cite{Kroll:2022roq}:
\begingroup\makeatletter\def\f@size{9}\check@mathfonts
\begin{eqnarray}
&&     \int \frac{d \tau}{2\pi}\,e^{i \tau (n \cdot P)x } \nonumber \\ &&
\langle \Delta^{++}(p_R, \lambda_R)|\bar{u}(-\tau n/2)i\sigma^{+j}
     d(\tau n /2)|N^p (p,\lambda_N)\rangle  
        \nonumber\\
         && = \frac{1}{(n \cdot P)} \bar{\cal U}_\delta(p_R, \lambda_R)\left\{G_{T1} \frac{p^\delta}{M_N} i\sigma^{+j} \right. \nonumber\\ &&
     + G_{T2}\,p^\delta\, \frac{P^+\Delta^j -\Delta^+ P^j }{M_N^3}     + G_{T3}\,p^\delta\,\frac{\gamma^+\Delta^j -\Delta^+\gamma^j }{2M_N^2} \nonumber\\
     &&
     + G_{T4}\,p^\delta\,\frac{\gamma^+P^j - P^+\gamma^j}{M_N^2}   +  G_{T5}\, (n^\delta\gamma^j-\gamma^\delta n^j)  \nonumber\\
     && \left.+ G_{T6}\,\frac{n^\delta \Delta^j-\Delta^\delta n^j}{M_N}
                          \right\}\,  \gamma_5\, u(p,\lambda_N) \nonumber\\
   && +  \frac{1}{(n \cdot P)}\,
   \left\{ G_{T7}\, (\bar{ \cal U}^+(p_R,\lambda_R)\gamma^j
                                                - 
                                                \bar{\cal U}^j(p_R,\lambda_R)\gamma^+) \right. \nonumber\\
      && + \left.   G_{T8}\,  \frac{\bar{\cal U}^+(p_R,\lambda_R)\Delta^j-\bar{ \cal U}^j(p_R,\lambda_R)\Delta^+}{M_N}
                         \right\}\,  \gamma_5\, u(p,\lambda_N)\,.
  \label{eq:trans-GPDs}
  \end{eqnarray}
 \endgroup
Isospin symmetry relates the $p \to \Delta^{++}$ GPDs to those of other $\Delta(1232)$ states \cite{Belitsky:2005qn}:
\begin{equation}
    G^{ud}_{p\Delta^{++}} = -\frac{\sqrt{3}}{2} G^{uu-dd}_{p\Delta^+} = -\sqrt{3} G^{du}_{p\Delta^0},
\end{equation}
where the flavor content of the GPDs is indicated.

The definitions (\ref{eq:even-pD-GPDs}), (\ref{eq:odd-pD-GPDs}), and (\ref{eq:trans-GPDs}) also hold for
any other octet-decuplet transition GPDs. Moreover, the flavor symmetry relates the various octet-decuplet GPDs
to each other \cite{Belitsky:2005qn}.

Four {vector GPDs}
$G_{1}$ - $G_{4}$
can be related to the Jones-Scadron electromagnetic form factors for the $N \to \Delta$ transition \cite{Jones:1972ky, Goeke:2001tz}:
\begin{equation}
\begin{split}
	\int^{1}_{-1}dx G_{1}(x;\xi;t) &\propto G^{*}_{M}(t), ~ 
	\int^{1}_{-1}dx G_{2}(x;\xi;t) \propto G^{*}_{E}(t), \\
	\int^{1}_{-1}dx G_{3}(x;\xi;t) &\propto G^{*}_{C}(t), ~ 
	\int^{1}_{-1}dx G_{4}(x;\xi;t) = 0,
\end{split}
\label{eq:eff_unpol}
\end{equation}
with the magnetic dipole, electric quadrupole, and the Coulomb quadrupole form factors $G^{*}_{M,E,C}(t)$.
The other four ($\widetilde{G}_{1}$ - $\widetilde{G}_{4}$) are 
{axial-vector GPDs, which}
can be related to the Adler
form factors of the $N \to \Delta$ transition ($C^{A}(t)$) \cite{Goeke:2001tz, Adler:1968tw, Adler:1975mt}:
\begin{equation}
\begin{split}
	\int^{1}_{-1}dx \widetilde{G}_{1}(x;\xi;t) &\propto C^{A}_{5}(t), ~ 
	\int^{1}_{-1}dx \widetilde{G}_{2}(x;\xi;t) \propto C^{A}_{6}(t), \\
	\int^{1}_{-1}dx \widetilde{G}_{3}(x;\xi;t) &\propto C^{A}_{3}(t), ~ 
	\int^{1}_{-1}dx \widetilde{G}_{4}(x;\xi;t) \propto C^{A}_{4}(t).
\end{split}
\label{eq:eff_pol}
\end{equation}

\subsection{Second resonance region}
The extension of the transition GPD formalism to the second $\pi N$ resonance region requires introducing parametrizations for the twist-$2$ $N \to N^*$ transition GPDs for
the case of the vector bilinear and axial-vector quark operators on the light cone
for the isospin-$\frac{1}{2}$ resonances 
$P_{11}(1440)$,
$D_{13}(1520)$
and
$S_{11}(1535)$  
with  and
the spin-parity quantum numbers, respectively,
$J^P = \frac{1}{2}^+, \frac{3}{2}^-, \frac{1}{2}^-$.

The parametrizations of $N \to P_{11}(1440)$, $D_{13}(1520)$, $S_{11}(1535)$ transition GPDs 
introduced in Ref.~\cite{Semenov-Tian-Shansky:2023bsy} 
for the case of the vector bilinear quark operator
were designed with the sets of spin-tensor structures consistent with the parametrizations of corresponding transition electromagnetic form factors 
detailed in Ref.~\cite{Tiator:2011pw}.
Similarly, for the axial-vector  bilinear quark operator, the parametrization of Ref.~\cite{Semenov-Tian-Shansky:2023bsy} tends to ensure the simple connection 
with the FFs occurring in the parametrizations of the transition matrix elements of the isoscalar $A_0^\nu(0)$ and isovector $A_3^\nu(0)$axial currents:  
\begin{eqnarray}
A_{ \{0, \, 3 \} }^\nu(0) \equiv \frac{1}{2} \left[ \bar u(0) \gamma^\nu \gamma_5 u(0) \pm \bar d(0) \gamma^\nu \gamma_5 d(0) \right].
\label{Def_axial_currents}
\end{eqnarray}
This provides a simple form of the normalization constraint for the first Mellin moments from  determining the dominant $N \to N^*$ axial FFs
with the help of the PCAC relation
\begin{eqnarray}
\partial_\nu \, A^\nu_3 = -f_\pi m_\pi^2 \, \Pi_3,
\label{eq:pcac}
\end{eqnarray}
where the pion decay constant $f_\pi \simeq 92.4$~MeV; and $\Pi_3$ stands for the $\pi^0$ field operator. 

As an example, we present the case of GPDs for the $N \to P_{11}$ transition.
Due to the identical spin-parity, the parametrization  has a form  similar to the  nucleon GPD case, with two 
{vector}
and two 
{axial-vector}
twist-$2$ GPDs.
The matrix element of the vector bilinear quark operator along the light cone can be parameterized as:
\begin{eqnarray}
&& 
  \int  \frac{d \tau}{2\pi} e^{i \tau (n \cdot P) x  } \nonumber 
  \\ &&
\sum_{q}  e_q^2
\langle R(p_R, \lambda_R) | 
\bar q \left(-\tau n/2 \right) \gamma^+   q \left( \tau n/2 \right)  
| N(p, \lambda_N) \rangle
\nonumber 
  \\
&&=    
\frac{1}{ (n \cdot P)} \bar{R} \left(p_R, \lambda_R \right) \left\{
H_1^{pP_{11}}(x, \xi, {t} 
) \nonumber \right. \\ &&   
 \times \left( n^\nu - \frac{n \cdot \Delta}{\Delta^2} \Delta^\nu \right)
\gamma_\nu  \nonumber \\ && \left.
 +  
 H_2^{pP_{11}}(x, \xi, 
 {t}
 )   \frac{i \sigma_{\nu \kappa} n^\nu \Delta^\kappa}{(M_R + M_N)}  \right\} u\left(p, \lambda_N \right), 
 \label{eq:NP11gpdunpol}
\end{eqnarray}
where the conventional DVCS  combination is weighted by the quadratic quark charges $e_q$; 
$ \bar{R} \left(p_R, \lambda_R \right)$ denotes the Dirac spinor of 
the final state $P_{11}$; {and $M_R$ is the resonance mass parameter}.
This definition results in the following normalization of the first Mellin moment of GPDs to the proton ($N=p$) and neutron ($N=n$) to $P_{11}$ transition electromagnetic FFs:
\begin{eqnarray}
\int_{-1}^{1} dx \, H_{1, 2}^{pP_{11}}(x, \xi, 
{t}
) &=& F_{1, 2}^{pP_{11}} ({t}
) + \frac{2}{3} F_{1, 2}^{nP_{11}} 
({t}
).
\nonumber \\
\end{eqnarray}

The transition electromagnetic FFs 
$F_1^{NP_{11}}$ 
and 
$F_2^{NP_{11}}$ 
are defined from the matrix element 
\begin{eqnarray}
&&\langle R(p_R, \lambda_R) | J_\text{em}^\nu (0) | N(p, \lambda_N) \rangle \nonumber \\
&&= \bar{R} \left(p_R, \lambda_R \right)
\Gamma_{\gamma N P_{11}}^{\nu}(p_R, p)  u\left(p, \lambda_N \right),
\label{eq:NP11em1}
\end{eqnarray}
where  the vertex $\Gamma_{\gamma N P_{11}}^{\nu}$ is parameterized as~\cite{Tiator:2008kd}:
\begin{eqnarray}
\Gamma_{\gamma N P_{11}}^{\nu} (p_R, p)
&=& F_1^{NP_{11}} (\Delta^2) \, \left[ \gamma^\nu -  \frac{(\gamma \cdot \Delta) \Delta^\nu}{\Delta^2}  \right] \nonumber\\
&+& F_2^{NP_{11}} (\Delta^2) \, \frac{i \, \sigma^{\nu \kappa} \Delta_\kappa}{(M_R + M_N)}.
\label{eq:NP11em}
\end{eqnarray}
For the phenomenological applications, we rely on the empirical MAID2008 analysis for the proton and the MAID2007 analysis for the neutron, as detailed in Ref.~\cite{Tiator:2011pw}.

Furthermore, in the case of the axial-vector operator, the parametrization for the $N \to P_{11}$  transition reads
\begin{eqnarray}
&&
\int  \frac{d \tau}{2\pi} e^{i \tau (n \cdot P)x  } \nonumber 
  \\ &&
\sum_{q}  e_q^2
\langle R(p_R, \lambda_R) | 
\bar q \left(-\tau n/2 \right) \gamma^+  \gamma_5 q \left( \tau n/2 \right)  
| N(p, \lambda_N) \rangle
\nonumber 
  \\ &&
=  
\frac{1}{(n \cdot P)} \bar{R} \left(p_R, \lambda_R \right) \left\{
\tilde H_1^{pP_{11}}(x, \xi, 
{t}
) \;  \gamma \cdot n \gamma_5    \right.
\nonumber \\ && \left.
 +  \tilde H_2^{pP_{11}}(x, \xi, 
 {t}
 ) \;  \frac{\Delta \cdot n}{(M_R + M_N)} \gamma_5  \right\} u\left(p, \lambda_N\right).  
 \label{eq:NP11gpdpol} 
\end{eqnarray}

To work out the normalization, we need to consider the transition matrix elements of the isovector and isoscalar axial 
currents (\ref{Def_axial_currents}).
For the  isovector axial current $N \to P_{11}$  transition it reads
\begin{eqnarray}
&& \langle R(p_R, \lambda_R) | A_3^\nu (0) | N(p, \lambda_N) \rangle
\nonumber \\
&&= \bar{R} \left(p_R, \lambda_R \right)
 \bigg\{ G_A^{NP_{11}}(\Delta^2) \, \gamma^\nu \gamma_5 \nonumber \\
 && 
 + G_P^{NP_{11}}(\Delta^2) \, \frac{\Delta^\nu \gamma_5}{(M_R + M_N)}  \bigg\} \,  \frac{\tau_3}{2} u\left(p, \lambda_N \right), \quad \quad
 \label{eq:NP11isovax}
\end{eqnarray}
with $\tau_3$ the third isospin Pauli matrix.,
Here, $G_A^{NP_{11}}$ and $G_P^{NP_{11}}$ are the corresponding isovector axial transition FFs. 
And, similarly, for the  isoscalar axial current $N \to P_{11}$ transition:
\begin{eqnarray}
&&\langle R(p_R, \lambda_R) | A_0^\nu (0) | N(p, \lambda_N) \rangle
 \nonumber \\
&&= \bar{R} \left(p_R, \lambda_R \right)
 \bigg\{ G_{A, 0}^{NP_{11}}(\Delta^2) \, \gamma^\nu \gamma_5 \nonumber \\ &&
+ G_{P, 0}^{NP_{11}}(\Delta^2) \, \frac{\Delta^\nu \gamma_5}{(M_R + M_N)}  \bigg\} \,  \frac{1}{2} u\left(p, \lambda_N \right), \quad \quad
 \label{eq:NP11isosax}
\end{eqnarray}
with $G_{A, 0}^{NP_{11}}$ and $G_{P, 0}^{NP_{11}}$ the corresponding isoscalar axial transition FFs.

The relations between the first moments of the 
{axial-vector}
GPDs 
(\ref{eq:NP11gpdpol}) 
for a given quark flavor entering the definition in Eq.~(\ref{eq:NP11gpdpol}) 
and the axial transitions FFs defined in Eqs.~(\ref{eq:NP11isovax}) and (\ref{eq:NP11isosax}) are then obtained as:
\begin{eqnarray}
\int_{-1}^{1} dx  \tilde H_{1}^{u, pP_{11}}(x, \xi, 
{t}
) &=& \frac{1}{2} \left( G_{A}^{NP_{11}} + G_{A, 0}^{NP_{11}} \right) 
({t}),
\nonumber \\
\int_{-1}^{1} dx  \tilde H_{1}^{d, pP_{11}}(x, \xi, 
{t}
) &= &\frac{1}{2} \left( - G_{A}^{NP_{11}} + G_{A, 0}^{NP_{11}} \right) 
({t}),
\nonumber \\
\label{eq:NP11gpdpolsr}
\end{eqnarray}
and analogous relations for $\tilde H_2^{u, pP_{11}}$ and $\tilde H_2^{d, pP_{11}}$ in terms of $G_{P}^{NP_{11}}$ and $G_{P, 0}^{NP_{11}}$.

The physical normalization for the $N \to P_{11}$ transition isovector axial FF can be obtained by use of the PCAC relation (\ref{eq:pcac})
together with an effective $\pi N P_{11}$ vertex derived from the effective Lagrangian 
\begin{eqnarray}
{\cal L}_{\pi N P_{11}} =  \left( \frac{f_{\pi N P_{11}}}{m_\pi} \right) \bar R  \gamma^{\mu} \gamma_5 \tau_i N \left( \partial_\mu \Pi_i \right) + \mathrm{h.c.} ,
\label{eq:LpiNP11}
\end{eqnarray}
with $\pi N P_{11}$ coupling constant $f_{\pi N P_{11}}$.
This results in the   generalized Goldberger-Treiman relation
for the $N \to P_{11}$ isovector axial FF:
\begin{eqnarray}
G_A^{NP_{11}}(0) = \left( \frac{f_{\pi N P_{11}}}{m_\pi} \right) 2 f_\pi.
\end{eqnarray}
Furthermore, using the pion-pole dominance for the axial FF $G_P^{NP_{11}}$ at small values of $\Delta^2$
yields:
\begin{eqnarray}
G_P^{NP_{11}}(\Delta^2) \approx\frac{(M_R + M_N)^2}{- \Delta^2 + m_\pi^2}  G_A^{NP_{11}}(\Delta^2).
\end{eqnarray}

The isoscalar $N \to P_{11}$ axial FFs $G_{A, 0}^{NP_{11}}$ and $G_{P, 0}^{NP_{11}}$ are not known.
As a possible best guess 
one may use for $G_{A, 0}^{NP_{11}}$ the same quark model relation as for the nucleon isoscalar axial FF:
\begin{eqnarray}
G_{A, 0}^{NP_{11}}(\Delta^2) \approx \frac{3}{5} G_{A}^{NP_{11}}(\Delta^2),
\end{eqnarray}
and parameterize $G_A^{NP_{11}}$ by a dipole form as in the nucleon case:
\begin{eqnarray}
 G_A^{NP_{11}}(\Delta^2) = 1/(1 - \Delta^2 / M_A^2)^2,
 \end{eqnarray}
with dipole mass $M_A \simeq 1.0$~GeV.
Furthermore,  the isoscalar FF is set to zero
\begin{eqnarray}
G_{P, 0}^{NP_{11}} \approx 0,
\end{eqnarray}
in line with pion-pole dominance.
A possible cross-check for such estimates can be provided by calculating these FFs within dynamical
quark-diquark approaches, which have proven to provide a good understanding of the data for the vector FFs $F_{1, 2}^{NP_{11}}$ for the 
$N \to P_{11}$ transition~\cite{Burkert:2017djo}.

The cases of $N \to D_{13}(1520)$ and $N \to S_{11}(1535)$
in \cite{Semenov-Tian-Shansky:2023bsy}
are treated in a similar manner. The parametrization of the leading twist 
vector and axial-vector GPDs for $N \to D_{13}(1520)$ involves $4+4$
invariant GPDs, while $N \to  S_{11}(1535)$ involves $2+2$ invariant GPDs.
The phenomenological normalization for the first Mellin moment vector transition GPDs is provided by the
relation to the transition electromagnetic FFs available {\it e.g.} from the
empirical MAID2008 analysis for the proton and the
MAID2007 analysis for the neutron detailed in Ref.~\cite{Tiator:2011pw}.
The normalization of the axial vector can be established relying on the PCAC, following a line similar to the case of $N \to P_{11}(1440)$ described above. 

{
\subsection{Definition of resonance GPDs}
\label{subsec:definition_resonance}
The concept of ``resonance GPDs'' and similar structures can be formulated rigorously in the context of the complex analytic properties of hadronic scattering amplitudes (S-matrix theory) \cite{chew1961s,Eden:1966dnq}.
Hadronic scattering amplitudes are analytic functions of the invariant variables constructed
from the 4-momenta of the particles, which can be analytically continued outside the physical region and have a definite physical meaning there. Resonances appear as pole-type singularities in the unphysical region of a two- or three-particle scattering process. E.g. the $\Delta$ resonance
appears as a pole in the $\pi N \rightarrow \pi N$ scattering amplitude in the $I = 3/2$
total isospin channel, upon analytic continuation in the invariant mass $s_{\pi N} \equiv (p_\pi + p_N)^2$ to the unphysical sheet. The pole is located at $\sqrt{s_{\pi N}} = \sqrt{s_\Delta} \equiv
M^{\rm pole}_\Delta - (i/2) \Gamma_\Delta$, where $M^{\rm pole}_\Delta \approx$ 1210 MeV is the pole mass and $\Gamma_\Delta \approx$ 100 MeV the width \cite{ParticleDataGroup:2022pth}. Formally,
\begin{align}
\langle \pi N | S | \pi N \rangle &= \textrm{function}(s_{\pi N}, ...) 
\nonumber \\
&= \frac{c^\ast c}{s - s_\Delta} + \textrm{less singular terms},
\label{delta_pole}
\end{align}
where the vertex $c$ in the numerator can be interpreted as the matrix element for the
$\pi N \rightarrow \Delta$ transition at the pole, $c = \langle \Delta | \pi N\rangle$.

Matrix elements of operators in the resonance state can be defined in a similar way.
E.g., to obtain the transition matrix element of a local QCD operator $\mathcal{O}(x)$
between $N$ and $\Delta$ states, one considers the transition amplitude
$\langle \pi N | \mathcal{O}(0) | N \rangle$ as a function of $s_{\pi N}$ (and other
variables characterizing the transition, such as the momentum transfer $t$).
According to general principles of S-matrix theory, the same pole Eq.~(\ref{delta_pole})
appears in the final state of the transition amplitude, the residue factorizes as
\begin{align}
\langle \pi N | \mathcal{O}(0) | N \rangle &= \textrm{function}(s_{\pi N}, ...) 
\nonumber \\
&= \frac{c^\ast c_\mathcal{O}^\ast}{s - s_\Delta} + \textrm{less singular},
\label{delta_matrix_element}
\end{align}
and the vertex $c_\mathcal{O}$ at the pole can be interpreted as the transition matrix
element of the operator between the $\pi N$ and $\Delta$ state,
$c_{\mathcal{O}} = \langle \Delta | \mathcal{O}(0) | \pi N\rangle$ 
(see Fig.~\ref{fig:gpd_resonance}). 
%
%
\begin{figure}[t]
\centering
\includegraphics[width = 0.77\columnwidth]{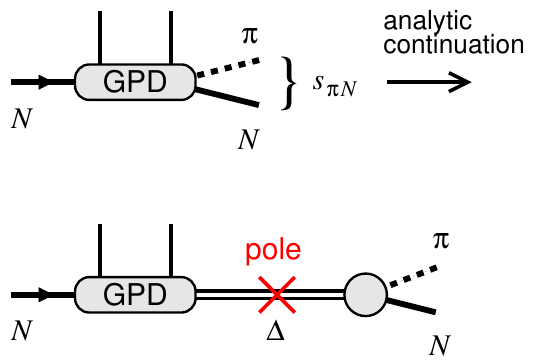}
\caption{{ Definition of the $\Delta$ resonance GPD as the pole term in the 
analytic continuation in the invariant mass $s_{\pi N}$ 
of the $N \rightarrow \pi N$ transition GPD}}
\label{fig:gpd_resonance}
\end{figure}

The method allows one to define and study the structure of the $\Delta$
resonance in analogy to that of a stable baryon; see Ref.~\cite{Pascalutsa:2006up}
for a review. Similar methods are used in the extraction of resonance structure
from Euclidean correlation functions in lattice QCD \cite{Briceno:2021xlc}.
Questions related to the analytic continuation in the invariant mass in the presence
of other variables (moving singularities, partial wave projection, etc.) have been
discussed for matrix elements of local operators and need to be studied further for
the extension to partonic operators and GPDs. An important theoretical simplification arises 
in the large-$N_c$ limit of QCD, where the $\Delta$ width is parametrically small,
and $\Delta$ structure can be discussed on the same basis as that of the nucleon
(see Sec.~\ref{subsec:ncexpansion}).
}

\section{Baryon resonance structure}
\label{sec:physics}

\subsection{3D tomography of baryon resonances}
\label{subsec:tomography}

GPDs provide a new set of tools for characterizing hadron structure in QCD.
Extending these concepts and methods from $N \rightarrow N$ to $N \rightarrow N^\ast$ transitions
and using them to resonance structure is an objective of the transition GPD program.

One major application of $N \rightarrow N$ GPDs is the 3D imaging of the quark/gluon structure of the nucleon.
In a frame where the momentum transfer between the nucleon states is in the transverse direction,
$\xi = 0$ and $\bm{\Delta}_T \neq 0$, the $N \rightarrow N$ GPD $H^q(x, \xi = 0, t = -\bm{\Delta}_T^2)$
($q$ denotes the quark flavor) can be represented as the Fourier transform of a transverse coordinate
distributions as (here $x > 0$),
\begin{align}
H^q (x, \xi = 0, t = -\bm{\Delta}_T^2)
= \int d^2 b \, e^{i \bm{\Delta}_T \cdot \bm{b}} \, f^q (x, \bm{b}),
\nonumber 
\\
-H^q (-x, \xi = 0, t = -\bm{\Delta}_T^2)
= \int d^2 b \, e^{i \bm{\Delta}_T\cdot \bm{b}} \, f^{\bar q} (x, \bm{b}).
\label{transverse_coordinate_representation}
\end{align}
The functions $f^q(x, \bm{b})$ and $f^{\bar q}(x, \bm{b})$ describe the distributions of quarks and antiquarks
with momentum fraction $x$ as a function of transverse position $\bm{b}$, in a nucleon state localized in transverse space at the
origin \cite{Burkardt:2000za,Burkardt:2002hr}. The representation provides a ``tomographic image'' of the nucleon's quark/antiquark structure
and reveals new information about the dynamical system, such as the average transverse radius of quarks/antiquarks
and the change of the transverse distribution with $x$.
Similar distributions describe the transverse
distribution gluons \cite{Kumericki:2009uq,Strikman:2003gz}, of polarized quarks and gluons (helicity, transversity), and the distortions of
the spatial distributions induced by transverse nucleon polarization (spin-orbit phenomena) \cite{Burkardt:2000za,Burkardt:2002hr,Burkardt:2005hp}.

Extending the transverse coordinate representation to $N \rightarrow N^\ast$ transition GPDs is a
subject of theoretical research. The definition of localized transverse coordinate states in light-front
quantization does not refer to the particle mass, so localized states can be defined for
$N \rightarrow N^\ast$ transitions in the same way as for $N \rightarrow N$.
The coordinate space densities associated with $N \rightarrow \Delta$ GPDs can be studied using methods based on the 
large-$N_c$ limit of QCD (see Sec.~\ref{subsec:ncexpansion}) \cite{Kim:2023yhp}.

A special case of the transverse coordinate representation is the transverse densities of electric charge
and magnetization in the nucleon \cite{Burkardt:2000za,Burkardt:2002hr,Miller:2007uy,Miller:2010nz}. The transverse charge density in the nucleon is given by the Fourier transform of
the $N \rightarrow N$ Dirac electromagnetic form factor,
\begin{align}
F_1 (t = -\bm{\Delta}_T^2) = \int d^2 b \, e^{i \bm{\Delta}_T \cdot \bm{b}} \, \rho_1 (\bm{b}).
\end{align}
The Dirac form factor is the integral of the GPD $H^q$ over $x$ (first moment), weighted with the quark charges
and summed over quark flavors 
\begin{align}
F_1 (t) = 
\sum_{q = u, d} e_q
\int_{-1}^1 dx H^q (x, \xi = 0, t),
\end{align}
and the transverse charge density is the integral of the difference of quark and antiquark distributions
\begin{align}
\rho_1 (\bm{b}) =  \sum_{q = u, d} e_q \left[ f^q (x, \bm{b}) - f^{\bar q} (x, \bm{b}) \right] .
\end{align}
In this way, the transverse charge density provides summary information on the transverse coordinate
distributions of quarks and antiquarks in the nucleon. Similarly, the transverse magnetization density
representing the Pauli form factor provides information on the transverse coordinate representation of the nucleon GPD $E$.
The transverse charge and magnetization densities can be extracted directly from the experimentally measured
electromagnetic form factors and provide interesting insight into nucleon structure; see Refs.~\cite{Venkat:2010by,  Miller:2011du,Granados:2013moa,Alarcon:2022adi}
for empirical and theoretical studies. The concepts can be generalized from $N \rightarrow N$ transitions to
$N \rightarrow N^\ast$ transitions. The $N \rightarrow N^\ast$ transition charge/magnetization densities
can be extracted directly from the transition 
form factors \cite{Wang:2024byt} and provide insight into the transverse spatial
structure of the $N \rightarrow N^\ast$ transition.

$N \rightarrow \Delta$ transition densities have been extracted from the $N \rightarrow \Delta$ transition
form factors \cite{Carlson:2007xd}. An interesting aspect is the appearance of a quadrupole deformation in the
transverse charge density due to the presence of a quadrupole form factor ($\Delta J = 2$)
in the $N \rightarrow \Delta$ transition.
The $N \rightarrow \Delta$ densities have been studied
theoretically in holographic and light-front quark models
\cite{Chakrabarti:2016lod}. The transverse densities in hyperon states have been studied using dispersion theory and EFT methods \cite{Alarcon:2017asr}.

The transverse coordinate representation Eq.~(\ref{transverse_coordinate_representation})
is defined at $\xi = 0$, where the transverse coordinate distributions represent proper particle densities
in the context of light-front quantization. The kinematics $\xi = 0$ is not accessible in experiment,
as actual scattering processes involve a longitudinal momentum transfer between the initial and final
baryon; this applies to $N \rightarrow N$ as well as $N \rightarrow N^\ast$ transitions.
It is generally not possible to extract $x$-dependent transverse distributions directly from the experiment.
Experimental information can be used to probe GPDs in other regions of the partonic variables.
The program of nucleon tomography with GPDs therefore relies essentially on theoretical models
predicting/constraining the behavior of GPDs. Lattice QCD calculations using the quasi/pseudo-distribution
formulation { can} 
direct information about the $x$-dependent GPDs at $\xi = 0$
for use in hadron tomography (see Sec.~\ref{subsec:lattice}).

%
%
\begin{figure}[t]
\centering
\includegraphics[width = 0.77\columnwidth]{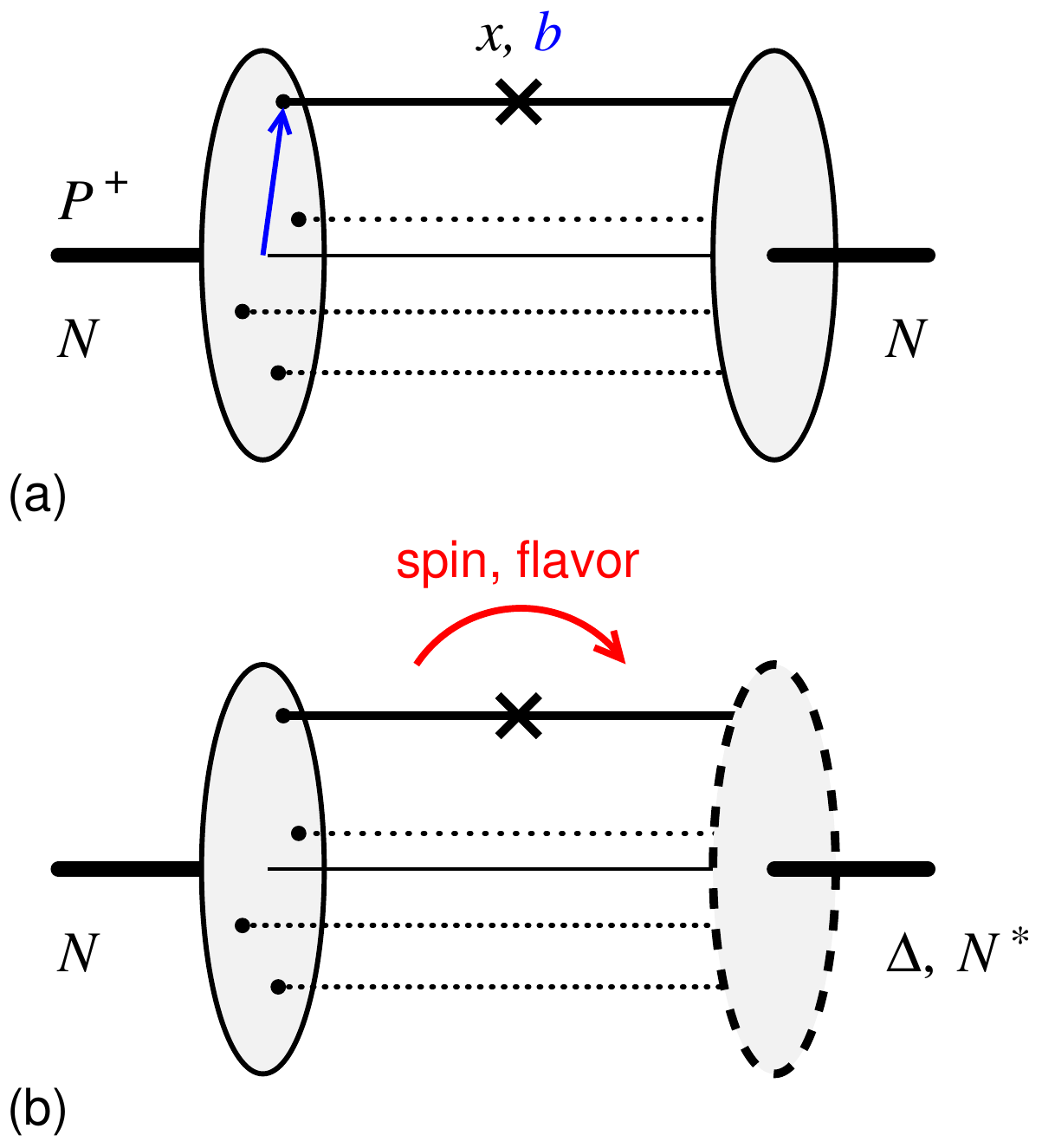}
\caption{{ Interpretation of transverse densities in light-front wave function
representation. (a) Conventional density $N \rightarrow N$. (b) Transition density $N \rightarrow \Delta, N^\ast$.
}}
\label{fig:density_wavefunction}
\end{figure}
{
The interpretation of the transverse densities associated with $N \rightarrow N^\ast$ transitions
can be developed within the light-front description of hadron structure at the wave function level \cite{Burkardt:2000za,Burkardt:2002hr}.
In this picture, the baryons are described by wave functions in quark/gluon degrees of freedom at fixed
light-front time $x^+ =$ const \cite{Brodsky:1997de}. The partonic QCD operators measure the number density of quarks
and gluons with a given light-cone momentum fraction $x$ at a given transverse position $\bm{b}$
(this interpretation appears in the transverse coordinate representation of the baryon matrix element
through the Fourier transform $\bm{\Delta}_\perp \rightarrow \bm{b}$). In the $N \rightarrow N$
case the matrix elements of the partonic operators describe conventional densities (expectation values),
with the same baryon state appearing in the initial and final state (see Fig.~\ref{fig:density_wavefunction}). In the $N \rightarrow N^\ast$ case the matrix elements describe transition densities, in which the operators are evaluated between different baryon initial and final states. It is important to note that in both cases the
partonic operators are diagonal in $x$ and $\bm{b}$. The difference between the diagonal and transition
densities is that in the latter the operator transfers quantum numbers (spin, flavor, orbital angular
momentum) which enable a transition to a different hadronic state. The light-front wave function
picture is used only for the interpretation of the transition densities; actual calculations
(renormalization, non-perturbative methods) or extraction from data are done at the level of the
second-quantized operator matrix elements described by the GPDs.
}

\subsection{Spin and angular momentum in baryon resonances}
Another major application of $N \rightarrow N$ GPDs is the characterization of the angular momentum
carried by the quark and gluon fields in the nucleon, including spin and orbital angular momentum.
This is	possible because the GPDs contain information on the nucleon matrix elements of the QCD
energy-momentum tensor (EMT), which enters into the definition of field-theoretical angular momentum in QCD.
Expanding the nonlocal QCD operator Eq.~(\ref{partonic_operator}) with
$\Gamma = \gamma^+ = n \cdot \gamma$ in powers of the light-like separation of the quark fields,	one obtains 
\begin{align}
&n_\alpha \, {\bar q} (-\tau n/2) [-\tau n/2, \tau n/2] \gamma^\alpha {q} (\tau n/2)
\nonumber \\[1ex]
&= n_\alpha \, {\bar q}  (0) \gamma^\alpha {q} (0)
+ \frac{\tau}{2} n_\alpha n_\beta \, {\bar q} (0) \gamma^\alpha \overleftrightarrow{\nabla}^\beta {q} (0)
+ \ldots
\end{align}
where
$\overleftrightarrow{\nabla}^{\beta} \equiv \frac{1}{2} (\overrightarrow{\partial}^\beta
- \overleftarrow{\partial}^{\mu}) - igA^{\beta}$ is the QCD covariant derivative. The local spin-2 operator in the second term is the quark part of the EMT ($q$ denotes the quark flavor),
\begin{align}
T_q^{\alpha\beta}(0) \equiv i {\bar q} (0) \gamma^\alpha \overleftrightarrow{\nabla} 
{q} (0).
\end{align}
The matrix elements of this operator between nucleon states can be obtained from the $x$-weighted integrals
(second moments) of the $N \rightarrow N$ GPDs.	The angular momentum of	quarks (and gluons) can	be computed
from its field-theoretical definition in terms of the EMT. As a	result,	the angular momentum of quarks and
gluons in the nucleon can be expressed as certain $x$-weighted integrals of the $N \rightarrow N$ GPDs
(spin sum rules) \cite{Ji:1996ek}. Several versions of these sum rules have been proposed, using different but equivalent definitions versions of the field-theoretical angular momentum;	the relation between them is now well understood \cite{Leader:2013jra}.
A field of study has emerged, exploring the angular momentum content of the nucleon based on GPDs using
theoretical methods \cite{Goeke:2001tz,Granados:2019zjw}, LQCD simulations \cite{Gockeler:2003jfa,LHPC:2007blg,LHPC:2010jcs,Bali:2018zgl,Alexandrou:2019ali}, 
and experimental data.

The extension of the notion of	QCD angular	momentum to $N \rightarrow N^\ast$ transitions is a subject
of ongoing theoretical research. Some principal questions need to be addressed	when generalizing the concept
of angular momentum to transitions between states with different masses and quantum numbers (spin, isospin).
The invariant formulation of the spin sum rule by Ji \cite{Ji:1996ek} uses specific properties of $1/2 \rightarrow 1/2$ transitions,
and the	Breit frame formulation	by Polyakov \cite{Polyakov:2002yz} assumes transitions between baryon states of equal mass.

%
%
\begin{figure}[t]
\centering
\includegraphics[width = 0.5\columnwidth]{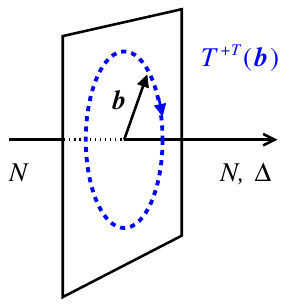}
\caption{Transition angular momentum derived from transition transverse density 
of energy-momentum tensor, 
Eq.~(\ref{J_momentum}), 
for $N \rightarrow \Delta$ transitions
\cite{Kim:2023xvw}.
}
\label{fig:density_am_ndelta}
\end{figure}
A framework that is well suited	for the generalization to $N \rightarrow N^\ast$ transitions is the formulation
of QCD angular momentum as a transverse density at fixed light-front time \cite{Adhikari:2016dir,Lorce:2017wkb,Granados:2019zjw}. A definition of $N \rightarrow N^\ast$
transition angular momentum based on this formulation has been proposed	in Ref.~\cite{Kim:2023xvw}.
One considers the transition matrix elements of the EMT between general baryon states with 
masses ${M}$ and ${M'}$ and 4-momenta $p$ and $p'$,
\begin{align}
\langle B'( p', \lambda' ) | \hat{T}^{\alpha\beta}_q (0) | B( p, \lambda) \rangle .
\label{T_matrix}
\end{align}
The baryons can have different spin $S' \neq S$. The spin states are described as light-front 
helicity states; the light-front helicities $\lambda$ and $\lambda'$ are directly related to 
the spin projection along the 3-axis in the baryon rest frame; $\lambda \leftrightarrow S_3, \lambda' \leftrightarrow S_3'$.
One analyzes the transition matrix element Eq.~(\ref{T_matrix}) in a class of frames where the momentum
transfer has $\Delta^+ = 0$ and the average momentum has $\bm{P}_T = 0$ (generalized Drell-Yan-West frame).
In these frames, the light-cone components of the baryon momenta are given by
(using the notation $p = [p^+,p^-,\bm{p}_T]$ with $p^\pm \equiv p^0 \pm p^3$)
\begin{align}
p = \left[ p^+, \frac{{M^2}  + |\bm{\Delta}_T|^2/4}{p^+}, -\frac{\bm{\Delta}_T}{2} \right] ,
\nonumber \\
p' = \left[ p^+, \frac{{M^{\prime 2}} + |\bm{\Delta}_T|^2/4}{p^+}, \frac{\bm{\Delta}_T}{2} \right] ,
\nonumber \\
\Delta = \left[ 0, \frac{{M^{\prime 2}} - {M^{2}}}{p^+}, \bm{\Delta}_T \right] .
\label{frame_lf}
\end{align}
The matrix element in Eq.~(\eqref{T_matrix}) thus becomes a function of $\Delta_{T}$,
with $t = - \bm{\Delta}_T^2 < 0$.
Taking the light-front component $+i$ of the EMT ($i$ = 1, 2) and performing the two-dimensional Fourier transform, one obtains the transverse coordinate density
\begin{align}
&T^{+i}_q (\bm{b}| S_3', S_3) \cr
&= \int \frac{d^2 \Delta_T}{(2\pi )^2} e^{-i\bm{\Delta}_T \cdot \bm{b}} \langle B'( p', \lambda' ) | \hat{T}^{+i}_q (0) |B (p, \lambda) \rangle .
\end{align}
The transition AM in the $z$-direction is then defined as
\begin{align}
& 2 S^z (S_3', S_3 ) \,
J_{B \rightarrow B'}^q
\nonumber \\
& \equiv \frac{1}{2 p^+} \int d^2 b \; \left[ \bm{b} \times \bm{T}^{+T}_q (\bm{b}|S_3', S_3) \right]^z ,
\label{J_def}
\end{align}
where $S^z (S_3', S_3)$ accounts for the kinematic spin dependence and $J_{B \rightarrow B'}^q$ contains the dynamical information (reduced matrix element). This definition of the AM in the transverse coordinate representation has the form of a vector product of position and momentum and permits a simple mechanical interpretation (see Fig.~\ref{fig:density_am_ndelta}); its properties and the equivalence with other definitions were established for $N \rightarrow N$ transitions in Refs.~\cite{Adhikari:2016dir,Lorce:2017wkb,Granados:2019zjw}). 
In the momentum representation of the matrix element, the transition AM is expressed as
\begin{align}
& 2 S^z (S'_3, S_3 ) \, 
J_{B \rightarrow B'}^q = \cr
&\frac{1}{2p^+}\left[ -i \frac{\partial}{\partial \bm{\Delta}_T}
\times \langle B'( p', \lambda' ) | \hat{\bm{T}}^{+T}_q (0) | B( p, \lambda) \rangle \right]^z_{\bm{\Delta}_T = 0}. \cr &
\label{J_momentum}
\end{align}
The kinematic spin vector is defined as (for transitions between baryon states with $|S' - S| = 0, 1$)
\begin{align}
&S^z (S'_{3}, S_{3} ) = \sqrt{S(S+1)} \, \sqrt{\frac{2S + 1}{2S' + 1}} \,
\langle S S_{3}, 1 0 | S' S'_{3}\rangle ,
\label{spin_vector_transition}
\end{align} 
where $\langle S S_{3}, 1 0 | S' S'_{3}\rangle$ are the SU(2) Clebsch-Gordan coefficients.
Equations~(\ref{J_def}) and (\ref{J_momentum}) provide a general definition of the QCD AM
associated with baryon transitions. Its implications should be explored in further research.

The transition AM definition Eq.~(\ref{J_def}) has been used to study the AM in $N \rightarrow \Delta$ transitions \cite{Kim:2023xvw}. 
Because of the isospin difference between the states, these transitions probe the isovector
component of the quark AM ($u - d$). The $N \rightarrow \Delta$ transition matrix elements of the EMT
can be analyzed using the $1/N_c$ of QCD, where $N$ and $\Delta$ states appear in the same representation
of the emergent SU(2$N_f$) spin-flavor symmetry and transitions between them are connected by the
symmetry (see Sec.~\ref{subsec:ncexpansion}). The main results of this analysis are:

(i) The isovector AM in the nucleon is leading in $1/N_c$; the isoscalar is subleading:
\begin{align}
J^{u+d}_{N\to N}  = \mathcal{O}(N_c^0),
\hspace{1em}
J^{u-d}_{p\to p}  = \mathcal{O}(N_c).
\label{jv_nc}
\end{align}
This explains the observed large flavor asymmetry of the quark AM in the nucleon. Note that this 
scaling is consistent with that of the quark spin contribution to the nucleon spin as given by 
the axial coupling, $g_A^{u+d} = \mathcal{O}(N_c^0)$ and $g_A^{u-d} = \mathcal{O}(N_c^1)$.

(ii) The isoscalar component of the AM in the nucleon and $\Delta$ are related by
\begin{align}
J^{{u+d}}_{N\to N} = J^{{u+d}}_{\Delta \to \Delta}  .
\end{align}
This provides insight into the spin structure of $\Delta$ resonance. Note that this relation
is consistent with the spin sum rule for the $\Delta$ state. 

(iii) The isovector AM in the nucleon, the AM in the $N \rightarrow \Delta$ transitions, and the
isovector AM in the $\Delta$ are related by
\begin{align}
J^{u-d}_{p\to p} = \frac{1}{\sqrt{2}} J^{{u-d}}_{p\to \Delta^+}
= 5 J^{u-d}_{\Delta^+ \to \Delta^+}.
\label{largenc_isovector}
\end{align}
This suggests that the $N \rightarrow \Delta$ transition AM is large and provides a way to probe the
isovector nucleon AM with $N \rightarrow \Delta$ transition measurements.

%
%
\begin{table}[t]
\setlength{\tabcolsep}{4pt}
\renewcommand{\arraystretch}{1.0}
\begin{tabular}{c|cc|ccc} 
\hline
\hline
Lattice QCD & $J^{u+d}_{p\to p}$ & $J^{u+d}_{\Delta^{+}\to \Delta^{+}}$ & $J^{u-d}_{p\to p}$ & $J^{u-d}_{p\to \Delta^{+}}$ &
$J^{u-d}_{\Delta^{+}\to \Delta^{+}}$   \\
\hline
  \cite{Gockeler:2003jfa}~$\mu^{2}=4 \, \mathrm{GeV}^{2}$ 
  & $0.33^\ast$ & $0.33$ & $0.41^\ast$ & $0.58$ & $0.08$ \\
  \cite{LHPC:2007blg}~$\mu^{2}=4 \, \mathrm{GeV}^{2}$
  & $0.21^\ast$ & $0.21$ & $0.22^\ast$ & $0.30$ & $0.04$ \\
  \cite{LHPC:2010jcs}~$\mu^{2}=4 \, \mathrm{GeV}^{2}$
  & $0.24^\ast$ & $0.24$ & $0.23^\ast$ & $0.33$ & $0.05$ \\
  \cite{Bali:2018zgl}~$\mu^{2}=1 \, \mathrm{GeV}^{2}$
  & $-$ & $-$ & $0.23^\ast$ & $0.33$ & $0.05$ \\
  \cite{Alexandrou:2019ali}~$\mu^{2}=4 \, \mathrm{GeV}^{2}$
  & $-$ & $-$ & $0.17^\ast$ & $0.24$ & $0.03$ \\
\hline 
\hline
\end{tabular}
\caption{Estimates of the isoscalar and the isovector AM for $p\to p$, $p\to\Delta^{+}$ and $\Delta^{+}\to\Delta^{+}$
obtained from lattice QCD data on $J^{u+d}_{p\to p}$ and $J^{u-d}_{p\to p}$ and the relations provided by the
leading-order $1/N_{c}$ expansion. Input values are marked by an asterisk~$^\ast$.} 
\label{tab:lattice}
\end{table}
Furthermore, within the $1/N_c$ expansion, one can obtain numerical estimates of the $N \rightarrow \Delta$
transition AM, using Eq.~(\ref{largenc_isovector}) and lattice QCD results for the isovector EMT 
$N \rightarrow N$ matrix elements \cite{Kim:2023xvw}. Results are summarized in Table~\ref{tab:lattice}.
The dominance of the isovector component of the quark angular momentum in $N \rightarrow N$
transitions were also confirmed by a calculation using the light-cone wave function of the
chiral soliton in the large-$N_c$ limit \cite{Kim:2023yhp}.

The $N \rightarrow \Delta$ transition matrix element of the EMT can be connected with the $N \rightarrow \Delta$ transition GPDs of the 
parity-even quark operator, Eq.~(\ref{eq:even-pD-GPDs})
\cite{Goeke:2001tz,Kim:2023xvw}. 
In the context of the $1/N_c$ expansion, the dominant $N \rightarrow \Delta$ transition GPD is the GPD $G_1$%
\footnote{In the
$h_{M,E,C,4}$ convention employed in 
Ref.~\cite{Semenov-Tian-Shansky:2023bsy}
the dominant  in $1/N_c$  
{vector}
$N \to \Delta$ GPD is $h_M$, which in leading order in $1/N_c$ corresponds to $\frac{4}{3} G_1$.},
which scales as
\begin{align}
G_1(x,\xi,t) &\sim N_c^3 \times \mathrm{function}(N_{c}x,N_{c}\xi,t).
\label{h_M_ncscaling}
\end{align}
In leading order of $1/N_c$, one obtains
\begin{align}
\int^{1}_{-1} dx \, x \, 
G_1
(x,\xi,0) &=
\frac{8}{3} 
J^{u-d}_{p\to \Delta^{+}} .
\label{H_M_AM}
\end{align}
The $N_c$ scaling of the second moment implied by 
Eq.~(\ref{h_M_ncscaling}) agrees with Eq.~(\ref{jv_nc}).

{
It should be stressed that the concept of transition angular momentum
(and the other properties of $N\rightarrow \Delta$ transition GPDs discussed in 
Sec.~\ref{sec:physics}) are objectively defined and can be discussed and interpreted
without reference to the large-$N_c$ limit \cite{Kim:2023xvw}.
The large-$N_c$ limit provides powerful theoretical constraints by connecting
$N$ and $\Delta$ through the dynamical spin-flavor symmetry (see Sec.~\ref{subsec:ncexpansion}). In this context a measurement or calculation
of an isovector $N \rightarrow N$ structure immediately constrains the
corresponding $N \rightarrow \Delta$ transition structure, and vice versa,
so that the two are not independent.
}
\subsection{Mechanical properties of baryon resonances}

A further major application of $N \rightarrow \Delta$ GPDs is the study of the so-called ``mechanical properties''
of the nucleon. The components of the EMT with 3-dimensional tensor and scalar character describe the
distributions of mass/energy and forces/pressure in a mechanical system \cite{Polyakov:2002yz,Polyakov:2018zvc,Lorce:2018egm,Burkert:2023wzr}. The nucleon matrix elements
of these EMT components	can again be connected with the $x$-weighted integrals of the GPDs (second moments).
The	second moment of the $N \rightarrow N$ GPD $H$ contains two EMT form factors\footnote{Because the EMT
describes the coupling of matter to gravity, these form factors are also known as the ``gravitational
form factors'' of the nucleon.}
\begin{equation}
	\int dx~x~H(x, \xi, t) = M_{2}(t) + \frac{4}{5} \xi^{2} d_{1}(t).
\end{equation}
$M_2(t)$ describes the distribution of mass/energy in the nucleon; $d_1(t)$ describes the distribution
of shear forces and pressure. An interpretation	of these structures has	been developed in terms	of
3-dimensional spatial densities	in the Breit frame ($\Delta^0 = 0$). Determining these form factors
and spatial distributions using theoretical and experimental methods data has become a major goal of
nucleon structure physics.

The extension of these concepts	to $N \rightarrow N^\ast$ transitions would provide new	tools for
resonance structure physics. The distributions of mass/energy and pressure/forces would allow for
new assessments of the size and spatial structure of resonances, correlated with their mass and spin.
One present limitation is that the interpretation of the EMT form factors in terms of spatial densities
uses 3D densities in the Breit frame and cannot readily be	generalized to transitions between states
with different masses. This could be overcome by using the framework of transverse densities at fixed
light-front time, as in	the definition of the transition AM, Eq.~(\ref{J_def}) \cite{Kim:2023xvw}.

In the large-$N_c$ limit, the baryon masses are $\mathcal{O}(N_c)$ and the baryon motion becomes effectively
non-relativistic so that the baryon mass does not play a role in transition matrix elements. In this situation,
the $N \rightarrow N^\ast$ transition EMT form factors (and even the $N^\ast \rightarrow N^\ast$ ones) permit
an interpretation in terms of Breit frame densities in the same	way as the $N^\ast \rightarrow N$ form factors.

The $d_1(t)$ form factor 
(the first coefficient of the Gegenbauer expansion of the so-called $D$-term \cite{Polyakov:1999gs})
is of special interest	because	it makes a determining contribution to the subtraction constant of the fixed-$t$ dispersion
relation for the DVCS amplitude within the	leading-twist approximation 
\cite{Goeke:2001tz}. As	such it	can be
extracted from the experimental data on the	beam spin asymmetry	(imaginary part of DVCS amplitude) and
cross-section (real part) without the need for a GPD parametrization. Several extractions have been performed
using existing experimental data, in particular	from JLab CLAS 6 GeV \cite{Burkert:2018bqq,Kumericki:2019,Moutarde:2019tqa}.
It would be interesting	to explore whether the DVCS dispersion relations could be generalized to
$N \rightarrow \Delta$ transitions, and	whether	there would be a connection with an $N \rightarrow \Delta$ D-term.
It should be noted that the amplitude of $N \rightarrow \Delta$ DVCS is suppressed in the high-energy limit
(non-diffractive process, quantum number exchange) so the need for subtraction in	a dispersion relation
is not apparent. It is also known that the isovector D-term is suppressed in the $1/N_c$ expansion \cite{Goeke:2001tz}. In $N \rightarrow N^\ast$ transitions to states with the same quantum numbers as the nucleon, the high-energy behavior of DVCS is the same as in $N \rightarrow N$, and a transition D-term might appear in the dispersion relation.


\subsection{Tensor charge and anomalous magnetic moment}
\label{subsec:tensor_charge}
{
The $N \rightarrow N$ tensor GPDs Eq.~(\ref{eqn:gpd-t}) give access to fundamental chiral-odd characteristics of the nucleon. The first moments of the tensor GPDs reproduce the form factors of the local tensor QCD operator $\bar q (0) \sigma^{\mu\nu} q(0)$. 
The first moments of $H_{T}$, $\bar E_{T}$, and $\tilde{H}_{T}$ are related to the tensor charge $\delta_{T}^{q}$~(monopole), the anomalous tensor magnetic moment $\kappa_{T}^{q}$~(dipole), and the quadrupole tensor moment $Q_{T}^{q}$~(quadrupole)~\cite{Diehl:2005jf, Burkardt:2005hp, Burkardt:2006ev, Ahmad:2008hp}: 
\begin{subequations}
\begin{align}
    \delta_{T}^{q} &= \int dx \, H^{q}_{T}(x, \xi = 0, t=0), \\	
    \kappa_{T}^{q} &= \int dx \, \bar E^{q}_{T}(x, \xi = 0, t=0), \label{eqn:tensorchargeGPD} \\
    Q_{T}^{q} &= \int dx \, 2\tilde{H}^{q}_{T}(x, \xi = 0, t=0);
\end{align}
\end{subequations}
the first moment of $\tilde{E}_{T}$ is zero due to time reversal invariance.
The possibility of obtaining information on these fundamental chiral-odd characteristics of the nucleon from GPD measurements in exclusive processes
is of great interest for nucleon structure. The tensor GPDs also describe 
the effect of transverse quark polarization on the
transverse spatial distribution of quarks in the nucleon, Eq.~(\ref{transverse_coordinate_representation}). This can be seen most directly by performing a multipole expansion of the tensor GPDs in the transverse
momentum transfer $\bm{\Delta}_\perp$ \cite{Kim:2024ibz}. The dipole and quadrupole 
tensor GPDs quantify the deviation of the transversely polarized quark distribution inside the nucleon from the spherical shape of the quark distribution. That an orbital quadrupole structure appears at leading-twist level is unique to the chiral-odd GPDs
and adds to the richness of transverse nucleon structure.

Both uses of the tensor GPDs can be extended to $N \rightarrow \Delta$ 
and $N \rightarrow N^\ast$ transitions. The first moments of the tensor
$N \rightarrow \Delta$ transition GPDs describe the transition form factors
of the local tensor operator and could provide the first information on
resonance excitation with a chiral-odd QCD operator. The transverse coordinate
representation of the $N \rightarrow \Delta$ transition GPDs reveals the
effect of quark transverse polarization on the overlap of the $N$ and $\Delta$ partonic wave functions (see Fig.~\ref{fig:density_wavefunction}). First studies of the
chiral-odd $N \rightarrow \Delta$ transition GPDs 
have been performed using the $1/N_c$ expansion (see Sec.~\ref{subsec:ncexpansion})
\cite{Schweitzer:2016jmd,Kroll:2022roq,Kim:2024ibz}. First experimental results
are available from exclusive pion production experiments (see Sec.~\ref{sec:current_results}).
}

\section{Theoretical methods for transition GPDs}
\label{sec:methods}

\subsection{Chiral dynamics in $N \rightarrow \pi N$ transitions}
\label{subsec:chiral_dynamics}
%
%
\begin{figure}[t]
\centering
\includegraphics[width = 0.77\columnwidth]{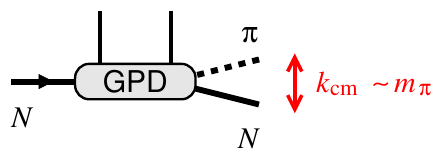}
\caption{$N \rightarrow \pi N$ transition GPD in the chiral regime}
\label{fig:gpd_chiral}
\end{figure}
The simplest case of transition GPDs are transitions between $N$ and $\pi N$ states in
the near-threshold region. If the center-of-mass momentum of the $\pi N$ system in the final state is ``small'' of the order $k \sim m_\pi$, the emission of the pion is governed by
chiral dynamics and can be computed from first principles (so-called soft-pion theorems) (see Fig.~\ref{fig:gpd_chiral}).
Chiral dynamics is the effective long-range dynamics emerging from the spontaneous
breaking of chiral symmetry in QCD, where the pion appears as a Goldstone boson,
the form of its	interaction with other hadrons is dictated by the chiral symmetry.
In particular, chiral symmetry implies that the QCD isovector axial current $J_{5V}^\mu$ produces low-momentum pions out of the vacuum and can be used as an interpolating QCD
operator for pion states. In nucleon transition matrix elements, this allows
one to derive relations between the $N \rightarrow \pi N$ transition matrix elements of
a QCD operator $\mathcal{O}$ and the $N \rightarrow N$ matrix elements of the
commutator $\mathcal{O}' = [\mathcal{O}, J_{5V}^\mu]$; schematically
\begin{align}
\langle	\pi N |	\mathcal{O} | N \rangle	\; \leftrightarrow \;
\langle N | \mathcal{O}' | N \rangle, \hspace{1em} \mathcal{O}' = [\mathcal{O}, J_{5V}^\mu]
\label{softpion}
\end{align}
These relations become predictive if the commutator can be reduced to an operator whose
matrix elements	are known, using the field equations. Application of this method to near-threshold $N \rightarrow \pi N$ transitions is demonstrated in Ref.~\cite{Pobylitsa:2001cz}. The $N \rightarrow \pi N$ transition GPDs measured in DVCS and meson production are considered in Refs.~\cite{Chen:2003jm,Guichon:2003ah,Birse:2005hh}.
Higher-order corrections in the pion momentum can be computed using chiral EFT \cite{Kivel:2004bb}. Applications to hard exclusive pion production with $N \rightarrow \pi N$ transitions are discussed in Ref.~\cite{Polyakov:2006dd}.
The soft-pion theorems are practically applicable in the S-wave of the $\pi N$ system.
In the P-wave, the strong $\Delta$ resonance limits the	useful range of the chiral expansion.
In this	channel, other methods can be applied, such as extensions of chiral EFT including
$\Delta$ degrees of freedom \cite{Alharazin:2023zzc,Alharazin:2023uhr}
or the $1/N_c$ expansion (see Sec.~\ref{subsec:ncexpansion}).

\subsection{$1/N_c$ expansion and $N\rightarrow \Delta$ transitions}
\label{subsec:ncexpansion}
\begin{figure}[t]
\centering
\includegraphics[width = 0.56\columnwidth]{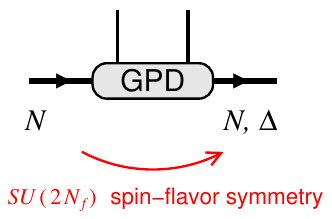}
\caption{$N \rightarrow N$ and $N\rightarrow \Delta$ 
transition GPDs in the large-$N_c$ limit}
\label{fig:gpd_largenc}
\end{figure}

The $1/N_c$ expansion is a general method for analyzing QCD in the nonperturbative domain and connecting it with meson and baryon properties. The large-$N_c$ limit corresponds to a semi-classical limit of the quantum field theory, in which dynamics simplifies in characteristic ways (it becomes ``string-like'') yet remains strongly coupled and generates meson/baryon spectra close to those observed in experiment
\cite{tHooft:1973alw,Witten:1979kh}. In the baryon sector, a dynamical spin-flavor symmetry emerges, whose ground-state representation consists of states
with equal spin and isospin, $S = I$ \cite{Dashen:1993jt,Dashen:1994qi,Jenkins:1998wy}. The $N$ and $\Delta$ appear as the states with $S = I = 1/2$ 
and $3/2$ in this representation.
Matrix elements of QCD operators between $N$ and $\Delta$ states
are therefore constrained by the spin-flavor symmetry (see Fig.~\ref{fig:gpd_largenc}). The transformation properties of QCD
operators under the emergent spin-flavor symmetry can be inferred from model-independent considerations
and are determined by their spin-flavor quantum numbers. The transition matrix between ground-state baryons is then connected by group-theoretical coefficients, similar to the matrix elements of spherical tensor operators between angular momentum states. Schematically,
\begin{align}
&\langle N (S_3', I_3') | \mathcal{O} | N(S_3, I_3) \rangle	
= 
C(\textstyle{\frac{1}{2}} \textstyle{\frac{1}{2}}| S_3'I_3', S_3 I_3) 
\langle \mathcal{O} \rangle,
\label{largenc_nn}
\\
&\langle \Delta (S_3', I_3') | \mathcal{O} | N(S_3, I_3) \rangle	
= 
C(\textstyle{\frac{3}{2}} \textstyle{\frac{1}{2}}| S_3'I_3', S_3 I_3) 
\langle \mathcal{O} \rangle,
\label{largenc_ndelta}
\\
&\langle \Delta (S_3', I_3') | \mathcal{O} | \Delta(S_3, I_3) \rangle	
= 
C(\textstyle{\frac{3}{2}} \textstyle{\frac{3}{2}}| S_3'I_3', S_3 I_3) 
\langle \mathcal{O} \rangle,
\label{largenc_deltadelta}
\end{align}
where $S_3/I_3$ denote the spin/isospin projections, and $C$ are group-theoretical 
coefficients given by products of SU(2) vector coupling coefficients, and $\langle \mathcal{O}\rangle$ is the reduced matrix element containing the dynamical information. 

The relations Eqs.~(\ref{largenc_nn})--(\ref{largenc_deltadelta}) can be used to derive
the $N \rightarrow \Delta$ and $\Delta \rightarrow \Delta$ transition matrix 
elements of QCD operators from the $N \rightarrow N$ matrix elements if the latter
are known (measured, calculated in Lattice QCD). The method has extensively been applied 
to matrix elements of local QCD operators such as vector and axial vector
currents, with very successful results \cite{Buchmann:2002mm,Jenkins:2002rj,Pascalutsa:2007wz}. It can also be applied to matrix elements
of nonlocal light-ray operators, to predict the $N\rightarrow \Delta$ transition
GPDs in terms of the $N\rightarrow N$ GPDs. Large-$N_c$ relations for the 
{vector and axial-vector}
transition GPDs are discussed in Ref.~\cite{Frankfurt:1999xe,Goeke:2001tz}
and applied to $N \rightarrow \Delta$ DVCS processes.
The extension to 
{tensor transition}
GPDs are discussed 
in Refs.~\cite{Schweitzer:2016jmd,Kroll:2022roq} and applied to hard exclusive 
pion production with $N \rightarrow \Delta$ transitions.

The $1/N_c$ expansion has other interesting applications to GPDs besides connecting $N$ and $\Delta$ transition matrix elements. It predicts the $N_c$-scaling of the reduced matrix element
$\langle \mathcal{O}\rangle$ is dependent on the spin-flavor quantum numbers of the operator and allows one to classify the matrix elements into leading and subleading ones. It also permits the calculation of $1/N_c$ corrections to the leading-order relations of Eqs.~(\ref{largenc_nn})--(\ref{largenc_deltadelta}). The $1/N_c$ expansion can also be combined with chiral dynamics 
to enable quantitative predictions. One formulation is given by effective field theories (EFTs) 
based on a combined chiral and $1/N_c$ expansion \cite{CalleCordon:2012xz,Fernando:2019upo}. Another formulation is the chiral 
quark-soliton model of baryons based on the dynamics of chiral constituent quarks \cite{Diakonov:1987ty}; 
this framework has been applied extensively to study 
GPDs and transition GPDs \cite{Petrov:1998kf,Goeke:2001tz}.

{
An important aspect of the $1/N_c$ expansion is that the width of the $\Delta$ resonance is
suppressed, so that its structure can be studied like that of a stable hadron. 
The $N$ and $\Delta$ masses scale as $M_{N, \Delta} = \mathcal{O}(N_c)$, and the mass splitting is
$M_{N, \Delta} = \mathcal{O}(N_c^{-1})$. In the standard $1/N_c$ expansion the pion mass is
$m_\pi = \mathcal{O}(N_c^0)$, and the $\Delta$ is stable. In the combined chiral and $1/N_c$ expansion 
with $m_\pi = \mathcal{O}(N_c^{-1})$, the $\Delta$ width is finite but parametrically small, $\Gamma_\Delta = \mathcal{O}(N_c^{-2})$
\cite{CalleCordon:2012xz,Fernando:2019upo}. The $1/N_c$ expansion can also be applied to
$N^\ast$ resonances in higher representations of the spin-flavor group; in this case
the mass splitting is $M_{N\ast} - M_N = \mathcal{O}(N_c^0)$
and the width is $\Gamma_{N\ast} = \mathcal{O}(N_c^0)$ \cite{Goity:2004pw,Goity:2004ss}.
}

\subsection{Light-cone sum rules}
\label{subsec:lightcone_sum_rules}
Helicity-conserving elastic and transition form factors at  very large momentum transfers  
can be calculated in QCD in terms of the light-cone distribution amplitudes (LCDAs)
that describe momentum fraction distributions of valence quarks at small transverse distances.
The problem is that the onset of the pQCD regime is postponed to very large $Q^2$ 
because the factorizable contribution involves a small factor $\sim (\alpha_s/(2\pi))^2$ and has to win over 
nonperturbative ``soft'' or ``end-point'' contributions that are suppressed by an extra power of $Q^2$ 
but do not involve small coefficients.
The light-cone sum rule (LCSR) technique \cite{Balitsky:1989ry} makes use of quark-hadron duality and 
dispersion relations to calculate such non-factorizable ``soft'' contributions
in terms of the same LCDAs that enter the pQCD calculation and avoid double counting. 
Thus, the LCSRs provide one with the most direct relation between the hadron form factors and LCDAs that
are available at present, with no other nonperturbative parameters. This approach was initially
developed to describe the weak decay $\Sigma \to p\gamma$ \cite{Balitsky:1989ry}. 
It is very flexible and has been applied to light- and 
heavy-meson decays (see, {\it e.g.}, \cite{Khodjamirian:2023wol} for a recent review), 
nucleon form factors~\cite{Braun:2006hz,Anikin:2013aka,Anikin:2016teg}, 
{
$N \rightarrow \Delta$ and $N^\ast$ transition form factors~\cite{Braun:2005be,Anikin:2015ita},}
and pion electroproduction near threshold \cite{Braun:2007pz}. The current state-of-the-art LCSRs 
have NLO accuracy. The LCSR technique proves to be very powerful, especially if
used in combination with lattice QCD inputs on the LCDAs, which are gradually becoming more and 
more accurate, see {\it e.g.}~\cite{RQCD:2019hps}.

{
LCSR methods could in principle be extended to compute the amplitudes of hard exclusive processes
and extract information on the GPDs. However, the analytic structure (spectral representation) of these amplitudes is much more complex than that of transition form factors and presents considerable challenges in this dispersive approach. Applications to GPDs and transition GPDs remain as a long-term prospect.}

\subsection{Lattice QCD calculations}
\label{subsec:lattice}

Aspects of GPDs have been investigated in lattice QCD for decades. Most of these calculations are based on the Mellin moments of GPDs, such as the electromagnetic and axial form factors of the proton, mainly because these are the most reliably extracted. Lattice calculations of these quantities have advanced significantly and now take into account sources of systematic uncertainties, such as physical pion mass calculations, excited states, infinite volume, and extrapolation to the continuum limit. Another crucial improvement is the inclusion of disconnected diagram contributions that enable the extraction of individual flavor quantities. Despite notable advancements, theoretical and computational constraints hinder the calculation of higher Mellin moments of GPDs, making the reconstruction of GPDs inherently challenging. Rather than relying solely on Mellin moments, alternative methods for determining the $x$-dependence of various distribution functions have emerged over time~\cite{Liu:1993cv, Ji:2013dva, Ji:2014gla, Detmold:2005gg, Braun:2007wv, Radyushkin:2017cyf, Orginos:2017kos, Chambers:2017dov, Ma:2017pxb}. In the past decade, significant progress has been made in extending these methods to calculate $x$-dependent GPDs~\cite{Alexandrou:2020zbe, Lin:2020rxa, Alexandrou:2021bbo, CSSMQCDSFUKQCD:2021lkf, Lin:2021brq, Bhattacharya:2022aob,Bhattacharya:2023ays,Bhattacharya:2023nmv,Bhattacharya:2023jsc}. Comprehensive reviews of recent progress in the field can be found in Refs~\cite{Cichy:2018mum, Ji:2020ect, Constantinou:2020pek, Cichy:2021lih}. 

{
Lattice QCD calculations can in particular provide information on the GPDs at zero skewness, $\xi = 0$, where they are needed for extracting the transverse spatial distribution of partons in the nucleon (see Sec.~\ref{subsec:tomography}).
As such they complement the information available from hard exclusive processes,
where the kinematics $\xi = 0$ is not directly accessible, and the GPDs are
sampled only at $x = \xi$ (imaginary part of exclusive amplitudes) or as
integrals over $x$ (real part) \cite{Goeke:2001tz,Diehl:2003ny,Belitsky:2005qn}.
Methods for the extraction of GPDs from combined lattice and experimental data
are being developed.

Lattice QCD methods can also be extended to compute transition matrix elements of 
QCD operators to multihadron states or resonances.
Calculations of meson and baryon transition form factors have been reported in
Refs.~\cite{Leinweber:1992pv,Aubin:2008qp,Alexandrou:2018jbt,ExtendedTwistedMass:2023hin}.
Methods for the treatment of excited hadron states in lattice QCD calculations have been developed,
using arrays of Euclidean correlation functions of operators projecting on ground and excited states
(generalized eigenvalue problem, distillation) \cite{Edwards:2011jj,xQCD:2019jke}; the extraction
of resonance signals in such calculations uses analyticity-based methods similar to those 
described in Sec.~\ref{subsec:definition_resonance}. 
Applications to resonance structure so far have been mostly 
in the meson sector \cite{Briceno:2016kkp,Sherman:2022tco}. 
Computations of transition matrix elements of low-spin local QCD operators
between nucleon and excited baryon states appear possible. The combination of excited-state 
and partonic-structure methods for computations of $x$-dependent transition GPDs appears 
very challenging and will require substantial further development.
}


\section{Transition processes: electron scattering}
\label{sec:observables}

Similar to the GPDs of the ground-state nucleon, transition GPDs are accessible in different lepton scattering reactions as well as hadronic reactions.
In leptoproduction experiments, transition GPDs can be accessed by the $N \to N^{*}$ deeply virtual Compton scattering (DVCS) and the $N \to N^{*}$ deeply virtual meson production (DVMP) processes.


\subsection{$N  \to \Delta,\, N^{*}$ DVCS and transition GPDs}
\label{sec:NtoNstarDVCS}

The non-diagonal DVCS reaction with a nucleon-to-resonance transition can be accessed through the study of the hard exclusive process
\begin{eqnarray}
 e^{-}(k)+N(p)  & \rightarrow & e^{-}\left(k^{\prime}\right)+\gamma\left(q^{\prime}\right)+R\left(p_R\right)  \nonumber \\
 & \rightarrow & e^{-}\left(k^{\prime}\right)+\gamma\left(q^{\prime}\right)+\pi\left(p_\pi\right)+N\left(p^{\prime}\right), \nonumber \\
\label{ND_hard_electroproducton_of_gamma}
\end{eqnarray}
in which the produced nucleon resonance $R$ further decays into a $\pi N$ system.
The first studies of $N \to \Delta(1232)$ non-diagonal DVCS were reported in Refs.~\cite{Guichon:2003ah,Guidal:2003ji} and a generalization for the second nucleon 
resonance region was recently considered in Ref.~\cite{Semenov-Tian-Shansky:2023bsy}.

\begin{figure}[t]
\includegraphics[width=\columnwidth]{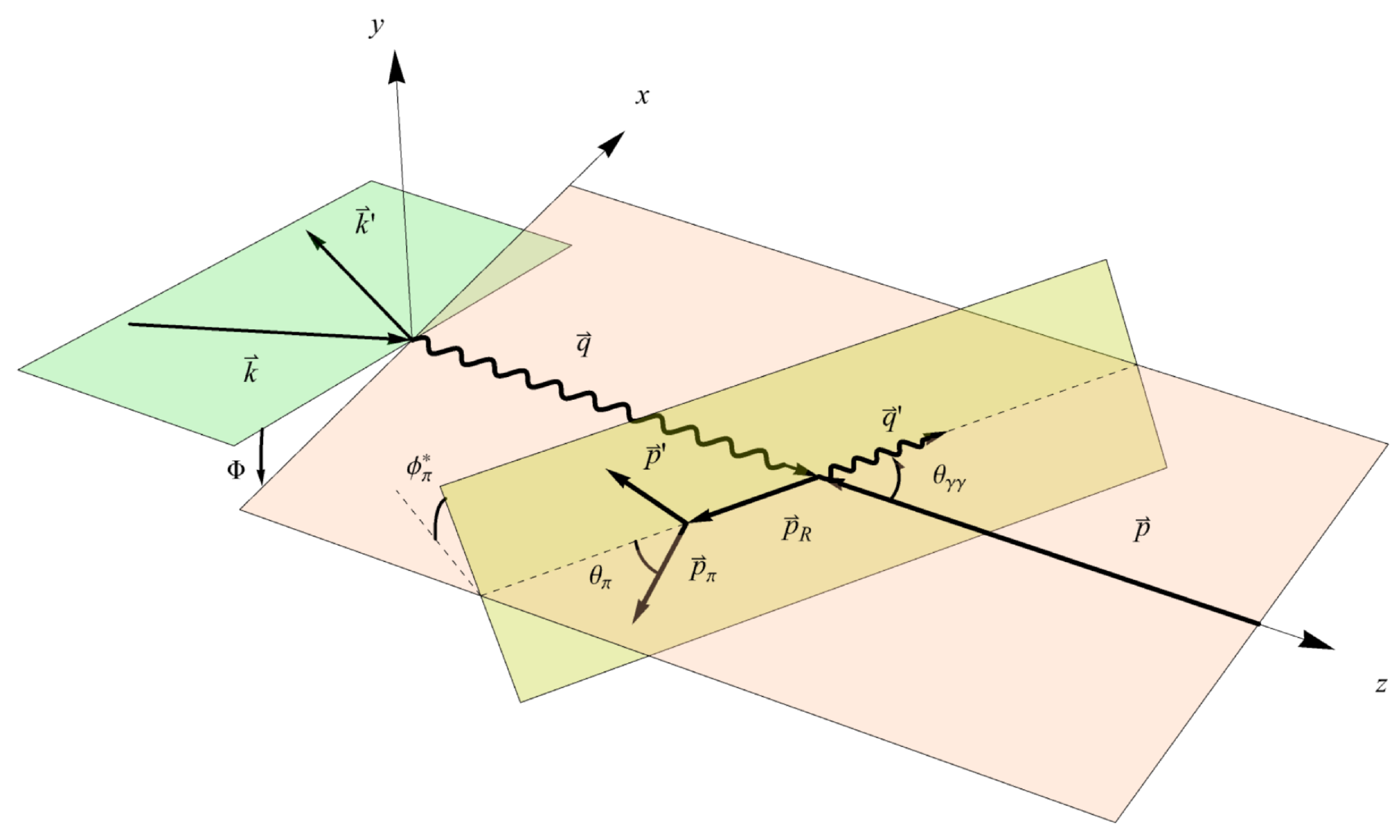}
\caption{Planes defining the scattering angles that characterize the $ e^- N \to e^- \gamma \pi N$  process. The angles $\Phi$ and $\phi^*_\pi$ are defined with respect to the $ xz$ plane, which is the scattering plane of the virtual photons with four-momenta $q$ and $q'$. This figure is taken from Ref.~\cite{Semenov-Tian-Shansky:2023bsy}.}
\label{fig:kin_plane}
\end{figure}

The process
(\ref{ND_hard_electroproducton_of_gamma}),
see Fig.~\ref{fig:kin_plane}
for the specification of kinematic quantities,
is described in terms of $8$ kinematic variables, which can be chosen as
\begin{equation}
\begin{aligned}
& s=(k+p)^2, \quad Q^2=-q^2 \equiv (k'-k)^2, \quad x_B=\frac{Q^2} {2 p \cdot q}, \\
& t=\Delta^2 \equiv (p_R-p)^2, \quad M_{\pi N}^2=p_R^2 \equiv (p'+p_\pi)^2, \\ & \Phi, \quad \theta_\pi^*, \quad \phi_\pi^*.
\end{aligned}
\end{equation}
Here $\Phi$ is the angle between the leptonic plane and the production plane spanned by the vectors
$\vec{q}$ and $\vec{q}'$
defined in the
$\gamma^*(q) N(p)$
Center-of-Mass System; 
and the angles
$\theta_\pi^*$, $\phi_\pi^*$
denote, respectively, the polar and azimuthal angles of the rest frame of the
$\pi(p_\pi) N(p')$
system.
%
%

Instead of the pion polar angle $\theta^*_\pi$, it is sometimes instructive to consider the invariant mass of the
$\pi \gamma$
system
$M_{\pi \gamma}^2=(p_\pi+q')^2$ as independent kinematical variable.

The sevenfold differential unpolarized cross section for the
$e^{-} N \rightarrow e^{-} \gamma \pi N$ reaction is expressed as
\begin{eqnarray}
&& \frac{d \sigma}{d Q^2 d x_B d t d \Phi d M_{\pi N}^2 d \Omega_\pi^*} 
=\frac{1}{(2 \pi)^7} \frac{x_B y^2}{32 Q^4 \sqrt{1+\frac{4 M_N^2 x_B^2}{Q^2}}}  \nonumber \\ &&
  \quad \times \frac{\left|\vec{p}_\pi^{\,*}\right|}{4 M_{\pi N}} \sum_i \sum_f\left|\mathcal{M}\left(e^{-} N \rightarrow e^{-} \gamma \pi N\right)\right|^2,
\label{Def_7fols_CS_NR}
\end{eqnarray}
where
$y \equiv p \cdot q / p \cdot k$;
$d\Omega_\pi^* \equiv d \cos \theta_\pi^* d \phi_\pi^*$
is the solid angle of the produced pion; and
$\left|\vec{p}_\pi^{\,*}\right|$
is the value of the pion $3$-momentum in the $\pi N$ rest frame.
The squared invariant amplitude
$\left|\mathcal{M}\left(e^{-} N \rightarrow e^{-} \gamma \pi N\right) \right|^2$
is averaged (summed) over the initial (final) particle helicities.

Another characteristic observable of the reaction
(\ref{ND_hard_electroproducton_of_gamma})
is the beam-spin asymmetry (BSA)  defined as
\begin{equation}
B S A=\frac{d \sigma^{+}-d \sigma^{-}}{d \sigma^{+}+d \sigma^{-}},
\label{Def_BSA_ND_DVCS}
\end{equation}
where
$d \sigma^{\pm}$
refer to the polarized cross sections with electron beam helicity~$\pm \frac{1}{2}$.

Analogously to the usual hard exclusive electroproduction of photons off a nucleon,
the amplitude of the process
(\ref{ND_hard_electroproducton_of_gamma}) obtains contributions from the Bethe-Heitler (BH) process, with
the final-state real photon emitted off the lepton lines; and the DVCS process, in which the photon is
produced from the hadron side. The BH contribution is a pure QED process, and the corresponding  amplitude
can be calculated exactly provided input on the nucleon-to-resonance transition electromagnetic
form factors.

\begin{figure}[t]
	\centering
		\includegraphics[width=0.40\textwidth]{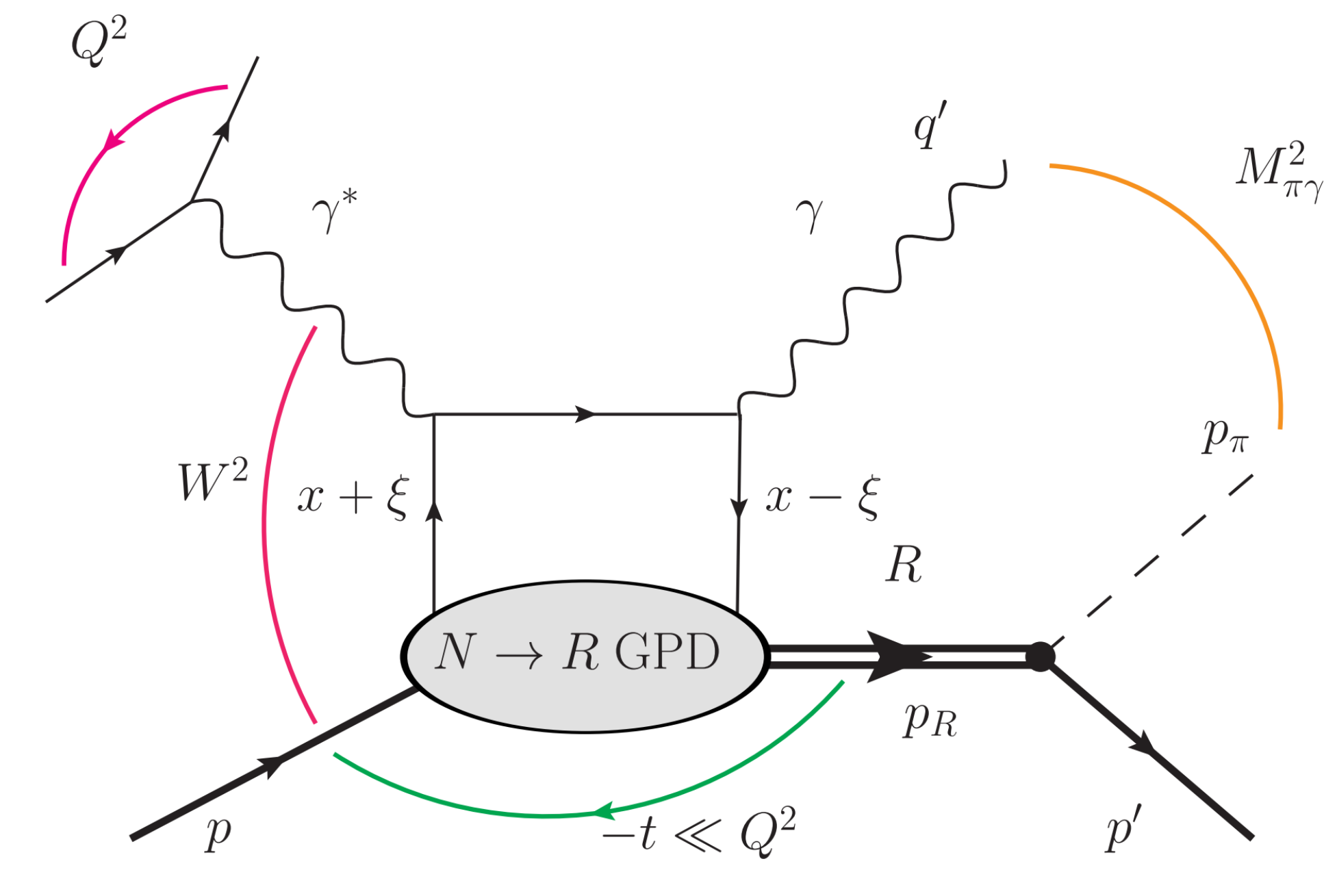}
	\caption{Handbag mechanism for the $N \to \Delta, \, N^{*}$ DVCS process. The diagram with the crossed real and virtual photon lines is not shown.
}
	\label{fig:NNstarDVCS_process}
\end{figure}

Similarly to the diagonal DVCS case, for which factorization is proven \cite{Ji:1998xh,Collins:1998be},
we assume the validity of the QCD collinear factorization theorem for the non-diagonal DVCS
in the generalized Bjorken kinematics
(large $Q^2$ and $W^2=(p+q)^2$; fixed $x_B$;  $-t$ and $W_{\pi N}^2$ of hadronic mass scale:
$-t, \, W_{\pi N}^2 \ll Q^2$).
Therefore, to the leading twist-$2$ accuracy, the reaction proceeds through the reaction mechanism of
Fig.~\ref{fig:NNstarDVCS_process};
and the corresponding amplitude is parameterized in terms of
$N \to R$ transition GPDs.

The interference between the BH and the non-diagonal DVCS leads to the interference term in the cross-section.
The BH process is usually expected to dominate the measured cross section for the kinematics accessible with current lepton scattering experiments
and has to be subtracted to extract the non-diagonal DVCS signal.

As a first example, we briefly review the description of the 
$e^- N \to e^- \gamma \Delta(1232) \to e^- \gamma \pi N$ process.
The corresponding invariant amplitude entering
the cross-section of Eq.~(\ref{Def_7fols_CS_NR}) is given by:
\begin{eqnarray}
&&\mathcal{M}(e^- N \to e^- \gamma \Delta \to e^- \gamma \pi N) = \bar C_{iso} \frac{f_{\pi N\Delta}}{m_\pi}
\left(- p_\pi\right)^\alpha
\nonumber \\
&&\times \, \bar{N} \left(p^\prime, s^\prime_N \right)
\frac{i P^{(3/2)}_{\alpha \beta}(p_R)}{M_{\pi N}^2 - M_\Delta^2 + i M_\Delta \Gamma_\Delta(M_{\pi N})}
\nonumber \\
&& \times \, \mathcal{M}_R^\beta(e^- N \to e^- \gamma \Delta),
\label{eq:NDelamplfull}
\end{eqnarray}
where 
$\mathcal{M}_R^\beta(e^- N \to e^- \gamma \Delta)$ 
is defined as the sum of the BH and DVCS amplitudes;
{
$M_\Delta$ is the resonance mass parameter, and 
$\Gamma_\Delta(M_{\pi N})$ stands for the energy-dependent 
$\Delta \to \pi N$ decay width
\begin{eqnarray}
&&
\Gamma_\Delta(M_{\pi N})\nonumber  \\ && = \frac{f_{\pi N\Delta}^2}{m_\pi^2}
\frac{1}{4 \pi}  \left( \frac{(M_{\pi N}+M_N)^2-m_\pi^2}{2 M_{\pi N}^2} \right) \frac{1}{3} \left|\vec{p}_\pi^{\,*}\right|^3,
\label{Gamma_Delta_energy_dependent}
\end{eqnarray}
with $\left|\vec{p}_\pi^{\,*}\right|$ denoting the absolute value of the decay pion $3$-momentum in $\pi N$ c.m. frame.
$f_{\pi N\Delta}$ is the $\pi N \Delta$ 
coupling constant of the effective $\pi N \Delta$
Lagrangian (see Eq.~(45) of 
Ref.~\cite{Semenov-Tian-Shansky:2023bsy}). From the PDG values 
\cite{ParticleDataGroup:2022pth}
$M_\Delta=1.232$~GeV;
$\Gamma_\Delta(M_\Delta) \simeq 0.117$~GeV we get
$f_{\pi N\Delta} \simeq 2.08$.}
The isospin factor 
$\bar C_{iso}$ 
takes on the values
$\bar C_{iso}=\sqrt{\frac{2}{3}}$ for $\Delta^+ \to \pi^0 p$;
and
$\bar C_{iso} =- \sqrt{\frac{1}{3}}$
for
$\Delta^+ \to \pi^+ n$.
Furthermore, in Eq.~(\ref{eq:NDelamplfull}),  
$P^{(3/2)}_{\alpha \beta}(p_R)$ denotes the
spin-$\frac{3}{2}$ projector, 
{\it cf.} Ref.~\cite{Pascalutsa:2006up}.

The amplitude 
(\ref{eq:NDelamplfull}) 
results in the specific angular distribution for the decay pion in the resonance rest frame:
\begin{eqnarray}
&&
\mathcal{M}(e^- N \to e^- \gamma \Delta \to e^- \gamma \pi N)
\nonumber \\ &&
= \bar C_{iso}
\frac{f_{\pi N\Delta}}{m_\pi}
 \sqrt{\frac{4 \pi}{3}}
 \,
\frac{i \, | \vec p^{\,*}_\pi | \, \left[(M_{\pi N} + M_N)^2 - m_\pi^2 \right]^{1/2}}{M_{\pi N}^2 - M_\Delta^2 + i M_\Delta \Gamma_\Delta(M_{\pi N})}
\nonumber \\ &&
\times \sum_{\lambda_R} \mathcal{M}(e^- N \to e^- \gamma \, \Delta (M_{\pi N}, \lambda_R))
\nonumber \\ &&
\times  \sum_{\lambda^\prime}
\, \langle \, 1 \lambda^\prime, \frac{1}{2} s_N^\prime  \, | \, \frac{3}{2} \lambda_R \, \rangle \, Y_{1 \lambda^\prime}(\Omega^\ast_\pi),
\label{eq:NDelamplfull2}
\end{eqnarray}
where
$Y_{l m_l}(\Omega^\ast_\pi)$
denote the spherical harmonic functions and 
$\langle \, s s_z, l m_l | J M \rangle$
are the Clebsch-Gordan coefficients for the 
$\Delta \to \pi N$ 
decay.

Eq.~(\ref{eq:NDelamplfull2}) 
is instrumental in computing the resonance angular distributions of the cross-section.
It is particularly instructive to present the cross-section
(\ref{Def_7fols_CS_NR})
integrated over the complete pion solid angle or integrated over the invariant mass of the 
$\pi N$ 
system 
$M_{\pi N}^2$.
 {Within the narrow resonance limit
$\Gamma_\Delta \ll M_\Delta$, assuming the Breit-Wigner form of the spectral function with the
energy-dependent decay width
$\Gamma_\Delta(M_{\pi N})$ of Eq.~(\ref{Gamma_Delta_energy_dependent}),
}
the unpolarized cross-section
(\ref{Def_7fols_CS_NR})
integrated both in $\Omega_\pi^*$ and $M_{\pi N}^2$
\begin{eqnarray}
&&
\int d M^2_{\pi N} \int  d \Omega^\ast_\pi
\frac{d \sigma}{d Q^2 d x_B d t d \Phi d M^2_{\pi N} d \Omega^*_\pi} \nonumber \\
&&\approx \frac{1}{(2 \pi)^4} \frac{x_B \, y^2}{32 \, Q^4 \sqrt{ 1 + \frac{4 M_N^2 x_B^2}{Q^2} }}\nonumber \\
&&\times
 \overline{\sum_i}\sum_f\left|\mathcal{M}(e^- N \to e^- \gamma \Delta(M_\Delta, \lambda_R)) \right|^2,
\label{eq:crossnarrow1}
\end{eqnarray}
where the sum runs over the final helicities of $e^-$, $\gamma$, and $\Delta$,
turns out to be fully consistent with the result for a stable particle~\cite{Goeke:2001tz}.

A very instructive piece of information can be revealed through working out the pion polar angular distribution integrated over the azimuthal angle $\phi_\pi^*$ within the narrow resonance approximation:
\begin{eqnarray}
&&
\int d M^2_{\pi N}
\frac{d \sigma}{d Q^2 d x_B d t d \Phi d M^2_{\pi N} d \cos \theta^*_\pi} \nonumber \\
&&\approx \frac{1}{(2 \pi)^4} \frac{x_B \, y^2}{32 \, Q^4 \sqrt{ 1 + \frac{4 M_N^2 x_B^2}{Q^2} }} \, \,
 \overline{\sum_{e, N}}\sum_{e, \gamma} (\bar C_{iso})^2 \nonumber \\
&&\times \bigg\{ \frac{1}{4} \left(1 + 3  \cos^2 \theta^\ast_\pi \right) \sum_{\lambda_R = \pm 1/2}
 \left|\mathcal{M}(e^- N \to e^- \gamma \Delta \right|^2
 \nonumber \\
&& \quad + \frac{3}{4} \sin^2 \theta^\ast_\pi  \sum_{\lambda_R = \pm 3/2}
 \left|\mathcal{M}(e^- N \to e^- \gamma \Delta \right|^2  \bigg\},
\nonumber \\
\label{eq:crossnarrow}
\end{eqnarray}
resulting in specific polar angular distributions for the $\Delta$ helicity states
$\lambda_R = \pm 1/2$ and $\lambda_R = \pm 3/2$. 
This gives rise to a peculiar ``autopolarization''
effect providing access to information on different polarization states of the produced $\Delta$-resonance. 
For a sufficiently detailed experimental angular resolution, this can open interesting options for
a precise partial wave analysis of $\Delta$-resonance production.

The cross-section
estimates of $N \to \Delta(1232)$ in Refs.~\cite{Guichon:2003ah,Guidal:2003ji} were based on a model for the $N \to \Delta$ transition GPDs dominant in the large-$N_c$
limit within the framework developed in
Ref.~\cite{Frankfurt:1999xe}.
The three dominant
{
proton-to-$\Delta^+$}
transition GPDs
$h_M$, $C_1$ and $C_2$
are related to the following combinations of nucleon isovector GPDs
\cite{Frankfurt:1999xe,Goeke:2001tz}:
\begin{equation}
\begin{aligned}
& h_M\left(x, \xi, 
{t}
\right)=\sqrt{2}\left[E^u\left(x, \xi, 
{t}
\right)-E^d\left(x, \xi, 
{t}
\right)\right]; \\
& C_1\left(x, \xi, 
{t}
\right)=\sqrt{3}\left[\tilde{H}^u\left(x, \xi, 
{t}
\right)-\tilde{H}^d\left(x, \xi, 
{t}
\right)\right]; \\
& C_2\left(x, \xi, 
{t}
\right)=\frac{\sqrt{3}}{4}\left[\tilde{E}^u\left(x, \xi, 
{t}
\right)-\tilde{E}^d\left(x, \xi, 
{t}
\right)\right].
\end{aligned}
\label{GPDs_N_to_Delta_Large_Nc}
\end{equation}
The control of the accuracy of this approximation is performed through a comparison of the first moments of
(\ref{GPDs_N_to_Delta_Large_Nc})
to values of the corresponding form factors, see discussion in Ref.~\cite{Pascalutsa:2006up}.

In Ref.~\cite{Semenov-Tian-Shansky:2023bsy} this model was employed to provide estimates of the cross-section for the
kinematical setup accessible with CLAS12@JLab. The phenomenological input included the nucleon GPD model of
Sec.~2.7.2 of Ref.~\cite{Pascalutsa:2006up}.
The GPD $\tilde{E}$ is determined by the pion pole contribution; see Sec.~2.4.2 of Ref.~\cite{Goeke:2001tz}.
For the relevant transition electromagnetic form factors, the results of the MAID2007 analysis reported in
\cite{Drechsel:2007if,Tiator:2011pw} were employed.

\begin{figure*}[t]
 \centering
\includegraphics[width=0.3\textwidth]{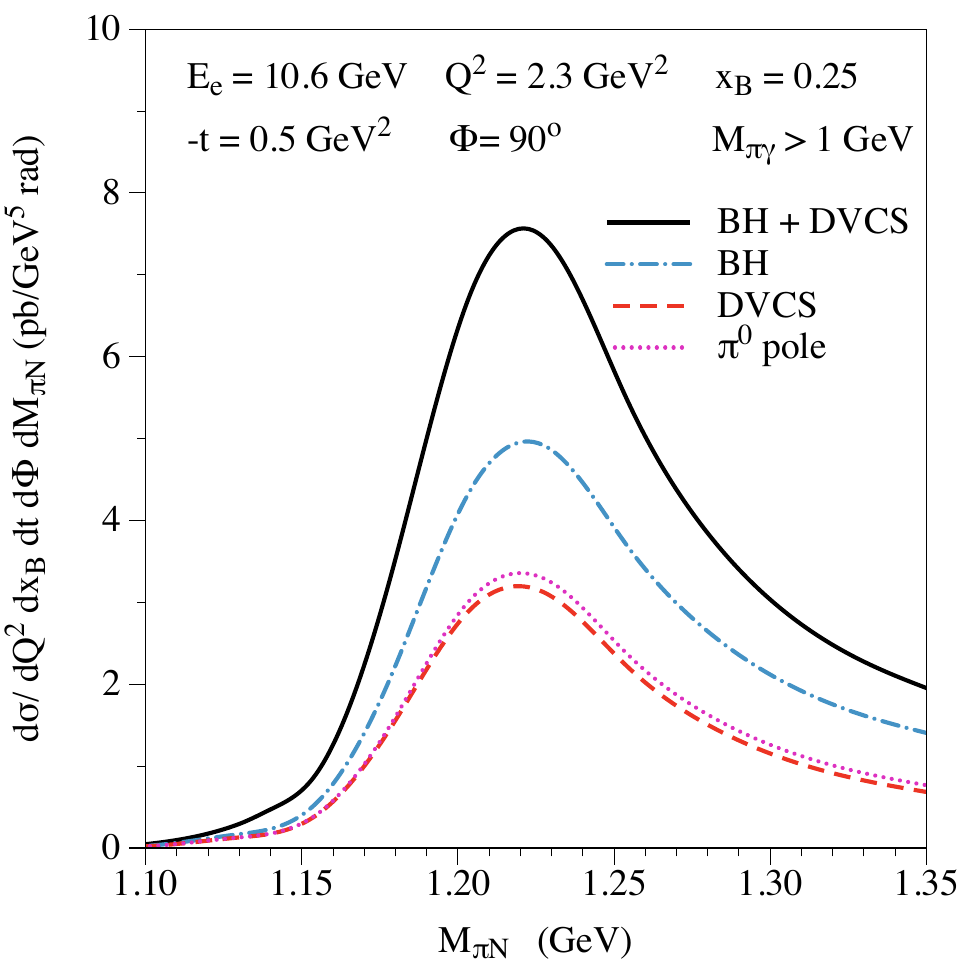}
\includegraphics[width=0.3\textwidth]{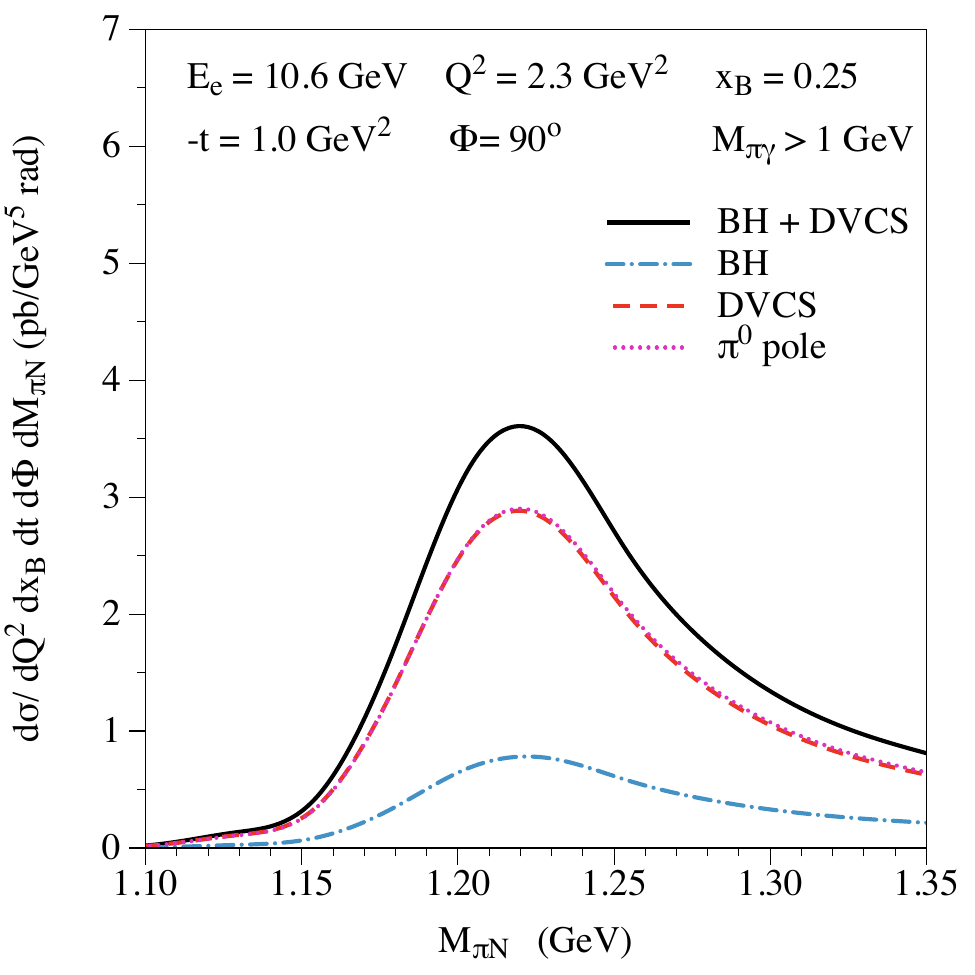}
\includegraphics[width=0.3\textwidth]{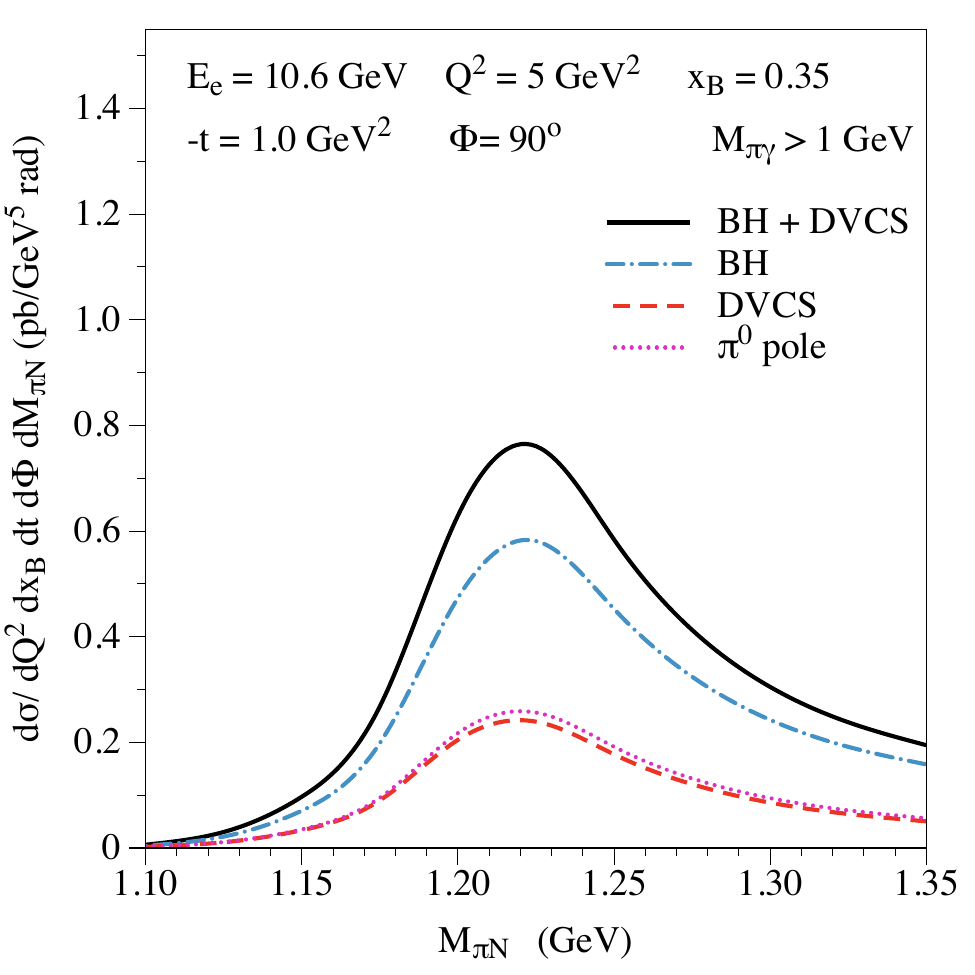}
\includegraphics[width=0.3\textwidth]{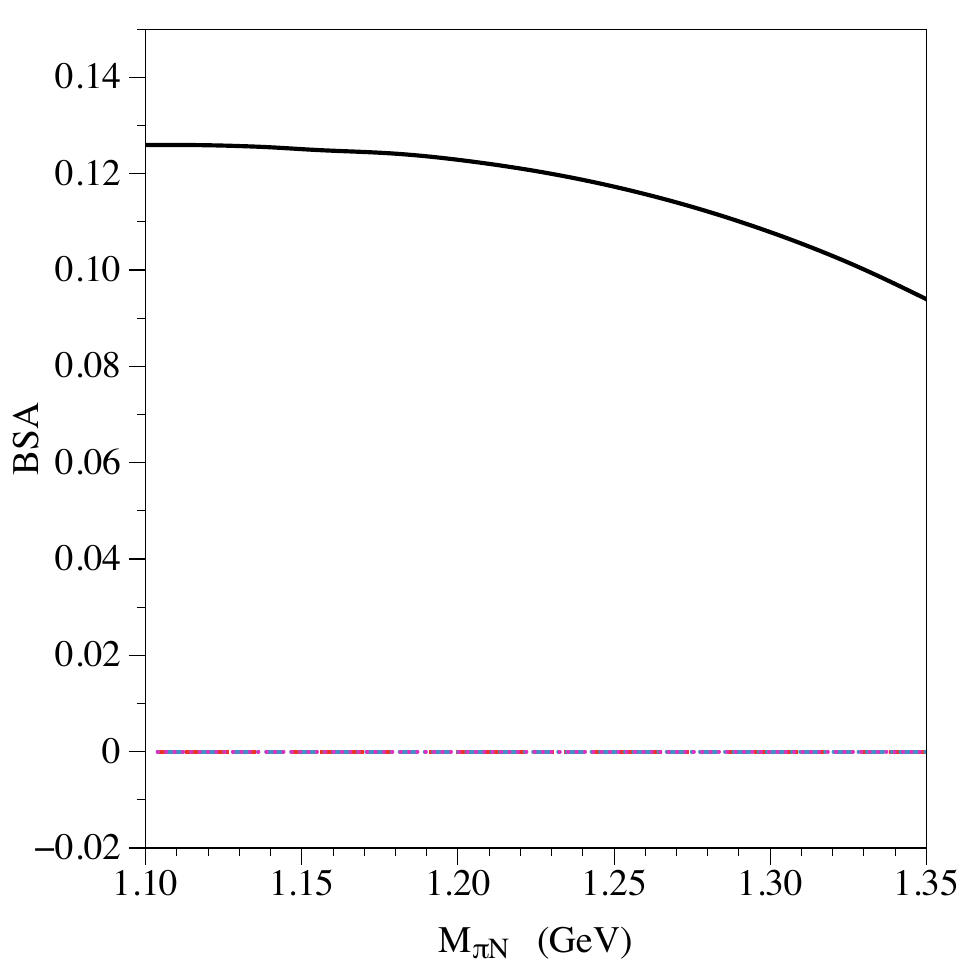}
\includegraphics[width=0.3\textwidth]{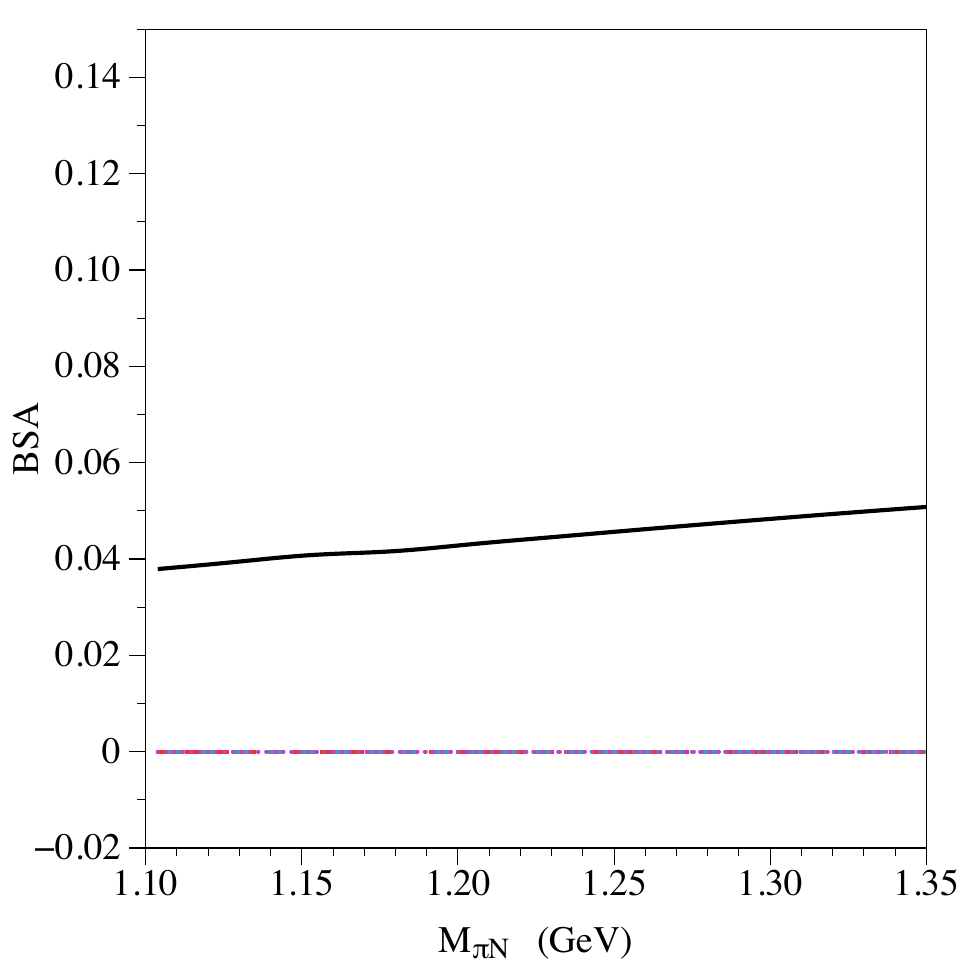}
\includegraphics[width=0.3\textwidth]{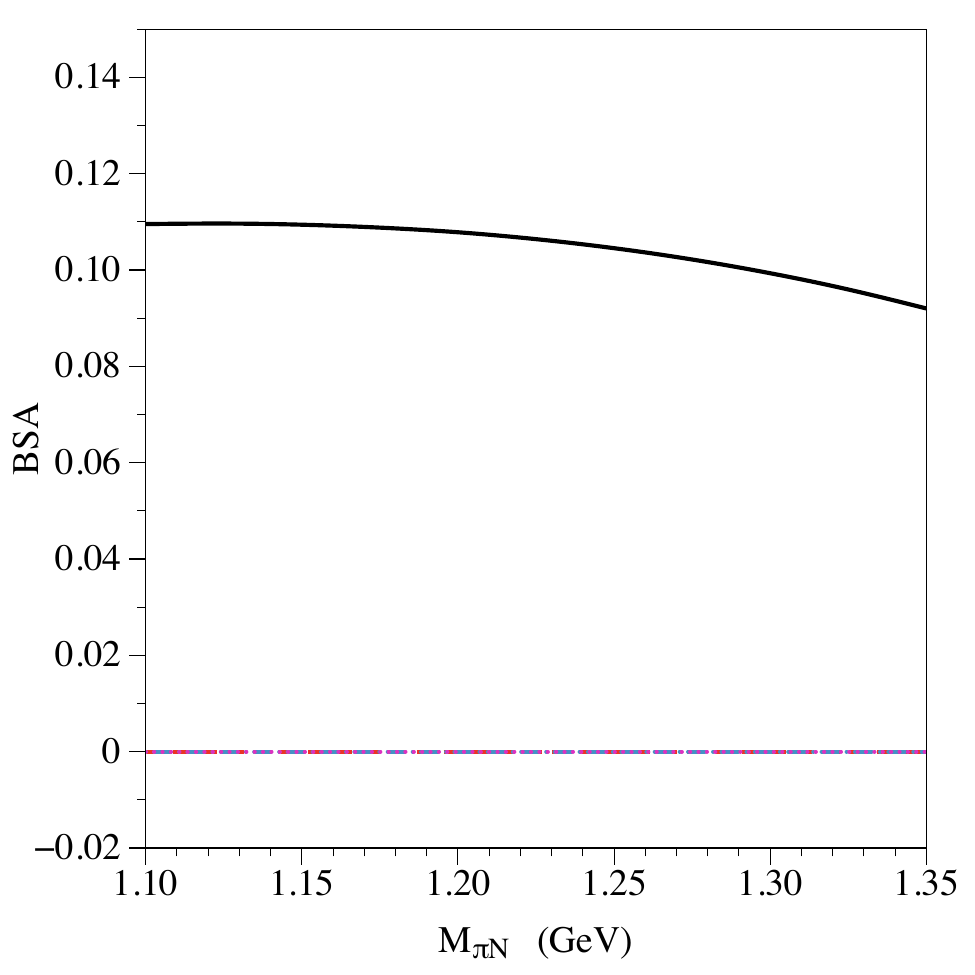}
\caption[]{Dependence on the invariant mass of the $\pi^+ n$ system ($M_{\pi N}$) of the $e^- p \to e^- \gamma \Delta(1232) \to e^- \gamma \pi^+ n$
cross section (upper panels)  integrated over the pion solid angle,
with the cut $M_{\pi \gamma} > 1$~GeV,  and
corresponding beam-spin asymmetry (lower panels) for three kinematical setups accessible with CLAS12@JLab.
Blue dashed-dotted curves: $p \to \Delta(1232)$ BH process;
red dashed curves: $p \to \Delta(1232)$ DVCS process;
black solid curves: BH + DVCS processes.
The magenta dotted curves show the $\pi^0$-pole contribution to the $p \to \Delta(1232)$ DVCS process separately. This figure is taken from Ref.~\cite{Semenov-Tian-Shansky:2023bsy}.
}
\label{Fig_CS_N_to_Delta_MpiN}
\end{figure*}

In the upper panels of Fig.~\ref{Fig_CS_N_to_Delta_MpiN} we present the dependence of the cross-section
(\ref{Def_7fols_CS_NR})
$e^- p \to e^- \gamma \Delta(1232) \to e^- \gamma \pi^+ n$
integrated over the pion solid angle
$d \Omega_\pi^*$
on the invariant mass of the
$\pi^+ n$ system ($M_{\pi N}$).
The kinematic cut
$M_{\pi \gamma} > 1$~GeV
is imposed to reduce the resonating background from the
$\gamma \pi$ subsystem
arising from the
$e^-p \to e^- \rho^+n \to e^- \gamma \pi^+ n$
process. The angle between the leptonic and production planes $\Phi$ is set to $90^\circ$ to maximize the BSA. 
We note that the non-diagonal
$N \to \pi N$
DVCS process in the
$\Delta(1232)$ region is dominated by the
$\pi^0$-pole contribution into the
$N \to \Delta$
{axial-vector}
transition GPD
$C_2$.
It is interesting that in the $\Delta(1232)$ resonance region, for $Q^2=2.3$~GeV$^2$, the
model estimates show the change from being dominated by the BH contribution for $-t = 0.5$~GeV$^2$  
to being DVCS-dominated for $-t = 1.0$~GeV$^2$. 
Such behavior is unlike the DVCS process on a nucleon in similar valence region kinematics, for which the BH dominates in both cases.
This can, probably, be attributed to an effect of kinematic power correction proportional to $t/Q^2$  and $M_N^2/Q^2$ 
\cite{Braun:2014sta}
that stays large 
for $Q^2=2.3$~GeV$^2$. For larger $Q^2=5$~GeV$^2$ the dominance of BH over the DVCS seems to be restored.    
The lower panels of Fig.~\ref{Fig_CS_N_to_Delta_MpiN} present the corresponding BSA
(\ref{Def_BSA_ND_DVCS}), as a function of $M_{\pi N}$.
It originates from the BH and DVCS interference, and in the lower range of $-t$ is estimated to be in the range of $10 \%$ with
a tendency to slightly decrease with growth of $-t$.

Ref.~\cite{Semenov-Tian-Shansky:2023bsy} also presented a generalization of the formalism for the second resonance region, including the contribution
of
$P_{11}(1440)$,
$D_{13}(1520)$
and
$S_{11}(1535)$ resonances
with isospin-$\frac{1}{2}$.  
This required the construction of phenomenological models for the
corresponding twist-$2$ 
{vector and axial-vector}
transition GPDs, see Sec.~\ref{sec:transGPDdef}: ($2$ of each for the spin-$\frac{1}{2}$ resonances and $4$ of each for the spin-$\frac{3}{2}$ resonance).
Let us briefly summarize the key ingredients and the phenomenological input.
\begin{itemize}
\item
The  
{vector}
transition GPDs were constructed relying on the constraints of the
corresponding first Mellin moments in terms of the transition electromagnetic form factors
$F_{1,2}^{N P_{11}}$,
$F_{1,2,3}^{N D_{13}}$,
and
$F_{1,2 }^{N S_{11}}$ 
available from the MAID2007 and  MAID2008 analyses detailed in 
Ref.~\cite{Tiator:2011pw}.
The 
$x$ 
and 
$\xi$ 
dependence of GPDs was provided by two types of models:
\begin{itemize}
\item I:  $\xi$-independent valence PDF type parametrization;
\item II: the Radyushkin double distribution Ansatz with a usual $b$-dependent profile with $b=1$ and the same valence PDF used as input.
\end{itemize}
In fact, model I turns out to be the  $b=\infty$ limit of model II.
The comparison of the results of the two models allows us to roughly quantify the sensitivity of the non-diagonal DVCS observables to 
the quark momentum fraction and skewness dependence of GPDs.

\item The 
{axial-vector}
transition GPDs
$\tilde{H}_1^{p P_{11}}$, $\tilde{H}_1^{p D_{13}}$,  $\tilde{H}_1^{p S_{11}}$
were modeled relying on the normalization provided by the PCAC through the generalized Goldberger-Treiman relations 
for the dominant axial transition FFs and adopting the dipole form for the $t$-dependence with the $x$ and $\xi$ dependence 
from the two types of models (I and II) described above.

\item Finally, the GPDs 
$\tilde{H}_2^{p P_{11}}$, $\tilde{H}_2^{p D_{13}}$, $\tilde{H}_2^{p S_{11}}$ 
were modeled assuming the pion pole dominance with the asymptotic form 
of the pion distribution amplitude.

\item GPDs $H_{4}^{pD_{13}}$, $\tilde H_{3}^{pD_{13}}$, $\tilde H_{4}^{pD_{13}}$, 
for which no phenomenological constraints are available, were neglected.

\end{itemize}

\begin{figure*}[h]
\centering
\includegraphics[width=0.3\textwidth]{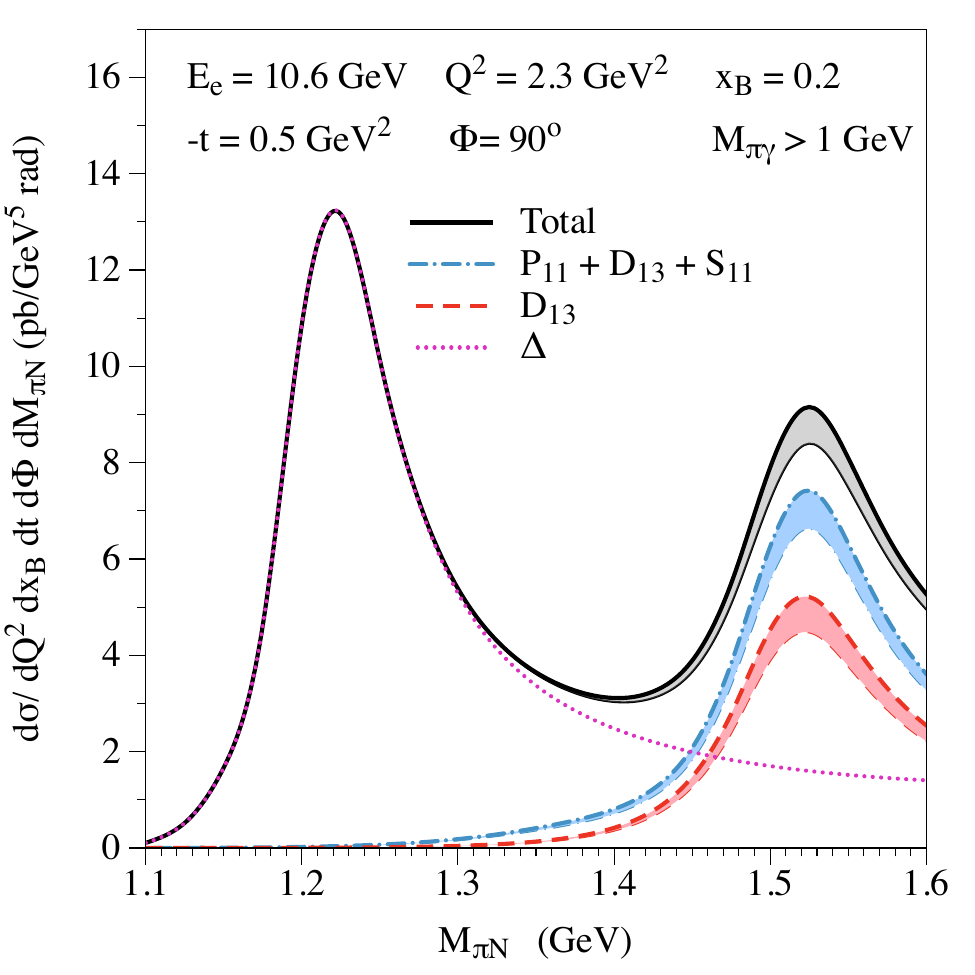}
\includegraphics[width=0.3\textwidth]{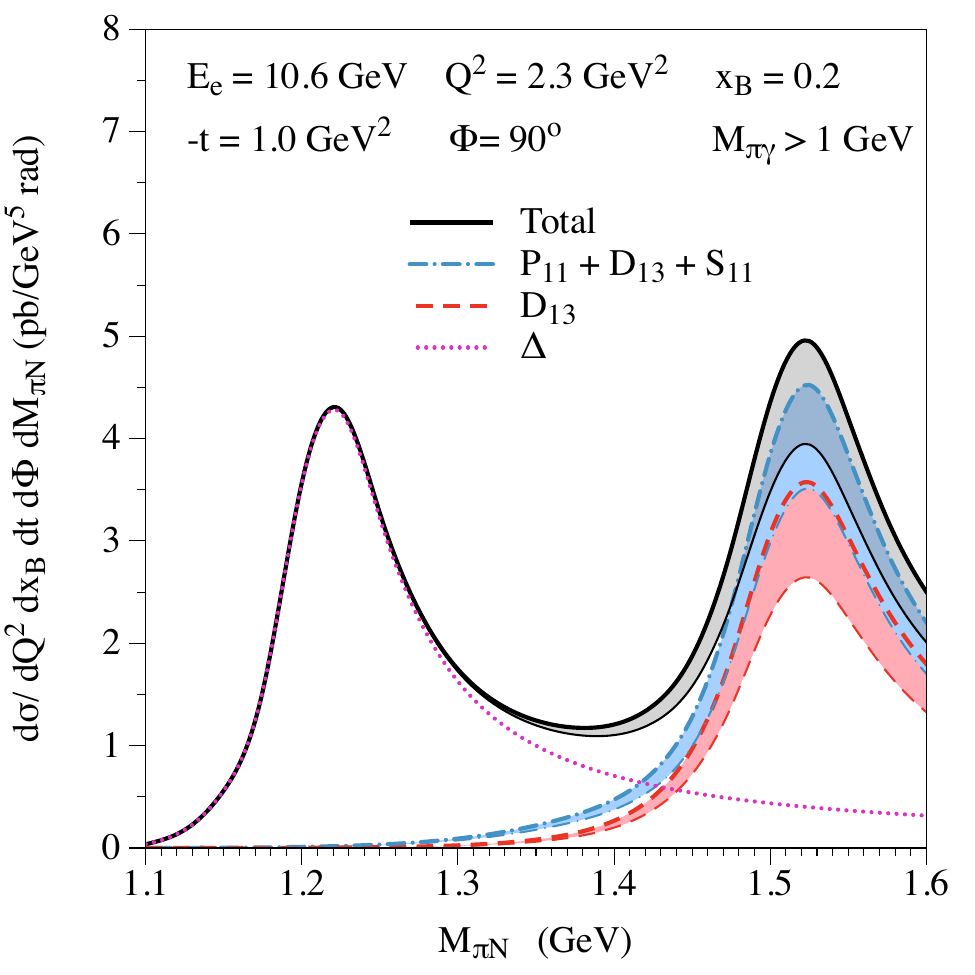}
\includegraphics[width=0.3\textwidth]{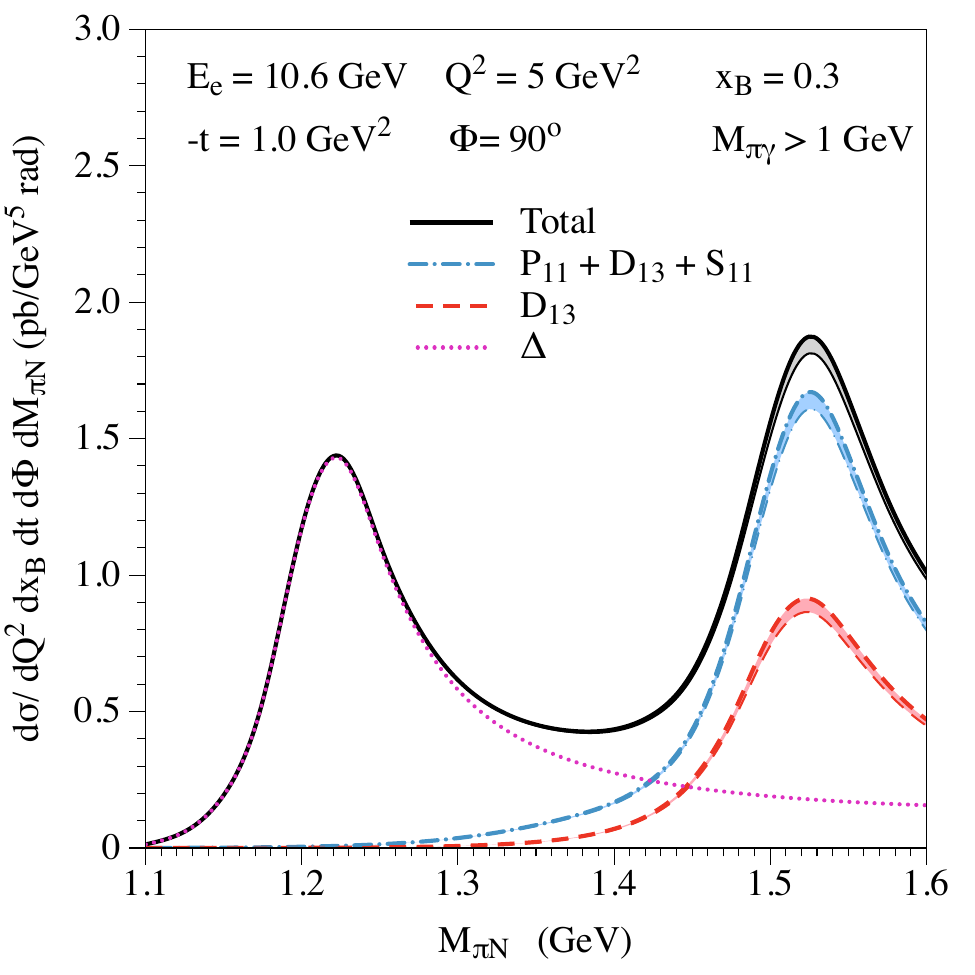}
\includegraphics[width=0.3\textwidth]{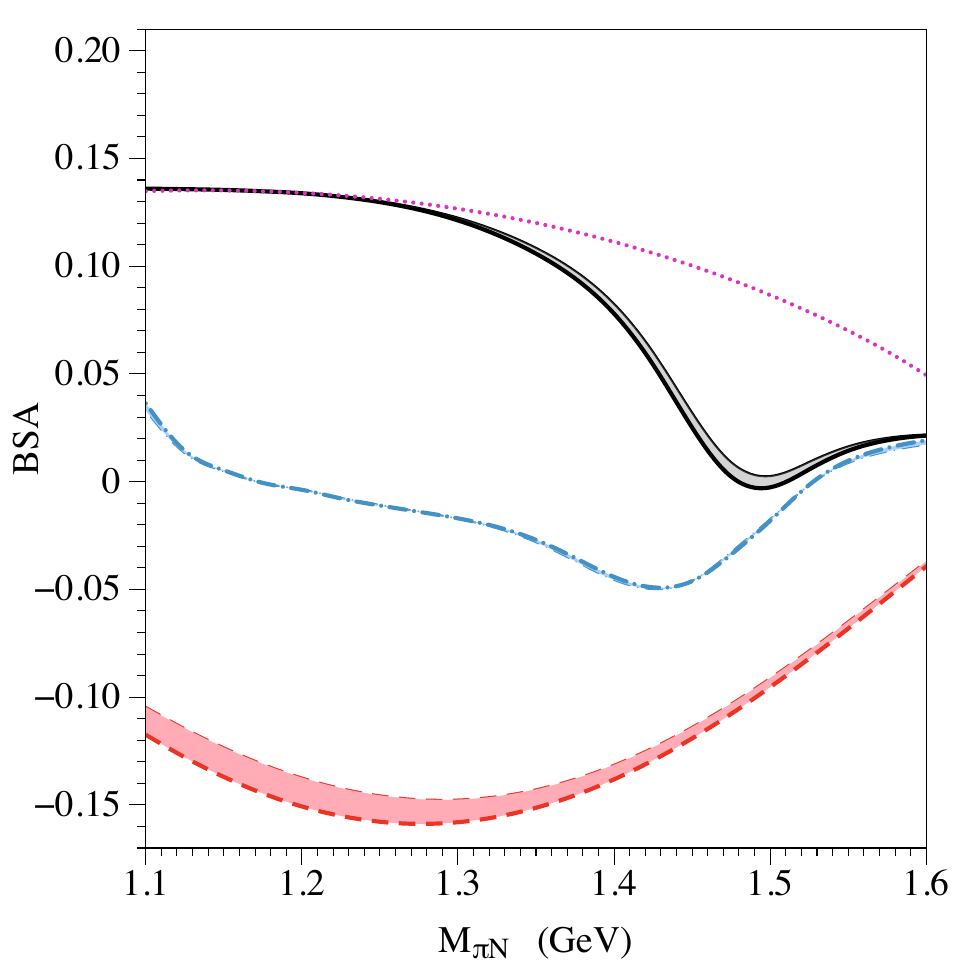}
\includegraphics[width=0.3\textwidth]{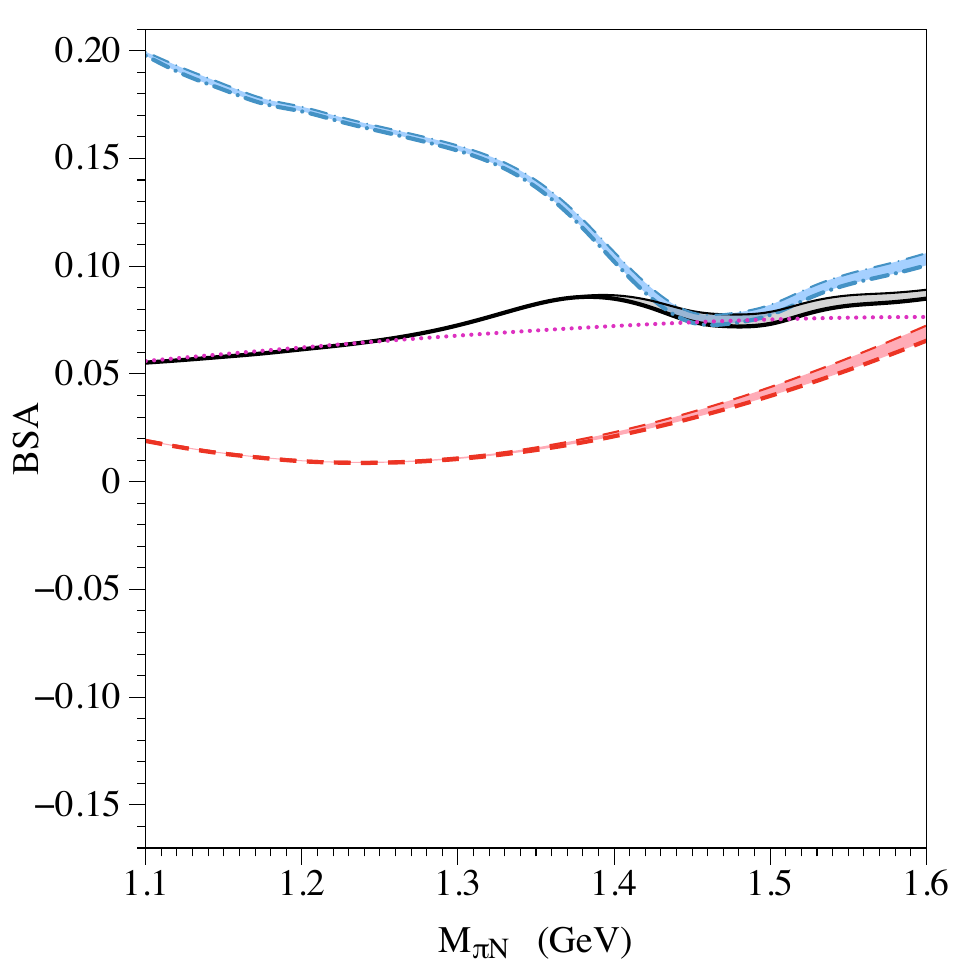}
\includegraphics[width=0.3\textwidth]{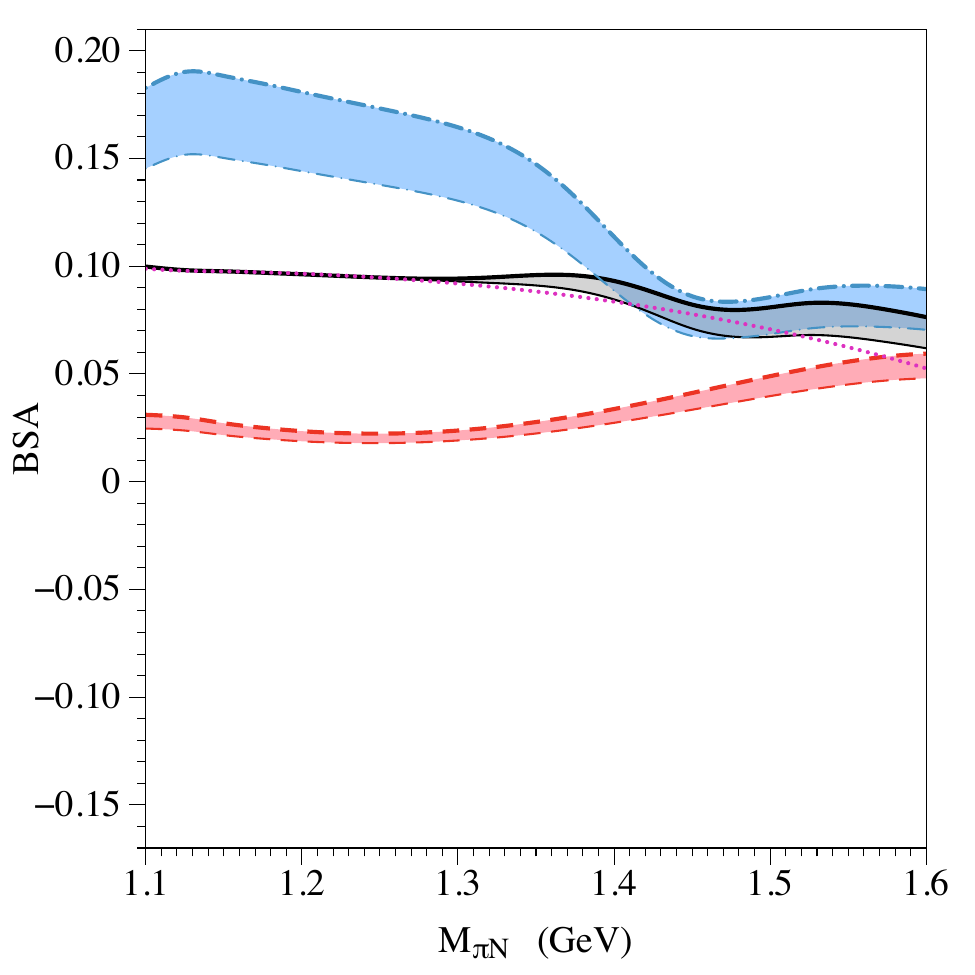}
\caption[]{Dependence on the invariant mass of the $\pi^+ n$ system ($M_{\pi N}$)
of the $e^- p \to e^- \gamma R \to e^- \gamma \pi^+ n$
cross section (upper panels) and corresponding beam-spin asymmetry (lower panels), integrated over the decay pion solid angle,
with the cut $M_{\pi \gamma} > 1$~GeV, for three kinematics accessible at CLAS12@JLab.
Magenta dotted curves: BH + DVCS process for $R = \Delta(1232)$;
red dashed curves and red bands: BH + DVCS process for $R = D_{13}(1520)$;
blue dashed-dotted curves and blue bands: BH + DVCS process for $R = P_{11}(1440) + D_{13}(1520) + S_{11}(1535)$;
black solid curves and grey bands: BH + DVCS process for $R = \Delta(1232) + P_{11}(1440) + D_{13}(1520) + S_{11}(1535)$. The thin (thick) curves represent the result of model I (model II) 
for $x$ and $\xi$ dependence of $N \to R$ GPDs, see a description in the text. The bands indicate the corresponding variation due to the modeling. This figure is taken from Ref.~\cite{Semenov-Tian-Shansky:2023bsy}. }
\label{fig:2ndres_5f}
\end{figure*}

Fig.~\ref{fig:2ndres_5f} presents the 
$M_{\pi N}$ 
invariant mass dependence in the first and second nucleon resonance regions of the 
$e^- p \to e^- \gamma \pi^+ n$ 
cross-section and corresponding BSA with the cut 
$M_{\pi \gamma} > 1$~GeV 
to minimize the possible contamination from the $\rho^+$ production channel.

We first notice from the cross-section behavior that, with increasing values of $-t$,  the 
second nucleon resonance region becomes more important relative to the $\Delta(1232)$ resonance region. 
This can be explained due to the  $\gamma^\ast N \Delta$ transition form factors  dropping faster with increasing 
$-t$ value in comparison with the corresponding ones for the 
$D_{13}(1520)$ 
and 
$S_{11}(1535)$ 
resonances. Moreover,
in the second nucleon resonance region, the 
$D_{13}(1520)$ 
excitation provides the largest contribution, followed by the 
$S_{11}(1535)$ 
resonance. On the other hand, the contribution of the 
$P_{11}(1440)$ 
excitation to the unpolarized cross section is only very small.
For the BSA, one notices that at 
$-t = 0.5$~GeV$^2$ 
it reaches a value of around 10\% in the $\Delta$-resonance region and shows a sharp drop in the second resonance region.
Fig.~\ref{fig:2ndres_5f} also shows that,
with an increasing value of $-t$, the BSA for the 
$D_{13}(1520)$ 
displays a sign change, resulting in a positive BSA in the second nucleon resonance region at $-t = 1$~GeV$^2$.

\subsection{$N \to N^{*}$ DVMP and  transition GPDs}
\label{sec:NtoNstarDVMP}

\begin{figure}[t]
	\centering
		\includegraphics[width=0.40\textwidth]{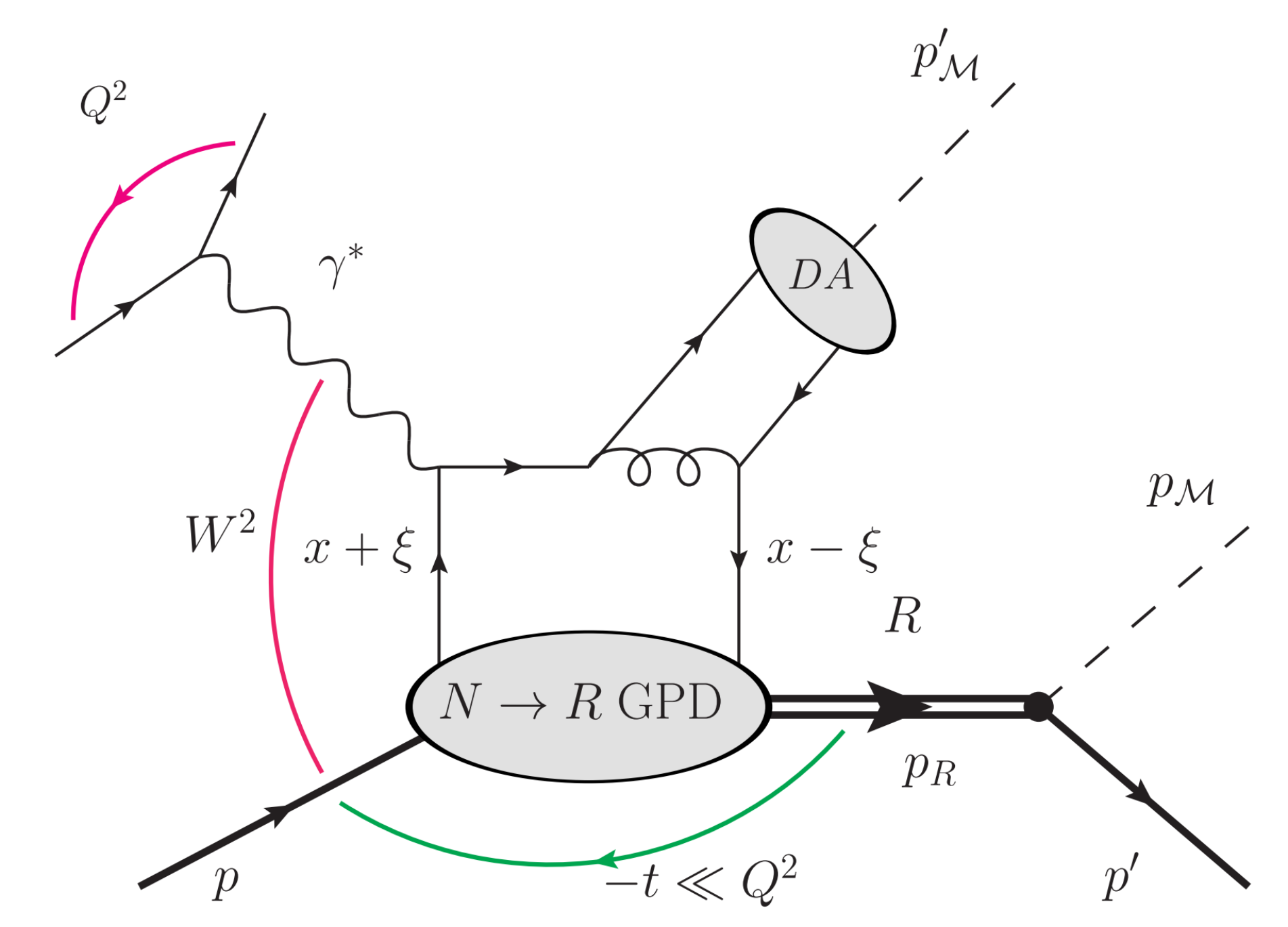}
	\caption{Schematic drawing of the $N \to N^{*}$ DVMP process.}
	\label{fig:NNstar_DVMP_processes}
\end{figure}

{While the $N \to N^*$ DVCS process, to the leading accuracy, is only sensitive to vector and axial-vector helicity non-flip 
transition GPDs, 
the $N \to N^{*}$ DVMP process
\begin{equation}
	\gamma^{*}~N \rightarrow \text{meson}~N^{*} \rightarrow~\text{meson}~[~N~\text{meson}~],
\end{equation}
which is shown schematically in Fig.~\ref{fig:NNstar_DVMP_processes}, can be used to access all twist-$2$ transition GPDs,
including the 
{tensor}
transition GPDs.}

Due to the possible different mesons that can be produced and due to the different decay channels of the resonance, a whole series of $N \to N^{*}$ DVMP processes can be studied, for example:
\begin{enumerate}
  \item ~~$\gamma^{*}~p ~~\rightarrow~~ \pi^{-} ~ N^{*++}$
  \item ~~$\gamma^{*}~p ~~\rightarrow~~ \pi^{0} ~ N^{*+}$
  \item ~~$\gamma^{*}~p ~~\rightarrow~~ \rho^{0} ~ N^{*+}$
  \item ~~$\gamma^{*}~p ~~\rightarrow~~ \pi^{+} ~ N^{*0}$
  \item ~~$\gamma^{*}~n ~~\rightarrow~~ \pi^{+} ~ N^{*-}$
\end{enumerate}
with a nucleon or a delta resonance $N^{*}$. Furthermore, deeply virtual Kaon production in relation to hyperons like the $\Lambda$ is possible and enables the study of transitions related to the strangeness sector. 
{
Some rough estimates of the magnitude of the corresponding cross sections can be made relying on the SU$(3)$ flavor symmetry, which relates proton-to-hyperon GPDs to a combination of standard proton GPDs \cite{Frankfurt:1999fp}, {\it e.g.}}
{
\begin{equation}
  H^{p \to \Lambda} = - \frac1{\sqrt{6}} \Big(2 H^u -H^d -H^s\Big),
\end{equation}
and similarly for $E^{p \to \Lambda}$ and $\tilde{H}^{p \to \Lambda}$.}
Predictions for the kaon-hyperon channels can be made analogously to those for pion production \cite{Kroll:2019wug}.

For DVMP reactions, the cross-section is given by \cite{Arens:1996xw}:
\begingroup\makeatletter\def\f@size{9}\check@mathfonts
\begin{eqnarray}
&&
\frac{d^4 \sigma}{ds dQ^2 d \phi dt} \propto  \bigg[ \frac{d \sigma_T}{dt} + \varepsilon \frac{d \sigma_L}{dt} \nonumber + \varepsilon \cos 2\phi \frac{d \sigma_{TT}}{dt} \\ &&
+\sqrt{2 \varepsilon (1+\varepsilon)} \cos \phi \frac{d \sigma_{LT}}{dt} 
+ h \sqrt{2\varepsilon(1-\varepsilon)} \frac{d\sigma_{LT'}}{dt} \sin\phi \bigg].
\label{Def_CS_Kroll}
\end{eqnarray}
\endgroup
Here the terms $\sigma_L, \sigma_T, \sigma_{LT}, \sigma_{LT'}, \sigma_{TT}$ correspond to the interaction of longitudinally ($L$) and transversely ($T$) polarized virtual photons, $\phi$ is the angle between the leptonic and hadronic planes, and $\varepsilon$ is the polarization of the virtual photon, representing the ratio of longitudinal to transverse photon flux.

As a first reaction, hard exclusive $\pi^{-}\Delta^{++}$ electroproduction:
\begin{equation}
	\gamma^{*}~p \rightarrow \pi^{-} \Delta^{++} \rightarrow~\pi^{-}~[~p~\pi^{+}~],
\end{equation}
has been theoretically described based on transition GPDs in Ref. \cite{Kroll:2022roq}. 
{The process is expected, but not proven yet, to factorize under the same conditions as the $N \to N^{*}$ DVCS process, and its 
  dynamics is supposed to be similar  to that of the pion electroproduction reaction
$\gamma^*N\to \pi N'$.}
From COMPASS and the JLab experiments, we learned that the transverse
cross-section is much larger than the longitudinal one for the $\pi^0p$ channel. For the charged pions, there
is a strong contribution from the pion pole to the longitudinal amplitudes. The pole is to be treated
as a one-boson-exchange contribution in the experimentally
accessible range of kinematics since its contribution evaluated from the GPD $\tilde{E}$
is too small \cite{Goloskokov:2009ia}. Based on this expectation and with the help of large-$N_c$ results 
\cite{Frankfurt:1999xe,Belitsky:2005qn}
first, admittedly rough, estimates of the $\pi^-\Delta^{++}$ cross sections have been given in \cite{Kroll:2022roq}.

The helicity amplitudes for
longitudinally polarized photons, which are fed by twist-2 contributions, reads ($e_0$ is the positron charge)
\begin{equation}
  {\cal M}^{tw2}_{0\lambda_R,0\lambda_N}=e_0 \int_{-1}^{1} dx {\cal H}^\pi_{0+0+}\Big[ A^\Delta_{\lambda_R+,\lambda_N +}-A^\Delta_{\lambda_R-,\lambda_N-}\Big],
\label{eq:M-Delta-tw2}
\end{equation}
where ${\cal H}^\pi_{0\lambda,0\lambda}$ is the leading-twist amplitude for the hard partonic subprocess
$\gamma^*q(\lambda)\to \pi q(\lambda)$ ($\lambda=\pm \frac{1}{2}$)  and
\begingroup\makeatletter\def\f@size{9}\check@mathfonts
\begin{eqnarray}
  &&A^\Delta_{\lambda_R\lambda,\lambda_N \lambda} = \int \frac{dz^-}{2\pi} e^{ixP^+z^-} \langle \Delta^{++}(p_R,\lambda_R)\mid   \nonumber\\
    &&\times \bar{u}(-z/2)\frac14 \gamma^+ (1+2\lambda)\gamma_5)d(z/2)
                                      \mid p(p,\lambda_N)\rangle_{z^+=0,z_\perp=0}.
\end{eqnarray}
\endgroup
To the difference of the $A^\Delta$s in Eq.~(\ref{eq:M-Delta-tw2}) only the odd-parity GPDs $\tilde{G}_i$
(\ref{eq:odd-pD-GPDs}) contribute.

The amplitudes for transversally polarized photons ($\mu=\pm 1$) read 
\begin{eqnarray}
&&
  {\cal M}^{tw3}_{0\lambda_R,\mu\lambda_N} \nonumber \\ && = e_0\int_{-1}^1 dx \Big[{\cal H}^\pi_{0-,\mu+} A^\Delta_{\lambda_R-,\lambda_N+}
    + {\cal H}^\pi_{0+,\mu-} A^\Delta_{\lambda_R+,\lambda_N-}\Big],  \nonumber \\ &&
\label{eq:M-Delta-tw3}
\end{eqnarray}
where
\begingroup\makeatletter\def\f@size{9}\check@mathfonts
\begin{eqnarray}
  &&A^\Delta_{\lambda_R-\lambda,\lambda_N \lambda}= \int \frac{dz^-}{2\pi} e^{ixP^+z^-}  \langle \Delta^{++}(p_R,\lambda_R)\mid   \nonumber \\
    &&\times -\lambda\frac{i}{2}\bar{u}(-z/2)(\sigma^{+1}-2\lambda \sigma^{+2}) d(z/2)\mid p(p,\lambda_N)\rangle_{z^+=0,z_\perp=0} \nonumber \\
\end{eqnarray}
\endgroup
to which the 
{tensor}
GPDs (\ref{eq:trans-GPDs}) contribute. The pion wave function appearing in
${\cal H}^\pi_{0-\lambda,\mu\lambda}$, is of twist-3 nature. The matrix elements $A^\Delta$ as well as the helicity
amplitudes  are explicitly given in terms of the transition GPDs in \cite{Kroll:2022roq}. The subprocess amplitudes,
${\cal H}^\pi$, can be found in \cite{Goloskokov:2009ia}. They have been calculated within the so-called modified
perturbative approach in which the quark transverse momenta and the Sudakov suppressions are taken into account.

In the large-$N_c$ limit, the following relations between the $p-\Delta^{++}$ transition GPDs and the proton to proton ones are found
\cite{Belitsky:2005qn,Frankfurt:1999xe}
\begin{equation}
  \tilde{G}_3= \frac32\Big( \tilde{H}^u-\tilde{H}^d\Big) \qquad \tilde{G}_4=\frac38\Big(\tilde{E}^u-\tilde{E}^d\Big)
  \end{equation}
and \cite{Kroll:2022roq}
\begin{equation}
\label{eqnHt}
  G_{T5} + \frac12 G_{T7} = -\frac32\Big( H_T^u-H_T^d\Big)\,.
\end{equation}

{
As it was already mentioned, we expect that the process
$\gamma^*p\to \pi^-\Delta^{++}$ is under control of similar dynamics
as electroproduction of an exclusive pion-nucleon final state. Since $H_T$
is the dominant tensor GPD for $\gamma^* N\to \pi^{\pm} N'$ 
\cite{Goloskokov:2009ia},
it is assumed that the GPD $G_{T5}$ (scenario I) or $G_{T7}$
(scenario II) is the dominant tensor GPD in $\gamma^*p\to \pi^-\Delta^{++}$.
All other tensor GPDs are neglected, as well as the odd-parity
helicity non-flip GPDs $\tilde{G}_1$ and $\tilde{G}_2$. Making use
of the parametrization of proton-proton GPDs advocated for
}
in \cite{Goloskokov:2007nt,Goloskokov:2011rd}, one is in the position to evaluate the helicity amplitudes
(\ref{eq:M-Delta-tw2}) and (\ref{eq:M-Delta-tw3}). In addition to the GPD contributions one has to take into account
also the pion-pole contribution, which reads \cite{Kroll:2022roq}
\begin{eqnarray}
&&
  {\cal M}^{pole}_{0\lambda_R,\mu\lambda_N} \nonumber \\ &&  = e_0 \frac{\varrho_{\pi\Delta}}{t-m_\pi^2} \bar{ \cal U}_\delta(p_R, \lambda_R) \frac{\Delta^\delta}{M_N}
u(p,\lambda_N) (q-2q')_\rho \epsilon^\rho(q,\mu),  \nonumber \\ && 
\end{eqnarray}
where
\begin{equation}
  \varrho_{\pi\Delta}= \sqrt{2} g_{\pi\Delta^{++}p} F_{\pi\Delta p}(t) F_\pi(Q^2)\,.
  \label{eq:coupling}
\end{equation}
The coupling of the pion to the proton and the $\Delta^{++}$ is described by a coupling constant, $g_{\pi\Delta^{++}p}$,
  and a $t$-dependent form factor $F_{\pi\Delta p}$. The last item in Eq.\ (\ref{eq:coupling}), $F_\pi(Q^2)$,
  is the electromagnetic form factor of the pion. The photon polarization vector is denoted by $\epsilon$. The photon
  and meson momenta are denoted by $q$ and $q'$, respectively. The longitudinal pole contribution dominates over the
  transversal ones.
  The transversal/longitudinal ratio of the pole contributions is proportional to $1/Q$ as is the case for the corresponding
  ratio of the amplitudes (\ref{eq:M-Delta-tw3}) and (\ref{eq:M-Delta-tw2}). 

\begin{figure}[t]
      \centering
      \includegraphics[width=0.23\textwidth]{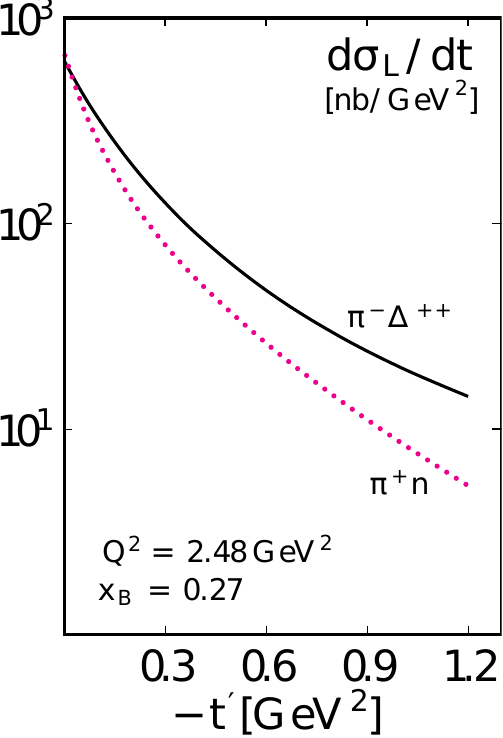}~~~
      \includegraphics[width=0.23\textwidth]{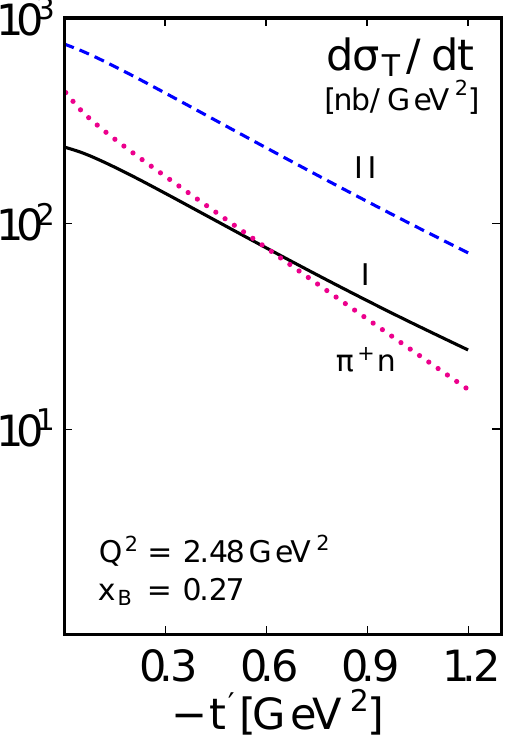}
      \caption{The longitudinal (left) and the transverse (right) cross sections of $\gamma^*p\to \pi^-\Delta^{++}$ versus $-t'=-t+t_0$. The solid (dashed) lines represent the predictions obtained for scenario I (II). For comparison, the dotted lines are the results for $\gamma^*p\to\pi^+n$ obtained with the same GPDs. The figure is taken from Ref. \cite{Kroll:2022roq}.}
      \label{fig:sigmaL}
\end{figure}

\begin{figure}[t]
      \centering
         \includegraphics[width=0.234\textwidth]{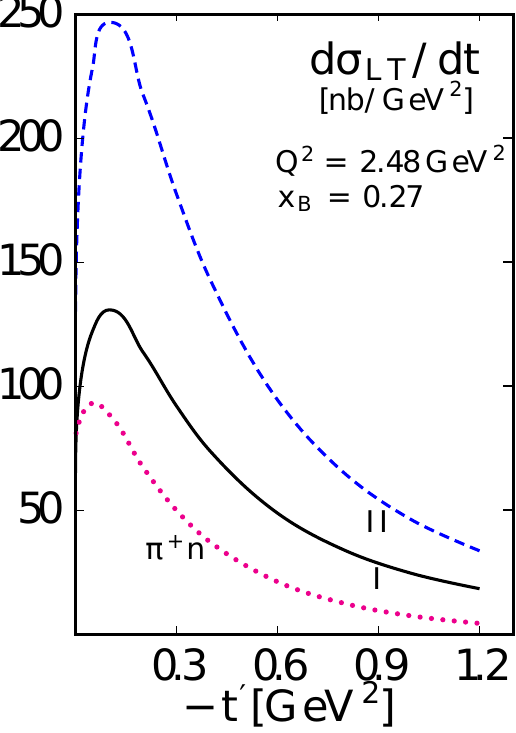}~~~
      \includegraphics[width=0.23\textwidth]{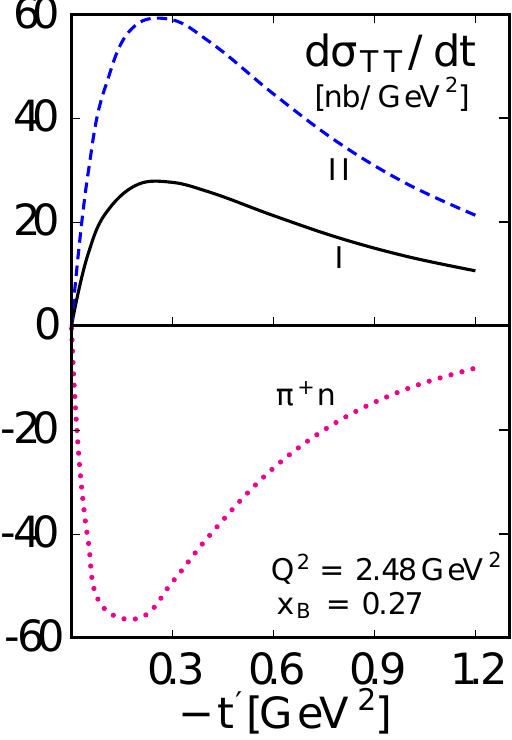}
      \caption{The longitudinal-transverse (left) and the transverse-transverse (right) interference
        cross sections of $\gamma^*p\to \pi^-\Delta^{++}$. For other notations, see Fig. \ref{fig:sigmaL}. The figure is taken from Ref. \cite{Kroll:2022roq}.}
        \label{fig:sigmaTT}
   \end{figure}

  With the above described model for the $p-\Delta$ GPDs and the pion-pole contribution, the partial cross sections
  \begingroup\makeatletter\def\f@size{9}\check@mathfonts
   \begin{eqnarray}
\frac{d\sigma_L}{dt} &=& \frac{\sum_{\lambda_R} |{\cal M}_{0\lambda_R, 0+}|^2}{16\pi (W^2-M_N^2)\sqrt{ {\lambda}(W^2,-Q^2,M_N^2)}}\,,\nonumber\\
\frac{d\sigma_T}{dt} &=& \frac{\sum_{\lambda_R}\Big[ |{\cal M}_{0\lambda_R,++}|^2 + |{\cal M}_{0\lambda_R,-+}|^2\Big]}
                                                   {32\pi (W^2-M_N^2)\sqrt{{\lambda}(W^2,-Q^2,M_N^2)}}\,,\nonumber\\  
\frac{d\sigma_{TT}}{dt} &=& -\frac{\sum_{\lambda_R} {\rm Re}\Big[ {\cal M}_{0\lambda_R,++}^* {\cal M}_{0\lambda_R,-+}\Big]}
                                             {16\pi (W^2-M_N^2)\sqrt{{\lambda}(W^2,-Q^2,M_N^2)}}\,,\nonumber\\
 \frac{d\sigma_{LT}}{dt} &=& -\sqrt{2}\frac{\sum_{\lambda_R} {\rm Re}\Big[ {\cal M}_{0\lambda_R,0+}^*
                                                 \big({\cal M}_{0\lambda_R,++}-{\cal M}_{0\lambda_R,-+}\big)\Big]}
                                             {32\pi (W^2-M_N^2)\sqrt{{\lambda}(W^2,-Q^2,M_N^2)}}\,.     \nonumber\\  
\label{eq:partial-cross-sections}
\end{eqnarray}
\endgroup
   for the $\pi^-\Delta^{++}$ channel are evaluated. The results at $Q^2=2.48\,{\rm GeV}^2$ and a Bjorken-$x$ of
   $x_B=0.27$ are shown in Figs.\ \ref{fig:sigmaL}, \ref{fig:sigmaTT} as a function of $t'=t-t_0$ (see Eq.\ (\ref{eq:t0}))
   and compared to the cross sections for the $\pi^+n$ channel. Note that the $t_0$-values are quite different
   in both processes ($t_0=-0.323\, (-0.088)\, {\rm GeV}^2$ for the $\pi^-\Delta^{++}$ ($\pi^+n$) channel at the quoted kinematics).
   The amplitudes in (\ref{eq:partial-cross-sections}) are the sum of the GPD and the pole contributions. The
   Mandelstam {kinematical} function is
   \begin{eqnarray}
     {\lambda}(W^2,{-}Q^2,M_N^2) = && W^4+Q^4+M_N^4+2 W^2 Q^2 \nonumber\\
     &&-2 W^2 M_N^2 +2 Q^2 M_N^2\,.
   \end{eqnarray}

As is to be seen from Fig.\ \ref{fig:sigmaL} the longitudinal cross section is large. The reason for this feature 
is a strong contribution from the pion pole. For the other partial cross-sections, there are substantial differences between the two scenarios. Particularly interesting is the opposite
sign of $d\sigma_{TT}$ for the $\pi^-\Delta^{++}$ and the $\pi^+n$ channels. The dominant contribution to $d\sigma_{TT}$
comes from an interference of a twist-3 helicity non-flip amplitude being proportional to the convolution $\langle H_T\rangle$
(in the large-$N_c$ limit for the $\pi^-\Delta^{++}$ channel), and a pion-pole contribution. 
The $Q^2$ and $x_B$ dependencies of the $\pi^-\Delta^{++}$ partial cross sections are similar to those of the $\pi^+n$
ones. It is to be stressed that the predictions are to be understood as rough estimates; their uncertainties are very large.

In \cite{Kroll:2022roq} also the asymmetry $A_{LL}$ has been calculated, which can be measured with longitudinally
polarized beam and target and is to be integrated upon the electroproduction cross-section
\begin{equation}
A_{LL} = \sqrt{1-\varepsilon^2}\,\frac1{2\sigma_0}\,\sum_{\lambda_R}\Big[ |{\cal M}_{0\lambda_R,++}|^2 -  |{\cal M}_{0\lambda_R,-+}|^2\Big]
\label{eq:ALL}
\end{equation}
where
\begin{equation}
\sigma_0 =\sum_{\lambda_R}\Big[|{\cal M}_{0\lambda_R,++}|^2 + |{\cal M}_{0\lambda_R,-+}|^2 + \varepsilon  |{\cal M}_{0\lambda_R,0+}|^2\big]\,.
\end{equation}
The magnitude of the
predictions are close to that one for the $\pi^+n$ channel but the two scenarios lead to opposite signs of $A_{LL}$.
This result is understandable: for scenario I the amplitudes ${\cal M}_{0\lambda_R,++}$ are large, whereas for scenario II
the amplitudes ${\cal M}_{0\lambda_R,-+}$ dominate. This feature provides an opportunity to discriminate between dominant
$G_{T5}$ and $G_{T7}$.

There are many other asymmetries, like the beam-spin one
\begin{equation}
  A_{LU}^{\sin{(\phi)}} \sim \frac1{\sigma_0} \sum_{\lambda_R}
  {\rm Im} \Big[ ({\cal M}_{0 \lambda_R,++}-{\cal M}_{0 \lambda_R,-+})^*{\cal M}_{0 \lambda_R,0+}\Big]\,.
\end{equation}
Most of these asymmetries depend on relative phases between different helicity amplitudes. At present, it is
not possible to estimate these asymmetries reliably.    

For vector-meson channels, for instance $\rho^0\Delta^{+}$, the same $A^\Delta$s hold. Only a plus sign appears
between the $A^\Delta$s in Eq.\ (\ref{eq:M-Delta-tw2}) instead of the minus sign. This projects out the 
{vector}
GPDs $G_i$ (\ref{eq:even-pD-GPDs}).


\subsection{Near-threshold pion production}
{
The simplest electromagnetic processes inducing $N \rightarrow \pi N$
are pion photo- and electroproduction on the nucleon
$\gamma N \to \pi N$, $\gamma^\ast N \to \pi N$, which have been studied 
extensively since the 1950s both experimentally and theoretically. 
In near-threshold kinematics the production processes can be connected 
with chiral dynamics in the $\pi N$ system (see Sec.~\ref{subsec:chiral_dynamics}).
}
The celebrated low-energy theorems (LET) \cite{Kroll:1953vq,Nambu:1962ilv,Nambu:1962lbq} 
relate the S-wave transverse $E_{0+}$ and longitudinal 
$L_{0+}$ multipoles at threshold to the nucleon electromagnetic and
axial form factors in the chiral limit $m_\pi = 0$.
The new insight gained from CHPT calculations \cite{Bernard:1992ys,Bernard:1995dp}
is that the expansion at small $Q^2$ has to be done with care as the 
limits $m_\pi \to 0$ and $Q^2 \to 0$ do not commute, in general. 
For large momentum transfers, the power counting of CHPT cannot be applied. However, 
the traditional derivation of LET using PCAC and the current 
algebra [see Sec.~\ref{subsec:chiral_dynamics} and Eq.~(\ref{softpion})]
is not affected as long as the emitted pion is ``soft'' with respect to the initial and 
final state nucleons simultaneously. The corresponding condition is, parametrically, $Q^2 \ll \Lambda^3/m_\pi$ 
\cite{Vainshtein:1972ih,Pobylitsa:2001cz} where $\Lambda$ is some 
hadronic scale, and might be violated at the highest $Q^2$ available in future experiments at JLab and EIC.

Such an interplay of chiral dynamics with ``hard'' QCD factorization picture of large-$Q^2$ reactions
is potentially very interesting since the generalized pion-nucleon LCDAs in the chiral limit
can be expressed in terms of the ``chirally-rotated'' nucleon LCDAs \cite{Pobylitsa:2001cz,Braun:2006td}, 
thus providing one with an additional handle to separate the different components in the nucleon wave function.  
In addition, if the structure of the corrections beyond LET can be quantified, 
pion electroproduction would allow one to determine the nucleon axial form factor at large momentum transfers, 
which is difficult to access by direct measurements.

The difficulty is that the onset of the pQCD regime in form factors
is generally believed to be postponed 
to very large momentum transfers, and at intermediate $Q^2\sim 1-10\,\text{GeV}^2$ one has to take into account 
non-factorizable ``soft'' or ``end-point'' contributions. 
The light-cone sum rule (LCSR) technique \cite{Balitsky:1989ry} 
(see Sec.~\ref{subsec:lightcone_sum_rules})
allows one to estimate such contributions
in terms of the same nucleon LCDAs that enter pQCD calculation using dispersion relations and duality.
It can be generalized and was applied to the pion electroproduction at threshold in Ref.~\cite{Braun:2007pz}.
The LCSR calculation (so far available in leading order only) suggests that the leading transverse $E_{0+}$ 
multipole receives moderate corrections w.r.t. the LET predictions in the whole studied $Q^2$ range, whereas  
the longitudinal $L_{0+}$ multipole receives large corrections and tends to change sign at large $Q^2$. 
These conclusions are, qualitatively, in agreement with the  existing CLAS measurements   
for charged~\cite{CLAS:2012ich} and neutral~\cite{CLAS:2012zia} pion production. 
More studies, both theoretical and experimental, are needed to make this comparison fully quantitative.


\section{First experimental results from JLab 12 GeV}
\label{sec:current_results}

Except for the information obtained from the relation to the transition form factors and the relation to the ordinary GPDs in the large $N_{c}$ limit, little is known about the transition GPDs so far. On the experimental side, recently, the first beam spin asymmetries for hard exclusive $\pi^-\Delta^{++}$ electro-production were published by the CLAS collaboration at JLab and further measurements on $N \to \Delta^{+}$ DVCS with CLAS and $\pi^+\Delta^{0}$ DVMP based on data from hall C at JLab are ongoing. This section will summarize the published and ongoing studies and their impact on constraining transition GPDs.


\subsection{$\pi^-\Delta^{++}$ electroproduction with CLAS12}
\label{sec:exp_DVMP1}

The first observable sensitive to $N\to\Delta$ 
{tensor}
transition GPDs and $N\to\Delta$ transition GPDs in general were provided by the CLAS collaboration with hard exclusive ($\gamma^{*}N \to \pi^{-}\Delta^{++} \to \pi^{-}[p\pi^{+}]$) electro-production beam-spin asymmetries, based on the scattering of longitudinally polarized 10.6~GeV electrons on an unpolarized hydrogen target in the CLAS12 detector \cite{Burkert:2020akg}, published in 2023 \cite{CLAS:2023akb}. 

\begin{figure}[t]
	\centering
		\includegraphics[width=0.45\textwidth]{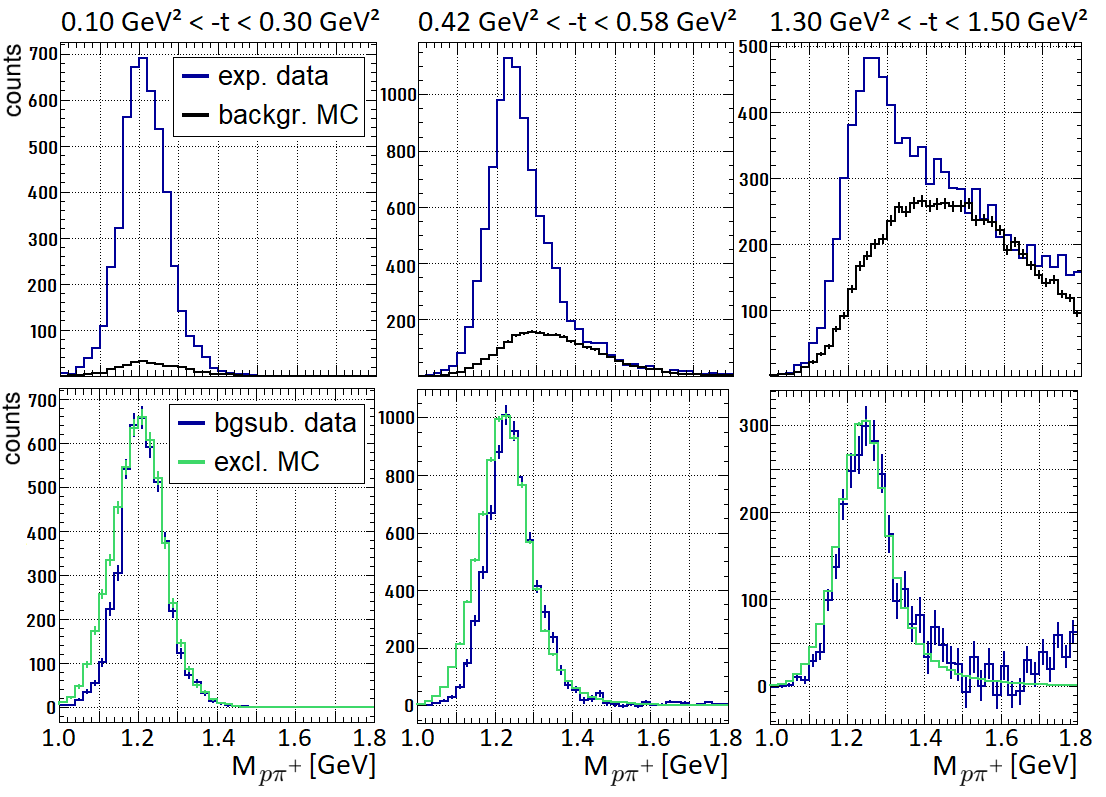}
	\caption{Upper row: $\Delta^{++}$ peak in the $p \pi^{+}$ invariant mass of the experimental data (blue, solid) in comparison to the non-resonant background obtained with the semi-inclusive deep inelastic scattering MC (black) for selected bins of $-t$ in the forward region ($Q^{2}$ = 2.48~GeV$^{2}$, $x_{B}$ = 0.27) after a cut on $M_{\pi^+ \pi^-} > 1.1$~GeV. Lower row: $\Delta^{++}$ peak in the same bins after the subtraction of the background (blue, solid) in comparison to the result from the exclusive MC (green). Figure in modified form taken from Ref. \cite{CLAS:2023akb}.}
	\label{fig:pimdelta_resonance_spect}
\end{figure}

For the study, the $e p \pi^- X$ final state was detected, and the $\pi+$ was reconstructed via a cut on the missing mass.
Kinematic cuts on W $>$ 2 GeV, $Q^2$ $>$ 1.5 GeV$^{2}$, y $<$ 0.75, and $-t$ $<$ 1.5 GeV$^2$ have been applied.
Furthermore, a cut on $M(\pi^+\pi^-)$~$>$~1.1~GeV is applied to reject the dominant background from 
$\gamma^{*}~p ~\rightarrow~ p ~ \rho^0 ~\rightarrow~ p~ \pi^+ ~\pi^-$. Fig. \ref{fig:pimdelta_resonance_spect} shows the $\Delta^{++}$ peak in the resonance spectrum of the $p\pi^+$ invariant mass and compares it to the background distribution from a full semi-inclusive deep inelastic scattering Monte Carlo simulation. In addition, the background-subtracted data is compared to a Monte Carlo simulation of the exclusive process.
It can be observed that the background contribution increases with increasing values of $-t$. However, the good agreement of the falling tail of the $\Delta^{++}$ peak after the background subtraction with the exclusive Monte Carlo shows that the background is well understood and under control.
To select the $\pi^-\Delta^{++}$ signal region, a cut on $M(p\pi^+)$~$<$~1.3~GeV is applied. 
For these events, the beam spin asymmetries are determined for 3 $Q^{2}-x_{B}$ bins and several $-t$ and $\phi$ bins. For each $-t$ bin, the structure function ratio $\sigma_{LT'}/\sigma_{0}$ is extracted based on a fit of the $\phi$ dependence of the beam spin asymmetry, and a background subtraction is performed for each kinematic bin. More details on the analysis can be found in Ref. \cite{CLAS:2023akb}.

\begin{figure}[t]
	\centering
		\includegraphics[width=0.45\textwidth]{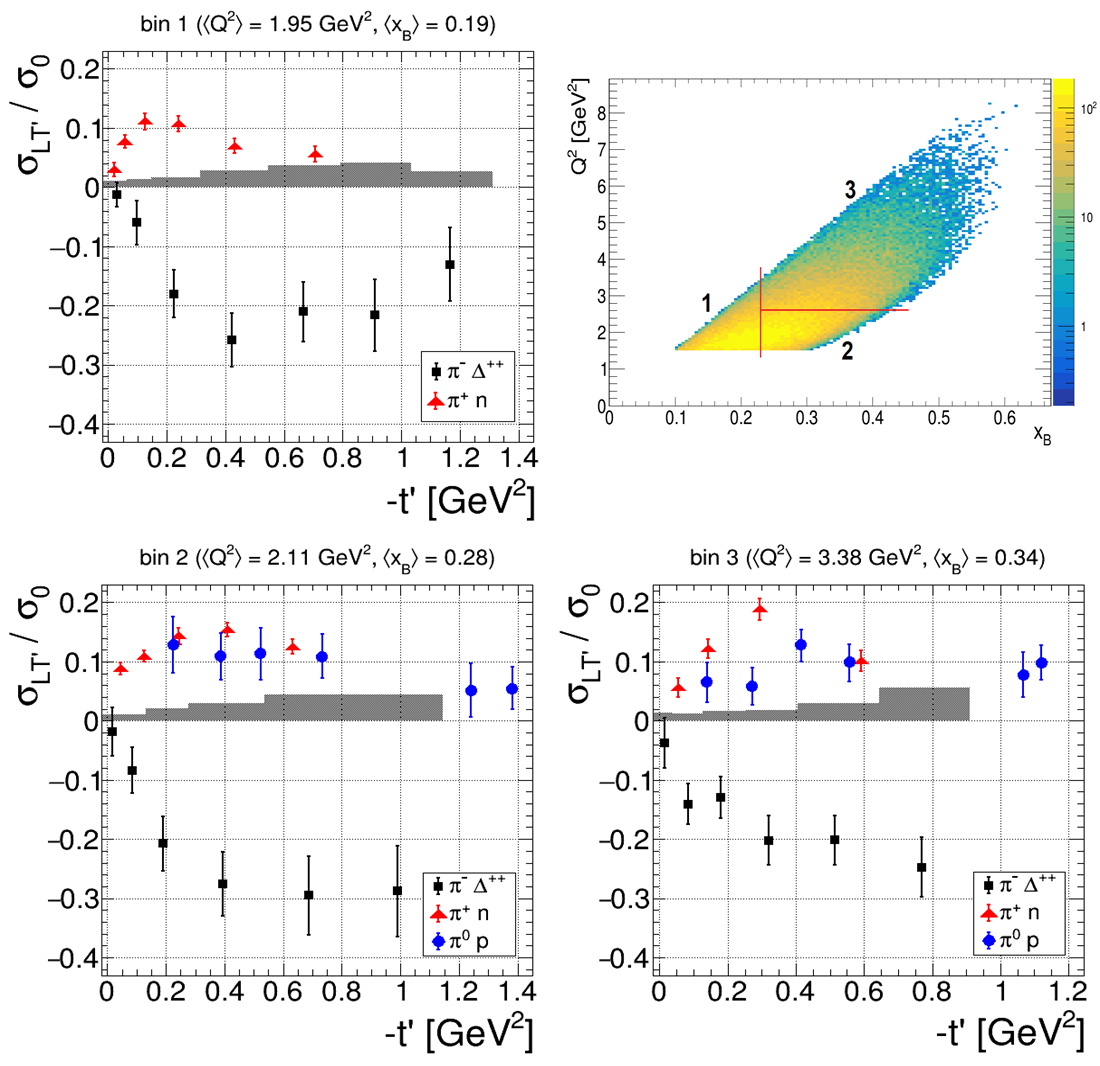}
	\caption{$\sigma_{LT'}/\sigma_{0}$ and its statistical uncertainty for $\pi^- \Delta^{++}$ (black squares, Ref. \cite{CLAS:2023akb}) as a function of $-t'= (|t|-|t_{min}|)$ and its systematic uncertainty (gray band), measured with CLAS12. For comparison, the CLAS12 results from the hard exclusive $\pi^+ n$ (red triangles, Ref. \cite{CLAS:2022iqy}) and $\pi^0 p$ (blue circles, Ref. \cite{CLAS:2023wda}) electro-production with similar kinematics are shown. The sub-figures correspond to the results for the different $Q^{2}$ and $x_{B}$ bins defined in the upper right insert. The mean kinematics are shown on top of each sub-figure. Figures in modified form taken from Ref. \cite{CLAS:2023akb}.}
	\label{fig:pimdelta_result}
\end{figure}

Fig. \ref{fig:pimdelta_result}, shows a comparison of the obtained structure function ratio $\sigma_{LT'}/\sigma_{0}$ for $\pi^{-}\Delta^{++}$ with results from $\pi^{+}n$ \cite{CLAS:2022iqy} and $\pi^{0}p$ \cite{CLAS:2023wda}.
The figure shows that the structure function ratio of $\pi^- \Delta^{++}$ shows an opposite sign than for $\pi^+ n$ and an approximately doubled magnitude compared to $\pi^+ n$. The different signs of the asymmetry can be related to an interaction of the virtual photon with different quarks ($u$, $d$) for the two reactions and the opposite sign of the quark polarization. However, the large absolute magnitude of $\sigma_{LT'}/\sigma_{0}$  for $\pi^{-}\Delta^{++}$, compared to $\pi^{+}n$ can be seen as a clear effect of the excitation process, which is encoded in the transition GPDs \cite{CLAS:2023akb}. 

The studied polarized structure function $\sigma_{LT'}$ is given by products of convolutions of 
{tensor}
and helicity non-flip transition GPDs with sub-process amplitudes and shows the following relation to the two dominant 
{tensor}
transition GPDs \cite{CLAS:2023akb}:
\begin{equation}
	\sigma_{LT'} \sim \sqrt{-t'}~~{\rm Im}\left[ G^{3}_{T_{5}} \cdot A +  c~G^{3}_{T_{7}} \cdot A' \right],
\end{equation}
with an unknown kinematic factor $c$ and helicity amplitudes for longitudinally polarized virtual photons $A$ and $A'$, which are determined by the helicity non-flip transition GPDs $\widetilde{G}_3$ and $\widetilde{G}_4$ within the large $N_{c}$ limit.
The 
{tensor}
transition GPDs $G^{3}_{T_{5}}$ and $G^{3}_{T_{7}}$ can be related to the ground state 
{tensor}
GPD $H_{T}$ in the large $N_{c}$ limit, as shown in 
Eq.~(\ref{eqnHt}). Therefore, they are expected to be sensitive to the tensor charge of the resonance following Eq.~(\ref{eqn:tensorchargeGPD}).

Based on CLAS12 data with a longitudinally polarized $NH_{3}$ target, it is planned to extend the study to target polarization observables like $A_{UL}$ and $A_{LL}$. Furthermore, the extraction of absolute cross sections ($\sigma_{0} = \sigma_{L} + \sigma_{T}$) and the $\cos(\phi)$ ($\sigma_{LT}$) and $\cos(2\phi)$ ($\sigma_{TT}$) modulations of the cross-section is planned based on data from an unpolarized proton target.


\subsection{Further $N \to N^{*}$ DVMP channels with CLAS12}
\label{sec:exp_DVMP2}

As shown in section \ref{sec:NtoNstarDVMP}, different $N \to N^{*}$ DVMP channels can be used to study transition GPDs, and a combination of data from different mesons in the final state will allow a flavor decomposition of the transition GPDs.

Currently, the feasibility of studying different channels, like 
\begin{itemize}
    \item~~ $e p \to e' \pi^+ \Delta^{0} \to e' \pi^+ p \pi^-$,
    \item~~ $e p \to e' \pi^+ \Delta^{0} \to e' \pi^+ n \pi^0$,
    \item~~ $e p \to e' \pi^0 \Delta^{+} \to e' \pi^0 n \pi^+$,
\end{itemize}
is under investigation. Since the electromagnetic calorimeter of CLAS12 is limited to a polar angle below 35$^{\circ}$, only resonance decays with a charged pion can be efficiently detected. However, if the DVMP meson is a charged pion, the resolution of CLAS12 is sufficient to consider the application of the missing mass technique to reconstruct the pion from the resonance decay.  

\begin{figure}[t]
	\centering
		\includegraphics[width=0.4\textwidth]{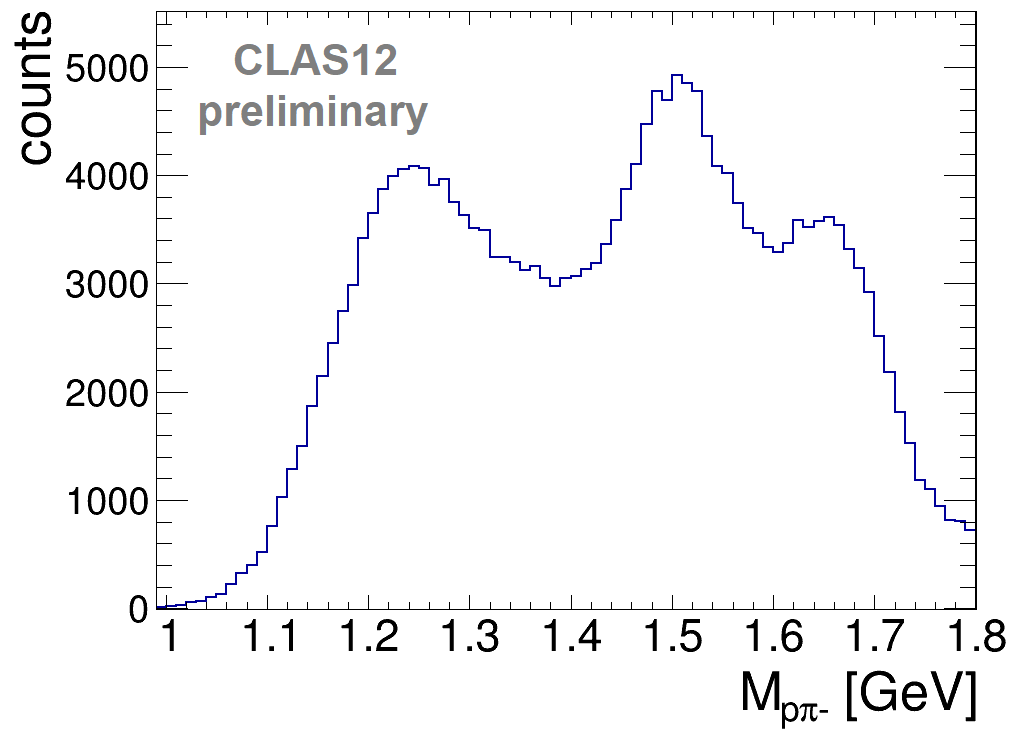}
	\caption{$p \pi^-$ invariant mass for the $e' \pi^+ \Delta^{0} \to e' \pi^+ p \pi^-$ DVMP process for $-t <$~1~GeV$^{2}$, after a cut on $M(\pi^+\pi^-)$~$>$~1.1~GeV to reject the exclusive $\rho$ events, measured with CLAS12.}
	\label{fig:DVMP_spectrum_CLAS12}
\end{figure}

As an example, Fig. \ref{fig:DVMP_spectrum_CLAS12} shows the $p \pi^-$ invariant mass for the $e' \pi^+ \Delta^{0} \to e' \pi^+ p \pi^-$ DVMP process.
It can be clearly seen that, in contrast to the $\pi^-[p\pi^{+}]$ final state, which was discussed in the last section and which can only be populated by $\Delta$ resonances, the $\pi^+ [p \pi^-]$ final state shows the complete resonance spectrum with 
nucleon and $\Delta$ resonances. Besides the $\Delta(1232)$ peak, the second and third resonance regions are clearly visible in Fig. \ref{fig:DVMP_spectrum_CLAS12}. All other $N \to N^*$ DVMP channels show a similar picture. Therefore, a clean separation of the $\Delta(1232)$ events is more difficult for these channels, and either a full fit of the spectrum, including non-resonant background, or a partial wave decomposition has to be performed to access the $\Delta(1232)$ events, and especially the higher resonances with a higher mass.


\subsection{$\pi^+\Delta^{0}$ electroproduction in JLab Hall C}
\label{sec:hallc_12gev}

Hall C at JLab \cite{blok2008charged} is a unique facility that gives access to high-precision DVMP reaction measurements over a wide range of kinematics.  The hall has a fixed target geometry with two focusing magnetic spectrometers, which detect the scattered electron and final state meson in coincidence. The final state hadron is reconstructed to high resolution through missing mass.  Hall C is also the only facility worldwide with the capability to perform high $Q^2$ Longitudinal/Transverse (L/T) separations of the reaction cross-section via the Rosenbluth separation process. This is done by taking data at two values of $\epsilon$ (virtual photon polarization factor) for each kinematic point, where  

\begin{equation}
\centering
    {\epsilon} = 1+ 2 \frac{(E_e - E_{e^{'}})^{2} + Q^2}{Q^2}\tan^{2}\frac{\theta_{e^{'}}}{2}
\end{equation}

\begin{figure}[t]
\centering
\includegraphics[width=0.48\textwidth]{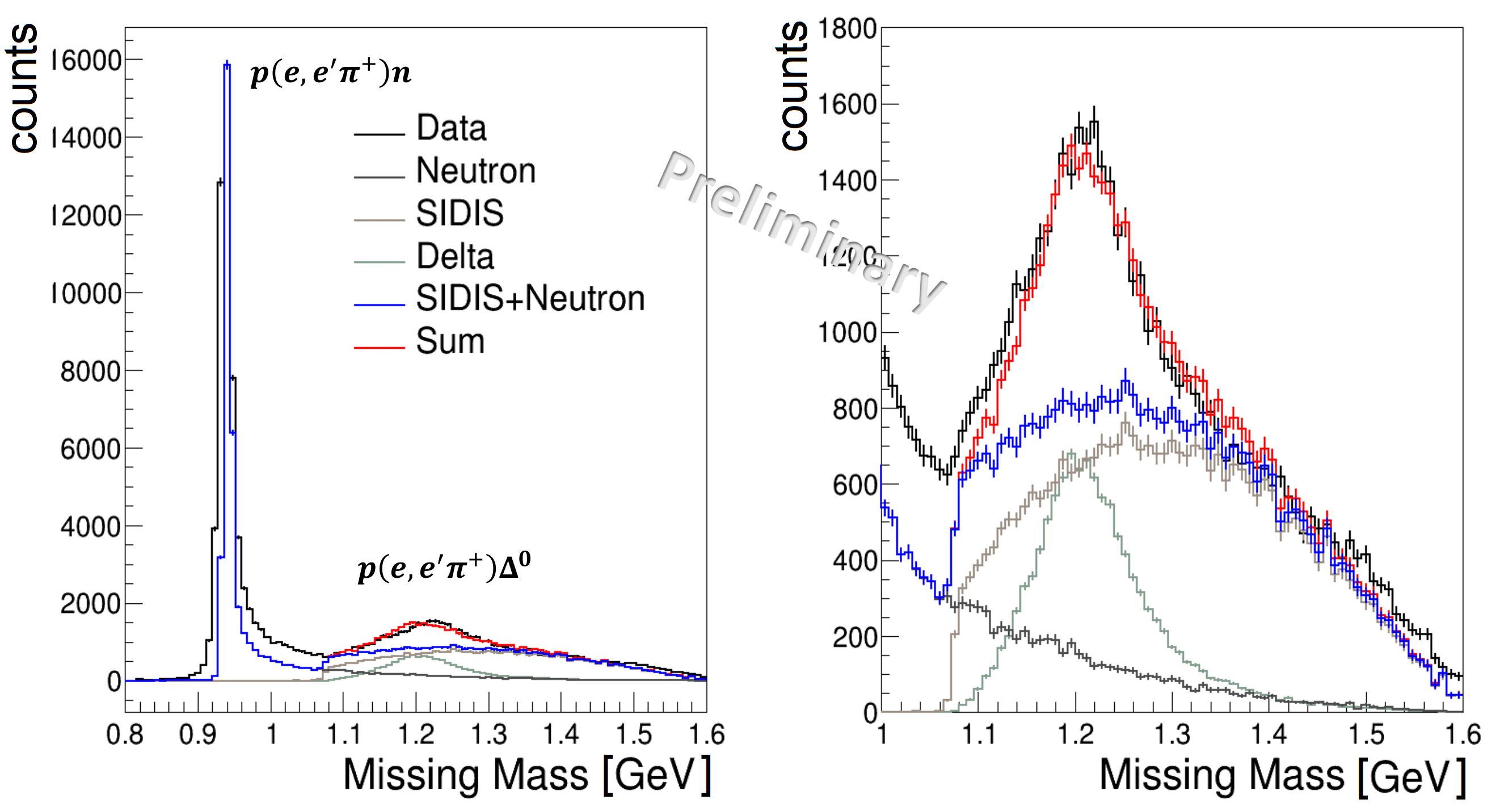}
\caption{Missing mass distribution for exclusive pion electroproduction reaction from Kaon-LT data ($Q^2=2.1$ GeV$^2$, $W=2.95$ GeV) from JLab hall C. All the curves except the data are from Monte Carlo simulations. The fits of the simulations to the data are preliminary.}
\label{mmfull}
\end{figure}

The Kaon-LT experiment (E12-09-011) \cite{JLABPAC:KaonLT} is the first L/T separation experiment after the 12 GeV upgrade to study exclusive $K^+$ electroproduction reaction $p(e,e^{\prime}K^{+})\Lambda$. The experiment also gathered high-statistics exclusive $\pi^+$ electroproduction data, as it is one of the background processes for the exclusive $K^+$ channel. The exclusive $\pi^+$ data include both the $p(e,e^{\prime}\pi^{+})n$ and $p(e,e^{'} \pi^{+})\Delta^0$ reactions. Table \ref{kaonkin} lists all settings collected to date during the Kaon-LT experiment. The BSA analysis is ongoing for the listed 10.6 GeV kinematic settings.  In this experiment, the meson spectrometer is rotated $3^{\circ}$ 
on either side of the $\vec{q}$-vector to get full $\phi$ coverage at fixed t, needed for the BSA measurement.

\begin{table}[t]
\begin{center}
\begin{tabular}{| c | c | c | c |}
 \hline
 $E_b$ (GeV) & $Q^2$ ($GeV^2$/$c^2$) & W (GeV) & $\epsilon_{high}$/$\epsilon_{low}$ \\ [0.5ex]
 \hline
 10.591/8.213 & 5.5 & 3.02 & 0.53/0.18 \\ 
 \hline
 10.591/8.213 & 4.4 & 2.74 & 0..72/0.48 \\
 \hline
 10.591/8.213 & 3.0 & 3.14 & 0.67/0.39 \\
 \hline
 10.591/6.187 & 3.0 & 2.32 & 0.88/0.57 \\
 \hline
 10.590/6.187 & 2.115 & 2.95 & 0.79/0.25 \\  
 \hline
 4.930/3.834 & 0.5 & 2.40 & 0.70/0.45 \\ [0.5ex]
 \hline
\end{tabular}
\caption{The kinematics settings acquired to date in the KaonLT experiment \cite{JLABPAC:KaonLT}.}
\label{kaonkin}
\end{center}
\end{table}

\begin{figure}[t]
\centering
\includegraphics[width=0.43\textwidth]{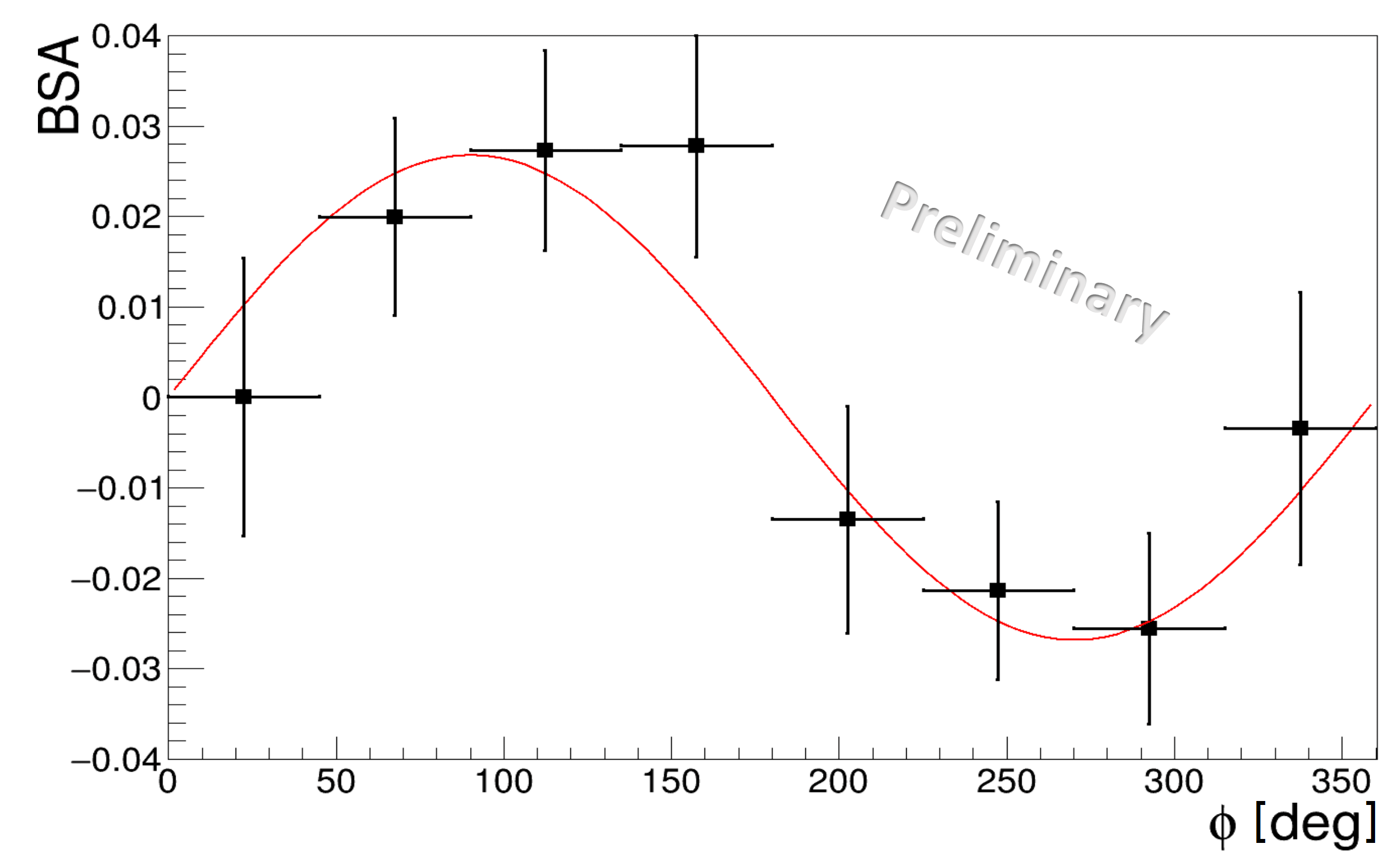}
\caption{Preliminary Beam Spin Asymmetry results for the $p(e,e^{\prime}\pi^{+})\Delta^0$ reaction from the KaonLT experiment ($Q^2=2.1$ GeV$^2$, $W=2.95$ GeV) in JLab hall C. Only statistical errors are shown. The final results will include three $\times$ more data after including all three meson spectrometer settings.}
\label{bsa}
\end{figure}

The Hall C spectrometers give access to very good missing mass resolution. Figure \ref{mmfull} shows the missing mass distribution for $p(e,e^{\prime}\pi^{+})X$ including experimental data and different simulated processes. The plot on the left gives the full range of missing mass, which clearly shows a clean $p(e,e^{\prime}\pi^+)n$ peak. The figure on the right is a zoomed-in version to focus on the $\Delta^0$ region. The $p(e,e^{\prime}\pi^{+})n$ sample has very little background underneath; therefore, it can be selected by using a missing mass cut around the peak. On the other hand, the $p(e,e^{\prime}\pi^{+})\Delta^0$ reaction has a Breit-Wigner distribution that sits atop other multi-pion processes. This requires a background shape study to cleanly isolate the $p(e,e^{\prime}\pi^{+})\Delta^0$ reaction. A shape study was performed for each $\phi$ bin for each beam helicity state by fitting the $\Delta^0$ Monte Carlo to the background-subtracted experimental data, and the BSA  was calculated.  Figure \ref{bsa} shows preliminary $p(e,e^{\prime}\pi^{+})\Delta^0$ BSA results for one kinematic setting. A parallel analysis of $p(e,e^{'} \pi^{+})n$ BSA is also in progress for the same kinematics. The $\Delta^0$ BSA paper will include the $\sigma_{LT'}$ ratio for both $n$ and $\Delta^0$ final states.

The future plan is to perform an L/T separation for all kinematics listed in Table \ref{kaonkin} for both $p(e,e^{\prime}\pi^{+})\Delta^0$, $p(e,e^{'} \pi^{+})n$ reactions.  This involves a more detailed understanding of all the experimental systematics as well as fine-tuning the simulation models. The L/T separated cross-section ratio for $p(e,e^{\prime}\pi^{+})\Delta^0$ and $p(e,e^{\prime}\pi^{+})n$ can give access to $N\rightarrow\Delta^0$ transition GPDs. The KaonLT experiment is anticipated to collect more data to complete the proposed kinematics. There is also the possibility of increasing the kinematic coverage with a future experiment using the Hall C experimental setup. An energy upgrade at Jefferson Lab will allow a further increase in kinematic coverage for the $p(e,e^{\prime}\pi^{+})\Delta^0$ L/T separation experiment.


\subsection{$N \to \Delta^{+}$ DVCS with CLAS12}
\label{sec:CLAS12}

 First feasibility studies for the non-diagonal DVCS process with limited statistics and kinematic coverage, not allowing a real background separation, have been performed based on CLAS data from the 6~GeV era in 2009 \cite{transGPDMor09}. 
 Studies of associated electroproduction of real photons, $e + p \rightarrow e' \gamma \pi N$, in the $\Delta(1232)$-resonance region were also performed with HERMES \cite{Duren:2014ola}, but similarly to the first CLAS studies, they also suffered from limited statistics and missing background separation.
 The 12~GeV upgrade of JLab finally allowed initial measurements of the $N \to N^{*}$ DVCS and DVMP processes at reasonably high $Q^{2}$ values and with sufficient statistics for the first time. For this process, the detection of all final state particles is important for a proper rejection of potential backgrounds.
For~~$\gamma^{*}~p \rightarrow~~ \gamma ~ \pi^+ ~ n$ $N \to N^{*}$ DVCS, two final states are possible based on the $N^{*+}$ decay:
\begin{enumerate}
  \item ~~$\gamma^{*}~p ~~\rightarrow~~ \gamma ~ N^{*+} ~~\rightarrow~~ \gamma ~ n ~ \pi^+$
  \item ~~$\gamma^{*}~p ~~\rightarrow~~ \gamma ~ N^{*+} ~~\rightarrow~~ \gamma ~ p ~ \pi^0$
\end{enumerate}
With a deuterium (neutron) target, also the following reactions can be studied:
\begin{enumerate}
  \item ~~$\gamma^{*}~n ~~\rightarrow~~ \gamma ~ N^{*0} ~~\rightarrow~~ \gamma ~ p ~ \pi^-$
  \item ~~$\gamma^{*}~n ~~\rightarrow~~ \gamma ~ N^{*0} ~~\rightarrow~~ \gamma ~ n ~ \pi^0$
\end{enumerate}
In all cases, the $N^{*}$ can be a nucleon or a $\Delta$ resonance.
For the very forward regime of the DVCS photon (low $-t$), the $N^{*}$ and its decay products are expected to be detected under central to backward lab angles. Therefore, the detection capabilities of neutral pions, originating from a resonance decay, are very limited with CLAS12 due to the missing electromagnetic calorimeter in the region of $\theta > 35^{\circ}$, leading to a limitation of the studies to the $\gamma ~ n ~ \pi^+$ final state for a proton target and to the $\gamma ~ p ~ \pi^-$ final state for a neutron target.

Based on data taken with a 10.2~GeV and a 10.6~GeV longitudinally polarized electron beam and a hydrogen target, a first study has been performed with CLAS12. A series of exclusivity cuts on the missing mass, missing energy, missing transverse momentum, and missing cone angle are applied to select exclusive events. In addition, kinematic cuts on W $>$ 2 GeV, $Q^2$ $>$ 1.5 GeV$^{2}$, y $<$ 0.8 and $E_{\gamma-DVCS}$ $>$ 2 GeV have been applied for the preliminary study.
Furthermore, a cut on $M(\pi^+\gamma)$~$>$~1~GeV is applied to reduce the dominant background from exclusive $\rho^+$ production ($\gamma^{*}~p ~~\rightarrow~~ \rho^+ ~ n$), with the $\rho^+$ decaying into $\pi^0 \pi^+$, with one undetected photon, and the much less frequent decay into $\gamma \pi^+$.

\begin{figure}[t]
	\centering
		\includegraphics[width=0.45\textwidth]{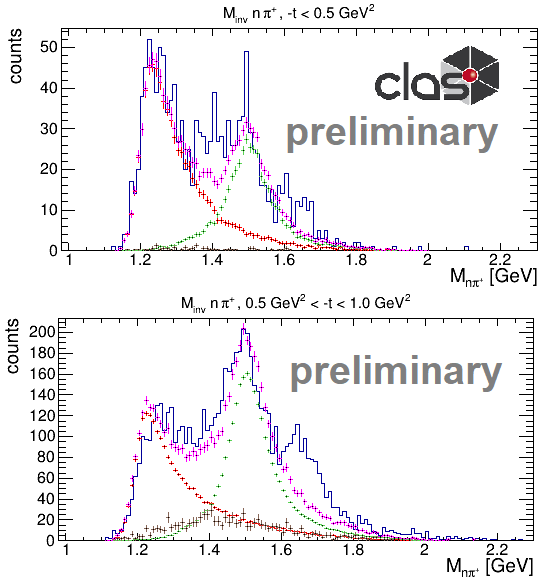}
	\caption{Preliminary measured resonance spectrum in the $n\pi^+$ invariant mass of the $e~p ~~\rightarrow~~ e^{\prime} ~ \gamma ~ n ~ \pi^+$ process from CLAS12 (blue). The kinematic and selection cuts described in the text are applied, and the $-t$ range is stated on top of the sub-figures. For comparison, MC simulations for the $N \to N^{*}$ DVCS process with the production of a $\Delta(1232)$ resonance (red) and for the production of the resonances in the second resonance region (green), following the model prediction in Fig. \ref{fig:2ndres_5f}, as well as MC simulations for the non-resonant contributions (brown), are shown. The magenta histogram provides the sum of all MC contributions.}
	\label{fig:DVCS_resonance_spectrum_CLAS12}
\end{figure}

Figure \ref{fig:DVCS_resonance_spectrum_CLAS12} shows the resonance spectrum of the $n\pi^+$ invariant mass after the listed kinematic cuts for two bins of $-t$, measured with CLAS12.
A clear peak from the delta resonance as well as from the second resonance region can be observed in both ranges of $-t$, while for the higher $-t$ bins, also the third resonance region exceeds the production threshold and becomes visible.

\begin{figure*}[ht]
	\centering
		\includegraphics[width=0.9\textwidth]{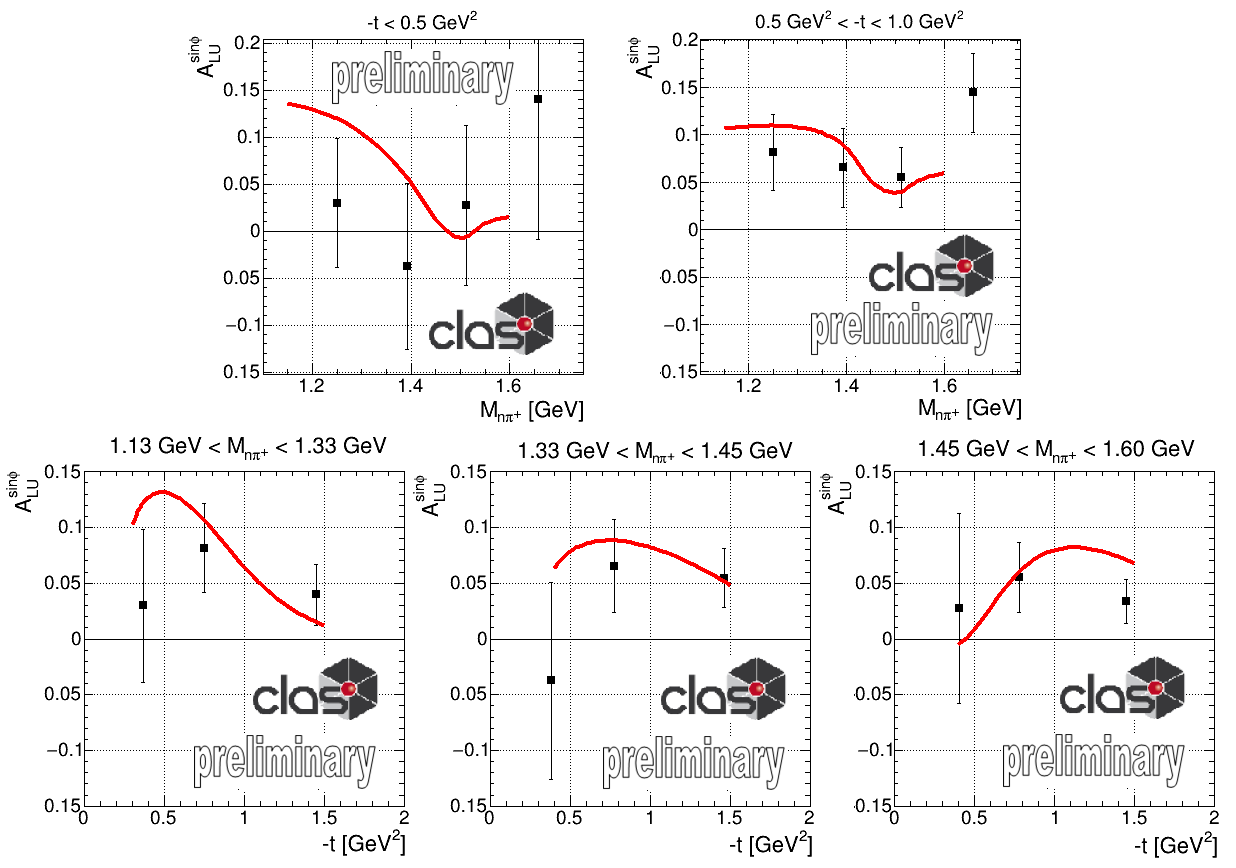}
	\caption{Preliminary results of the $A^{\sin\phi}_{LU}$ moment of the beam-spin asymmetry from the $e~p ~~\rightarrow~~ e^{\prime} ~ \gamma ~ n ~ \pi^+$ process, with the cuts described in the text, as a function of the resonance mass $M_{n\pi^{+}}$ in two bins of $-t$, as stated above the figures (upper row), and as a function of $-t$ in the region around the $\Delta(1232)$ resonance (1.13 GeV $< M_{n\pi^{+}} <$ 1.33 GeV, lower row left), the Roper resonance (1.33 GeV $< M_{n\pi^{+}} <$ 1.45 GeV, lower row center) and the second resonance region (1.45 GeV $< M_{n\pi^{+}} <$ 1.60 GeV, lower row right). Backgrounds from the $e~p ~~\rightarrow~~ e^{\prime} ~ \pi^{0} ~ N^{*} ~~\rightarrow e^{\prime} ~ \pi^{0} ~ n ~ \pi^+$ process, with one photon from the $\pi^0$ decay misidentified as the DVCS photon are not subtracted yet. The red line shows the prediction for the $N \to N^{*}$ DVCS process from the transition GPD-based model described in section \ref{sec:NtoNstarDVCS}, integrated over the CLAS12 kinematic distributions.}
	\label{fig:DVCS_BSA_Delta_CLAS12}
\end{figure*}
Figure \ref{fig:DVCS_BSA_Delta_CLAS12} shows a preliminary extraction of the $A^{\sin\phi}_{LU}$ moment of the BSA as a function of the resonance mass in different bins of $-t$ and as a function of $-t$ in different regions of the resonance mass $M_{n\pi^{+}}$, in comparison to the transition GPD-based theoretical predictions described in section \ref{sec:NtoNstarDVCS}.

The study is currently ongoing, and further backgrounds like the $N \to N^{*}~\pi^0$ DVMP process, with one photon from the $\pi^0$ decay misidentified as the DVCS photon, need to be considered and subtracted. The contamination from $N \to N^{*}~\pi^0$ DVMP was found to be in the overall order of $\approx$~30~\% for the shown data sample and shows an asymmetry with the same sign but a smaller magnitude than the $N \to N^{*}$ DVCS process. A subtraction is therefore expected to lead to a slight increase of the measured $A^{\sin\phi}_{LU}$ moments.
Considering these uncertainties, the preliminary results for distributions and beam spin asymmetries show a promising agreement with the transition GPD-based theory predictions described in section \ref{sec:NtoNstarDVCS} and shown in Fig. \ref{fig:DVCS_BSA_Delta_CLAS12}.

Similar to the $N \to N^{*}$ DVMP processes described in sections \ref{sec:exp_DVMP1} and \ref{sec:exp_DVMP2}, also for the $N \to N^{*}$ DVCS processes, studies with further target polarizations as well as the extraction of the unpolarized cross-section terms are possible based on CLAS12 data.


\section{Transition processes: hadron scattering}
\label{sec:transGPDhadrons}

\subsection{Meson-induced exclusive Drell-Yan process}

Besides the discussed exclusive lepton scattering reactions, 
the exclusive hard reactions with the hadron beam are expected 
to be sensitive to the GPDs. The most promising reaction is the meson-induced 
exclusive Drell-Yan process \cite{Berger:2001zn,Sawada:2016mao}, 
which can be studied at J-PARC or CERN COMPASS/AMBER with the pion and kaon beams. 
In addition, the primary proton beam at J-PARC could be used for the GPDs 
\cite{Kumano:2009he}. These processes are described in detail below.


Recently, JLab CLAS reported the first measurement of timelike Compton Scattering (TCS) process~\cite{CLAS:2021lky}. The comparison of the measured polarization asymmetry of TCS with model predictions suggests the universality of GPDs in both spacelike and timelike processes. In the same analogy between TCS and DVCS processes, the meson-induced exclusive process is the complementary timelike process to access the nucleon GPDs, with respect to the DVMP process.

\begin{figure}[t]
	\centering
 \includegraphics[width=0.40\textwidth]{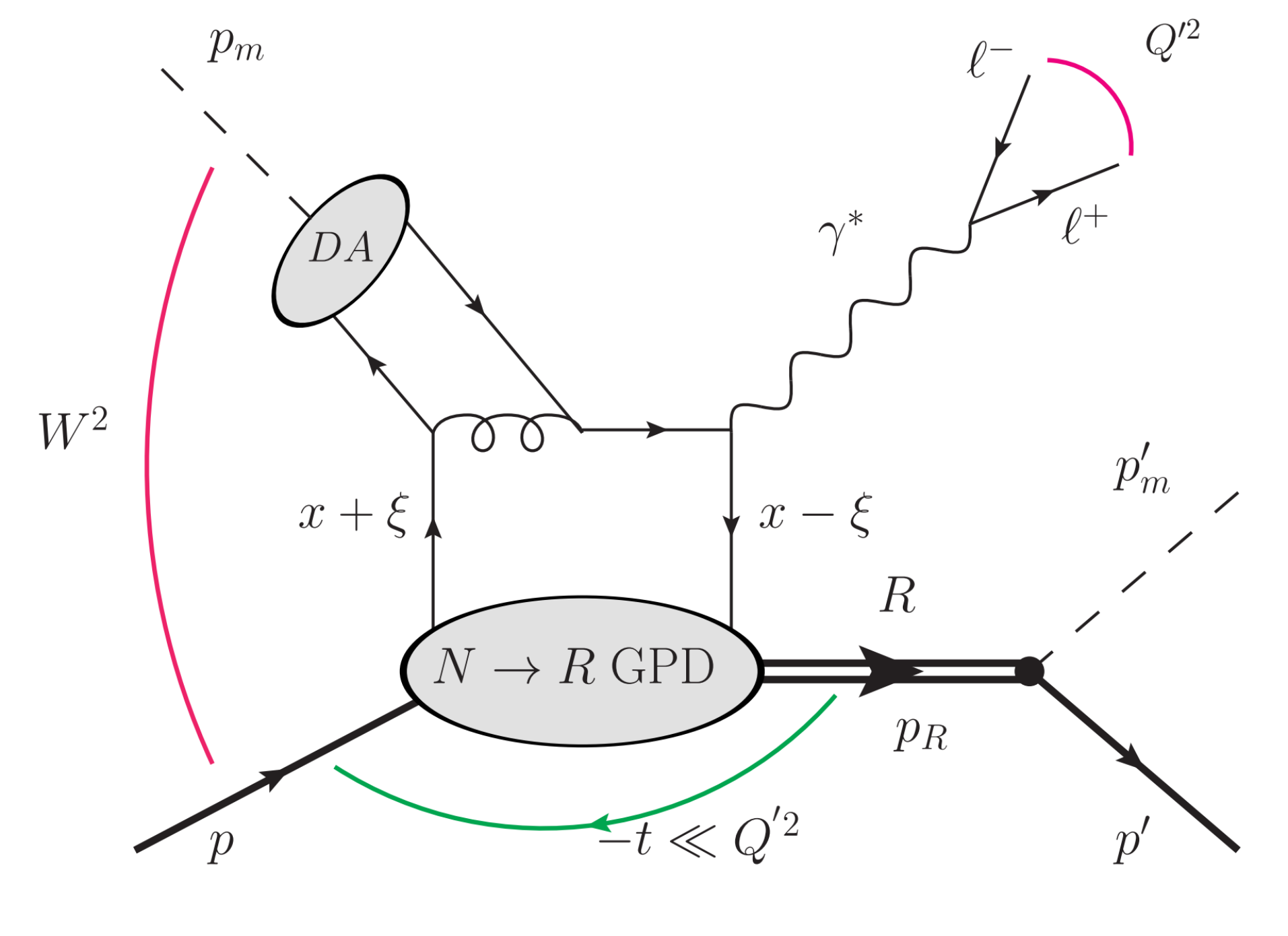}
		\caption{Schematic drawing of the exclusive Drell-Yan process. For a ground-state nucleon in the final state, the process is sensitive to the ordinary GPDs, while a baryon resonance in the final state provides sensitivity to the transition GPDs. }
	\label{fig:GPD_hadr_exclDY}
\end{figure}

The exclusive meson-induced Drell-Yan process $m N \rightarrow \ell^{+} \ell^{-} B$, as shown in Fig.~\ref{fig:GPD_hadr_exclDY}, describes the interaction of a meson beam ($m$) with a nucleon target ($N$) during which a lepton pair ($\ell^{\pm}$) and a ground state nucleon or a baryon resonance ($B$) are produced. 
The process is expected to factorize if the momentum transfer squared $t$ between the initial and final state baryon is small and the four-momentum squared $Q^{\prime 2}$ of the final state virtual photon is large. However, potential large contributions of non-factorized processes still need to be evaluated in more detail and considered for the analysis and interpretation of the measurements, as shown in Refs.~\cite{Tanaka:2017ccr}.

The exclusive Drell-Yan process $\pi(p_m) N(p) \to \gamma^* (q') N(p') \ (\to \ell^+ \ell^- N)$ is described by the following variables: 
the virtual-photon momentum squared $Q'^2 =q'^2$, 
the invariant momentum transfer $t=(p'-p)^2$,
the scaling variable $\tau$ given by
\begin{equation}
\tau = \frac{Q'^2}{2 p\cdot p_m} 
\simeq \frac{Q'^2}{s-M_N^2} 
\simeq \frac{Q'^2}{s} ,
\label{eqn:tau}
\end{equation}
where $s=(p+p_m)^2$ and $M_N$ is the nucleon mass.
The variable $\tau$ is similar to the Bjorken variable
in the DVCS 
and DVMP,
and it is related to the pion momentum as 
$q \simeq s/(2M_N) \simeq Q'^2/(2M_N\tau)$.
The skewness variable $\xi$ is approximately given by $\tau$ as 
\begin{equation}
\xi \simeq \frac{Q'^2}{2s - Q'^2} = \frac{\tau}{2 - \tau}.
\label{eqn:xi2}
\end{equation}

Using these kinematical variables, we write the 
$\pi^- N \to \ell^+ \ell^- N$ 
cross-section up to twist-4 terms as
\cite{Goloskokov:2009ia,Goloskokov:2015zsa}
\begin{align}
\frac{d\sigma}{dt dQ'^2 d\cos\theta d\phi} 
&= \frac{3}{8\pi} \biggl( \sin^2\theta \frac{d\sigma_L}{dt dQ'^2} 
+ \frac{1+\cos^2\theta}{2} \frac{d\sigma_T}{dt dQ'^2} 
\nonumber \\ 
& \ \hspace{-1.7cm}
+ \frac{\sin2\theta \cos\phi}{\sqrt{2}} \frac{d\sigma_{LT}}{dt dQ'^2} 
+ \sin^2\theta \cos2\phi \frac{d\sigma_{TT}}{dt dQ'^2} \biggr),
\label{eqn:cadall}
\end{align}
where the $\theta$ and $\phi$ are polar and azimuthal angles
of a lepton direction in the final state.
The cross-section $d\sigma_L$ ($d\sigma_T$)
is for the longitudinally (transversely)-polarized 
virtual photon.
The $d\sigma_{LT}$ and $d\sigma_{TT}$
are the longitudinal-transverse interference 
and transverse (helicity $h=1$)-transverse ($h=-1$)
interference terms.
The cross-section $d\sigma_L$ is leading twist,
$d\sigma_{LT}$ is twist three,
and $d\sigma_T$ and $d\sigma_{TT}$ 
are twist-four terms.

For the reaction $\pi^{-} N \rightarrow \ell^{+} \ell^{-} N$, Refs.~\cite{Berger:2001zn,Sawada:2016mao} show that the longitudinal part of the cross section $\sigma_{L}$ can be related to the GPDs $\widetilde{H}$ and $\widetilde{E}$. The transverse part of the cross sections $\sigma_{T}$ and the $LT$ and $TT$ interference terms $\sigma_{LT}$ and $\sigma_{TT}$ are also sensitive to transversity GPDs as shown in Ref.~\cite{Goloskokov:2015zsa}. For reactions with a baryon resonance in the final state, this can be related to the corresponding transition GPDs. However, a formalism is not available at the present stage.

In this process, the relevant amplitude is expressed as the convolution of the short-distance partonic annihilation processes with the two parts of long-distance nature, associated with the nucleon GPD and the pion DA. 

The factorization has been proven for the DVMP processes at the
leading twist, including the exclusive electroproduction of a pion,
$\gamma^* N \to \pi N$~\cite{Collins:1996fb}.  The amplitude can be written in terms of the hard-scattering processes at the parton level, combined with the pion distribution amplitude (DA), $\phi_\pi$, and also the nucleon GPDs, $\tilde{H}$ and $\tilde{E}$.
By interchanging the initial $\gamma^*$ and final $\pi$
in the exclusive electroproduction of pion, and replacing the spacelike momentum of $\gamma^*$ by the timelike momentum,
the factorization at twist-2 is argued to be applicable to the exclusive Drell-Yan process, $\pi(q) N(p) \to \gamma^*(q') N(p')$, with the same universal non-perturbative input~\cite{Berger:2001zn}.

The appropriate kinematical region is of large timelike virtuality $Q'^2 = q'^2$ at fixed $t=(p'-p)^2$ and fixed scaling variable $\tau \equiv Q'^2 / (2 p\cdot q )$ where the $q$, $p$ and $p'$ are the momenta of the pion, initial, and final nucleons, respectively. At the large $Q'$ scaling limit, the corresponding leading-twist cross-section of $\pi^- p \to \gamma^* n$ as a function of $t$ and $Q'^2$ is expressed in terms of convolution integrals $\tilde{\cal H}^{du}$ and $\tilde{\cal E}^{du}$, as follows~\cite{Berger:2001zn}
\begin{align}
\left.\frac{d\sigma_L}{dt dQ'^2}\right|_{\tau}
&= \frac{4\pi \alpha_{\rm em}^2}{27}\frac{\tau^2}{Q'^8} f_\pi^2\, \Bigl[ (1-\xi^2) |\tilde{\cal H}^{du}(\tilde{x},\xi,t)|^2 \nonumber \\
&- 2 \xi^2 \mbox{Re}\ \bigl( \tilde{\cal H}^{du}(\tilde{x},\xi,t)^* \tilde{\cal E}^{du}(\tilde{x},\xi,t) \bigr) \nonumber \\
&-  \xi^2 \frac{t}{4 M_N^2}|\tilde{\cal E}^{du}(\tilde{x},\xi,t)|^2 \Bigr],
\label{eq_dcross}
\end{align}
where $\alpha_{\rm em}$ is the fine structure constant, $f_\pi$ is the pion decay constant, and the variable $\tilde{x}$ is 
\begin{equation}
\tilde{x}=-\frac{(p_m+q')^2}{2(p+p') \cdot (p_m+q')}
 \simeq - \frac{Q'^2}{2s - Q'^2} = -\xi .
\label{eqn:barx}
\end{equation}

The convolution integral $\tilde{\cal H}^{du}$ involves two soft
objects: the nucleon GPD $\tilde{H}^{du}$ for the $p \to n$ transition and the twist-2 pion DAs $\phi_{\pi}$.  Using $\tilde{H}^{du} (x,\xi,t) = \tilde{H}^u (x,\xi,t) - \tilde{H}^d (x,\xi,t)$ to relate the transition GPD with the usual proton GPDs $\tilde{H}^q$ for quark
flavor $q=u,d$, the expression of $\tilde{\cal H}^{du}$ is given, at the leading order in $\alpha_s$, by~\cite{Berger:2001zn}
\begin{align}
  \tilde{\cal H}^{du}(\tilde{x},\xi,t) &= \frac{8}{3} \alpha_s \int_{-1}^1
  dz\, \frac{\phi_\pi(z)}{1-z^2} \nonumber \\ 
  &\times \int_{-1}^1 dx
  \Bigl( \frac{e_d}{\tilde{x}-x- i\epsilon} - \frac{e_u}{\tilde{x}+x- i\epsilon}
  \Bigr) \nonumber \\
  & \bigl( \tilde{H}^{d}(x,\xi,t) - \tilde{H}^{u}(x,\xi,t)
  \bigr),
\label{eq_Hdu}
\end{align}
where $e_{u,d}$ are the electric charges of $u,d$ quarks in units of the elementary charge, and $\alpha_s$ is the running coupling constant of QCD.
The corresponding expression of $\tilde{\cal E}^{du}$ is given by (\ref{eq_Hdu}) with $\tilde{H}^q$ replaced by the proton GPDs $\tilde{E}^q$.  Because of the pseudo-scalar nature of the pion, the cross sections (\ref{eq_dcross}) receive the contributions of $\tilde{H}$ and $\tilde{E}$ only, among the GPDs in
Eqs.~(\ref{eqn:gpd-n}) and ~(\ref{eqn:gpd-p}).
The pion distribution function $\phi_\pi (z)$ is 
defined in the region $-1 < z <1$ as
$\int_{-1}^1 dz \phi_\pi (z) =1$.
Use of the same pion beam also provides an opportunity
to access pion-to-nucleon transition distribution amplitudes
\cite{Pire:2021hbl}
through investigation of backward charmonium production
in pion-nucleon collisions $\pi^- p \to J/\psi \, n$ 
\cite{Pire:2016gut}.

To ensure proper factorization and sensitivity to the partonic
structure, $Q'$ has to be larger than 1 GeV and $|t|$ is large
enough. Both of these require a large center-of-mass energy
($\sqrt{s}$) for the reactions. Nevertheless, the estimated hard
exclusive Drell-Yan cross sections drop dramatically with an increase of $\sqrt{s}$. Furthermore, the experimental determination of exclusiveness for the measured reactions through a missing-mass technique favors the measurement of lepton tracks from the Drell-Yan process in an open apparatus, and thus the charged multiplicity has to be low enough for good tracking. Considering various experimental factors and constraints, it is found that measuring the exclusive Drell-Yan process in the coming high-momentum beamline at J-PARC with a 10-20 GeV $\pi^-$ beam ($\sqrt{s}=$ 4-6 GeV) is unique and optimized.

The planned E50 experiment at the high-momentum beam line~\cite{Noumi:2013yxa, Shirotori:2015eqa} of J-PARC will provide a spectrometer system of a large acceptance and good momentum resolutions. Thanks to the relatively low track density in the energy regime of J-PARC and the high-granularity tracking chambers of the E50 experiment, the measurement of the Drell-Yan process could be operated without the installation of a hadron absorber in front of the spectrometer. Without the multiple-scattering effect in the hadron absorber, a good momentum determination of muon tracks can be achieved so that the exclusive Drell-Yan process can be characterized via the missing-mass technique.

To identify the muon track, a muon identification ($\mu$ID) system after the spectrometer is necessary. The current design of the $\mu$ID  system consists of hadron absorber layers made of 20-cm concrete and 230 cm iron to absorb incoming hadrons, a large-size tracker resistive plate chambers (RPC) upstream of the absorber (2.4x1.8  m$^2$ active area with a few mm spatial resolutions), and another RPC downstream of the absorber (3.5x2.5 m$^2$ active area with 5cm spatial resolution). The thickness of concrete and iron shall be optimized considering the stopping power for low-momentum tracks and penetrating efficiency for high-momentum ones.

\begin{figure}[ht!]
\centering
\hspace{-0.5cm}
\subfigure[]
{\includegraphics[width=0.2\textwidth]{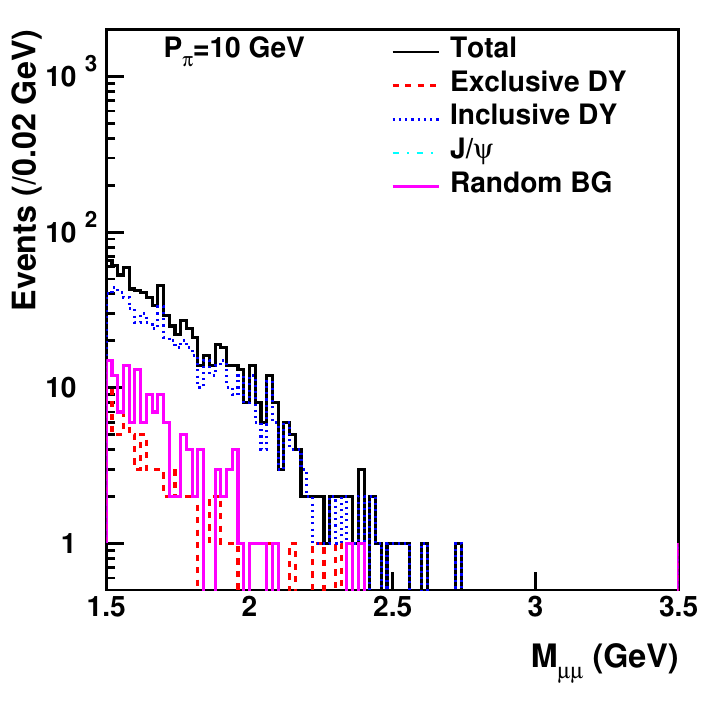}}
\hspace{-0.5cm}
\subfigure[]
{\includegraphics[width=0.2\textwidth]{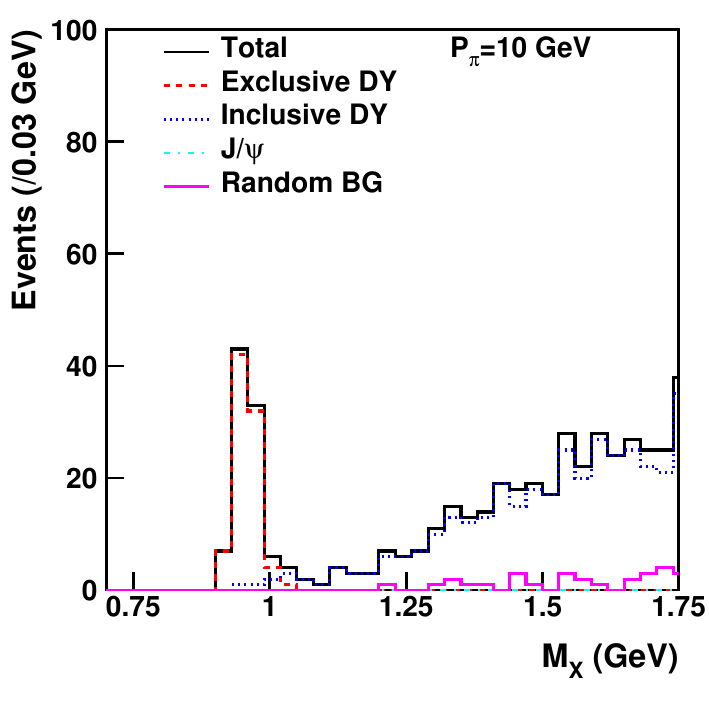}}
\hspace{-0.5cm}
\subfigure[]
{\includegraphics[width=0.2\textwidth]{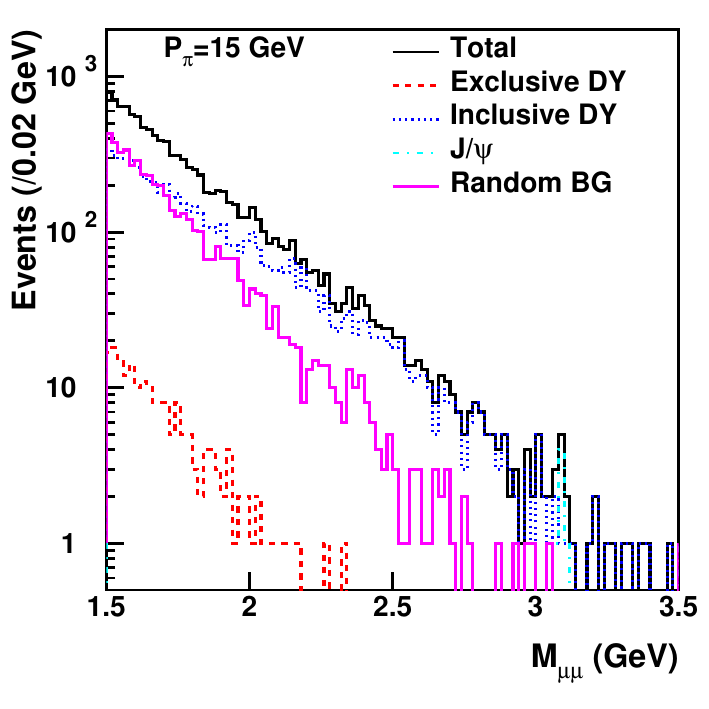}}
\hspace{-0.5cm} 
\subfigure[] 
{\includegraphics[width=0.2\textwidth]{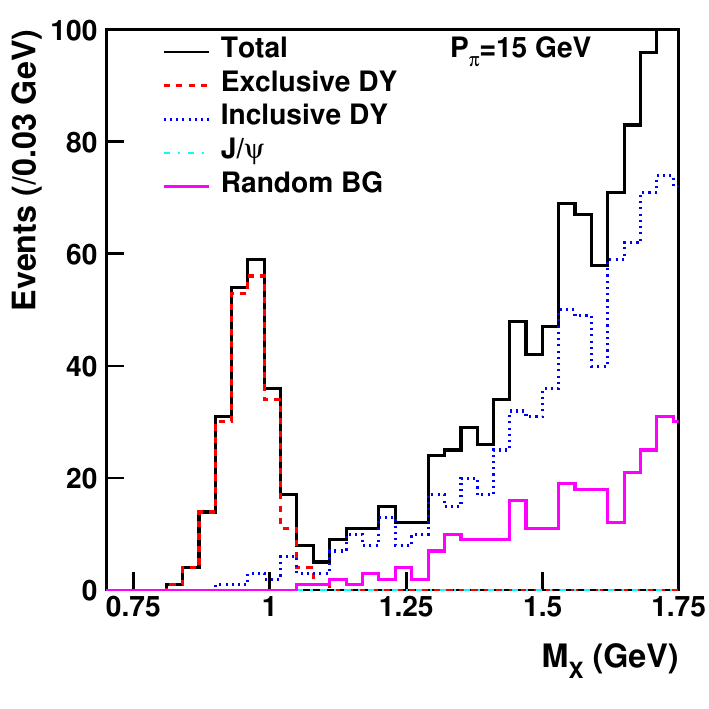}}
\hspace{-0.5cm} 
\subfigure[] 
{\includegraphics[width=0.2\textwidth]{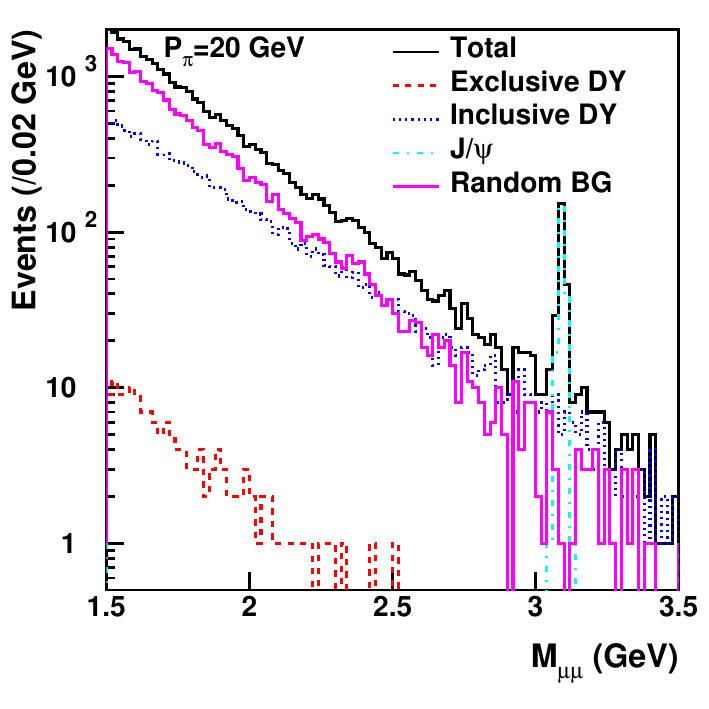}} \hspace{-0.5cm} 
\subfigure[] 
{\includegraphics[width=0.2\textwidth]{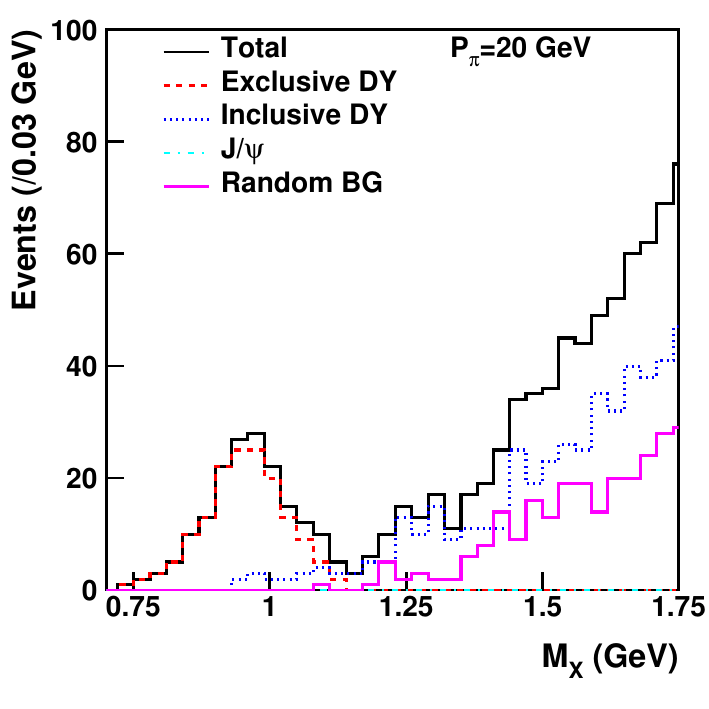}}
\caption{The Monte-Carlo simulated invariant mass $M_{\mu^{+}\mu^{-}}$ and missing-mass $M_{X}$ spectra of the $\mu^{+}\mu^{-}$ events with $M_{\mu^{+}\mu^{-}} >  1.5$ GeV and  $|t-t_{0}|<0.5$ GeV$^2$ for $P_{\pi}$=10, 15, and 20 GeV. Lines with different colors denote the contributions from
  various sources. The GK2013 GPD parameterization was used as an input for
  the evaluation of the exclusive Drell-Yan process.}
\label{fig:invm_mm}
\end{figure}

Using GK2013 GPDs~\cite{Kroll:2012sm} for the exclusive Drell-Yan process, a feasibility study for 50-day beam time was done in Ref.~\cite{Sawada:2016mao}. The simulated invariant mass $M_{\mu^{+}\mu^{-}}$ and missing-mass $M_{X}$ spectra of the $\mu^{+}\mu^{-}$ events with $M_{\mu^{+}\mu^{-}} > 1.5$ GeV and $|t-t_{0}|<0.5$ GeV$^2$ for $P_{\pi}$=10, 15, and 20 GeV are shown in Fig.~\ref{fig:invm_mm}, where $t_{0}$ ($= -4 M_N^2 \xi^2 / (1-\xi^2)$) is the limiting value of $t$ corresponding to the scattering angle in the center-of-mass system $\theta^{CM}=0$. Lines with different colors denote the contributions from various sources: exclusive Drell-Yan (red, dashed), inclusive Drell-Yan (blue, dotted), $J/\psi$ (cyan, dash-dotted), and random background (purple, solid), respectively. Signals of $J/\psi$ are only visible in the invariant mass distributions for $P_{\pi}$=15 and 20 GeV. 

The study shows that the exclusive Drell-Yan events could be identified by the signature peak at the neutron mass ($M_n \sim 0.940$ GeV) in the missing-mass spectrum for all three pion beam momenta, and the differential cross sections as a function of $t$ are good enough to differentiate two different GPD parameterizations. In terms of statistics and missing mass resolution, it was found that the measurement with a 15-GeV pion beam would be optimal.

\subsection{Proton-induced 2 $\rightarrow$ 3 processes}

GPDs can also be investigated by using the primary proton beam
through the $2 \to 3$ process $NN \to N M B$ \cite{Kumano:2009he},
where $M$ is a meson and $B$ is a baryon.
It is especially interesting that this process probes
the ERBL kinematical region of the GPDs.

The cross sections for $NN \to N \pi B$ are estimated by
using the pion- and rho-pole contributions to the GPDs
in the ERBL region \cite{Kumano:2009he}. 
Here, $B$ is a nucleon or $\Delta$.
We consider the process where the pion and the final nucleon 
have large and nearly opposite transverse momenta 
and a large invariant mass. Namely, the Mandelstam variables
satisfy the hard condition 
$s', |t'|, |u'|\gg {M_N^2}$ with $t'/s'={\rm const}$.
Under such a kinematical condition, an intermediate exchange
could be considered as a $q\bar q$ state. 
The $q\bar q$ attached to the nucleon is described by the GPDs 
in the ERBL region.

\begin{figure}[t]
\centering
\includegraphics[width=0.33\textwidth]{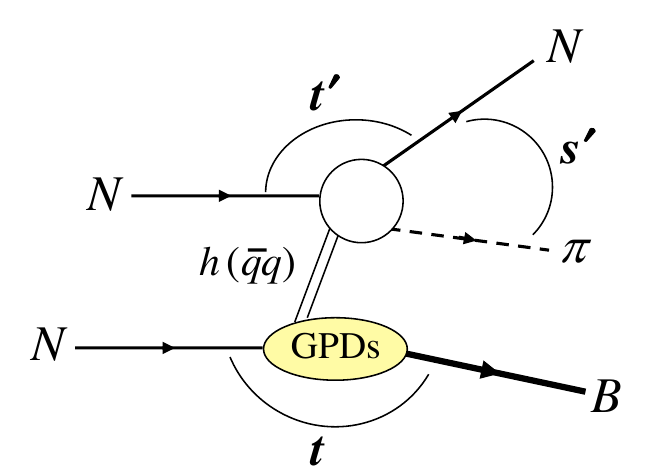}
\caption{Factorized process for the $NN \to N\pi B$ reaction
under the kinematical condition $s', |t'|, |u'|\gg M_N^2$ with $t'/s'={\rm const}$.}
\label{fig:gpd-2-3}
\end{figure}

The cross-section $NN \to N \pi B$ is described by the process
in Fig.\,\ref{fig:gpd-2-3}.
The process is factorized into two terms, the GPD part 
and the scattering of the intermediate hadronic $q\bar q$ state $h$ 
with the nucleon, as
$ {\cal M}_{NN \to N\pi B} 
 = {\cal M}_{N \to h B}  \, {\cal M}_{h N\to \pi N}$.
The transverse sizes of the projectile nucleon ($N$) and
the outgoing nucleon ($N$) and pion ($\pi$)
near the interaction point are of the order of
$\sim 1/\sqrt{|t'|}$. Therefore, the factorization and
the color transparency phenomena are closely related
\cite{Collins-1998}.
The factorization was shown for the DVMP process \cite{Collins-1998},
and such studies were also done recently, including $2 \to 3$ processes
\cite{Qiu:2022bpq, Qiu:2022pla}. Detailed factorization studies
would be needed for the $NN \to N \pi B$ processes.

The $NN \rightarrow N \pi B$ cross-section is expressed
in the factorized form, as \cite{Kumano:2009he}
\begin{align}
& \frac{ d\sigma _{NN \rightarrow N \pi B} }
       {dt \, dt'}
= \int_{y_{\min}}^{y_{\max}} dy \ 
\frac{s}{16 \, (2 \pi)^2 \, M_N \, p_N}
\nonumber \\
& \ \ 
\times 
\sqrt{ \frac{(ys-t-M_N^2)^2 -4M_N^2 t }
            {(s-2M_N^2)^2   -4M_N^4} }
\, \frac{d\sigma_{M N \to \pi N}(s'=ys,t')}{dt'}
\nonumber \\
& \ \ 
\times 
 \sum_{\lambda_N, \, \lambda_B}
 \frac{1}{[\phi_M (z)]^2}
 |{\cal M}_{N \to B}|^{2} .
\label{eqn:X-NN3}
\end{align}
where the variable $y$ is defined by $y =s' /s$,
and $d\sigma_{M N \to \pi N}/(dt')$ is 
the meson-nucleon elastic scattering cross-section.
This part can be estimated using the 
Brookhaven National Laboratory (BNL) data
by the E755 and E838 collaborations.
The GPDs are contained in the matrix element ${\cal M}_{N \to B}$.
If the final baryon $B$ is a nucleon, the matrix-element part
is given by the vector and axial terms as
$ |{\cal M}_{N \to N'}|^2 = 
  |{\cal M}_{N}^{V}|^2 + |{\cal M}_{N}^{A}|^2  $.
They are given by
\begin{align}
& 
\sum_{\lambda_{N}, \lambda_{N'}} |{\cal M}_{N}^{V}|^2
 = I_{N}^{\ 2} \bigg [ 8 (1-\xi^2) \{ H(x,\xi,t) \}^2
\nonumber \\[-0.30cm]
& 
+ 16 \xi^2 H(x,\xi,t) E(x,\xi,t) 
- \frac{t}{M_N^2} (1+\xi)^2 \{ E(x,\xi,t) \}^2 \bigg ] ,
\label{eqn:MV-N}
\end{align}
\vspace{-1.00cm}
\begin{align}
& 
\sum_{\lambda_{N}, \lambda_{N'}} |{\cal M}_{N}^{A}|^2
 = I_{N}^{\ 2} 
\bigg [ 8 (1-\xi^2) \{ \widetilde H(x,\xi,t) \}^2
\nonumber \\[-0.30cm]  
& \ \ \                  
+ 18 \xi^2 \widetilde H(x,\xi,t) \widetilde E(x,\xi,t) 
- \frac{2 \, t \, \xi^2}{M_N^2}
  \{ \widetilde E(x,\xi,t) \}^2 \bigg ] .
\label{eqn:MA-N}
\end{align}
Here, $I_N$ is the isospin factor for the nucleon defined by
$I_N = \langle 1/2 || \widetilde T || 1/2 \rangle
 \langle
 \frac{1}{2} m_N,\, 1 m \big |
 \frac{1}{2} m_N^\prime 
\rangle 
 /\sqrt{2}$,
and its actual values are
$I_N = 1, \ \sqrt{2} \ \ \text{for $p \rightarrow p$, $n$}$,
respectively. The GPDs $H$, $E$, $\tilde H$, and $\tilde E$
contribute to the cross-section.
In particular, they are estimated by considering 
the pion and rho as for the intermediate state $h$.
The $NN \to N \pi \Delta$ cross section is calculated
in the same way with the transition GPDs of $N \to\Delta$
\cite{Kumano:2009he}.

\begin{figure}[t]
\centering
\includegraphics[width=0.33\textwidth]{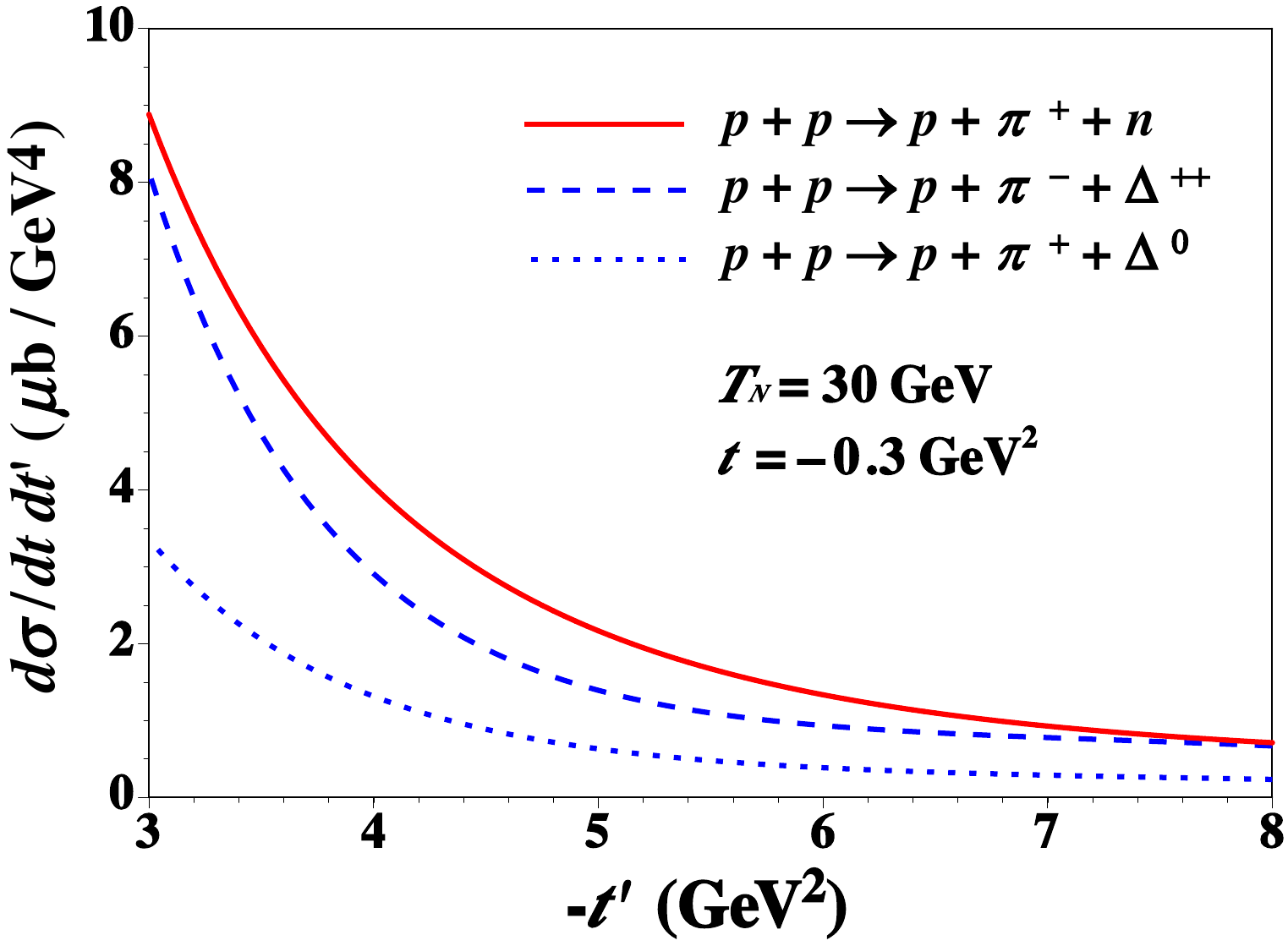}
\caption{The cross sections $d\sigma/(dt dt')$ are shown 
for $p + p \rightarrow p + \pi^+ + n$,
$p + p \rightarrow p + \pi^- + \Delta^{++}$,
and $p + p \rightarrow p + \pi^+ + \Delta^0$
as a function of $t'$.}
\label{fig:2-3-cross}
\end{figure}

The GPD measurements by the $2 \to 3$ processes are possible 
at hadron accelerator facilities, for example, at J-PARC.
It is a unique opportunity to investigate the ERBL kinematical region
of the GPDs, so such experimental measurements are complementary
to the DVCS and DVMP experiments at lepton accelerator facilities.
The $pp \to p \pi^+ B$ ($B=\Delta^0,\,n$) cross sections were 
calculated by considering the J-PARC kinematics.
In Fig.\ref{fig:2-3-cross}, the differential cross-sections 
$d\sigma/(dt dt')$ are shown for the three processes,
$p + p \rightarrow p + \pi^+ + n$,
$p + p \rightarrow p + \pi^- + \Delta^{++}$,
and $p + p \rightarrow p + \pi^+ + \Delta^0$,
as a function of $t'$ by taking 
the incident proton-beam energy 30 GeV 
and the momentum transfer $t=-0.3$ GeV$^2$ \cite{Kumano:2009he}.
Here, only the pion- and rho-pole contributions to the GPDs were
included as the nucleon GPDs for estimating the cross sections.

There are advantages to studying these reactions.
The cross sections are large, and they are 
of the order of $\mu \,$barn/GeV$^2$,
these processes are simple to measure experimentally,
the vacuum quantum number $0^+$ does not contribute
in the intermediate state, and
the $\Delta$ cross sections are as large as the nucleonic ones
for measuring the $N\to\Delta$ transition GPDs
in addition to the nucleon GPDs.

\subsection{Exotic hadron production using GPDs}

In recent years, there have been reports on exotic hadrons.
However, it is not straightforward to find a firm evidence
on the exotic nature. The GPDs of exotic hadrons,
including transition GPDs, could provide clear information
on the exotic structure by considering the following points.
The simplest form of the quark GPD $H$ could be written 
in the separable form, as \cite{Guidal:2004nd}
\begin{align}
H_q^h (x,\xi=0,t)= f_n (x) \, F_n^h (t, x) ,
\label{eqn:gpd-paramet1}
\end{align}
where $n$ indicates the number of valence quarks, 
$h$ is a hadron, 
$f_n (x)$ is a quark distribution function,
and $F_n^h (t, x)$ is a transverse form factor at $x$.

\begin{figure}[t]
\centering
\includegraphics[width=0.23\textwidth]{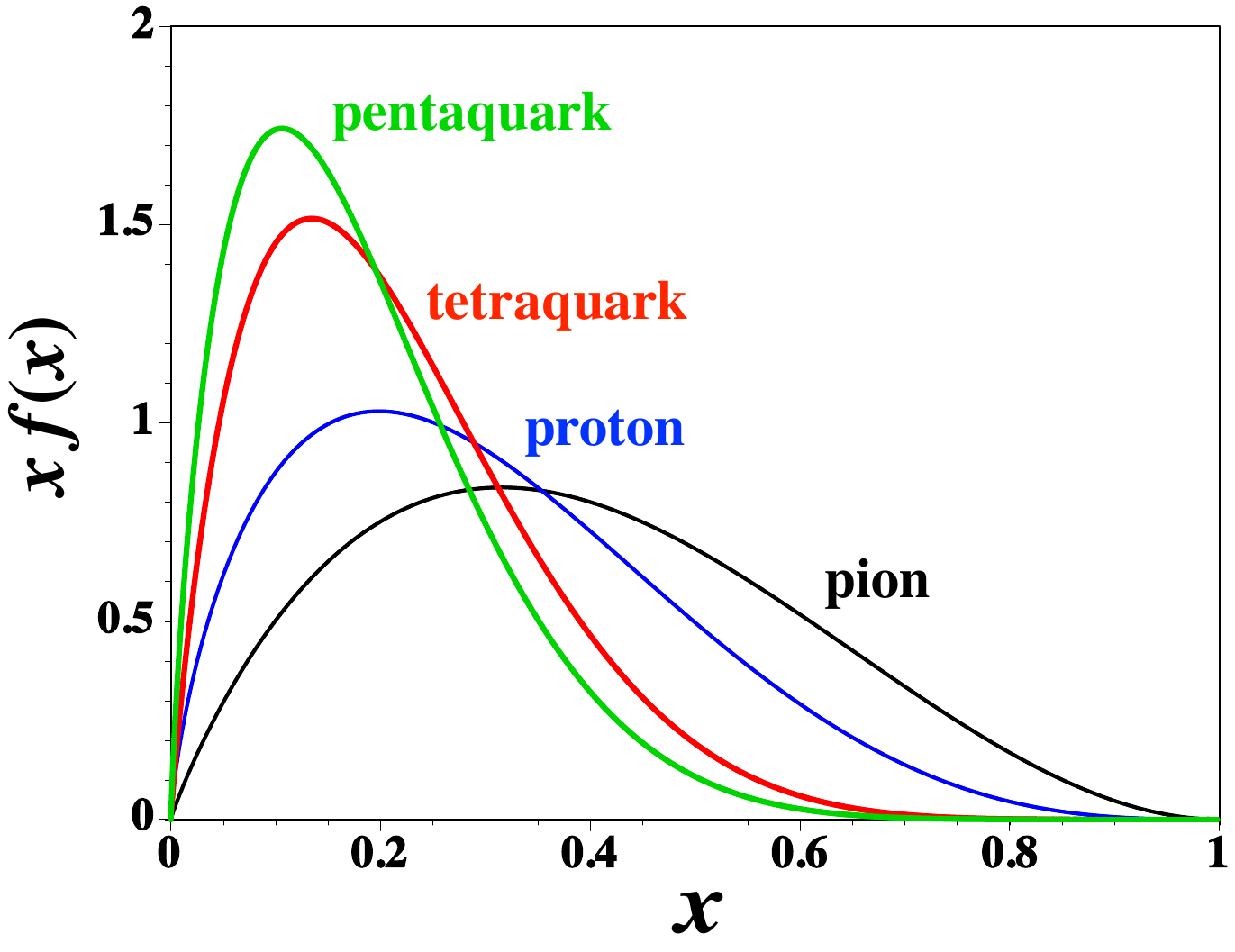}
\hspace{0.20cm}
\includegraphics[width=0.23\textwidth]{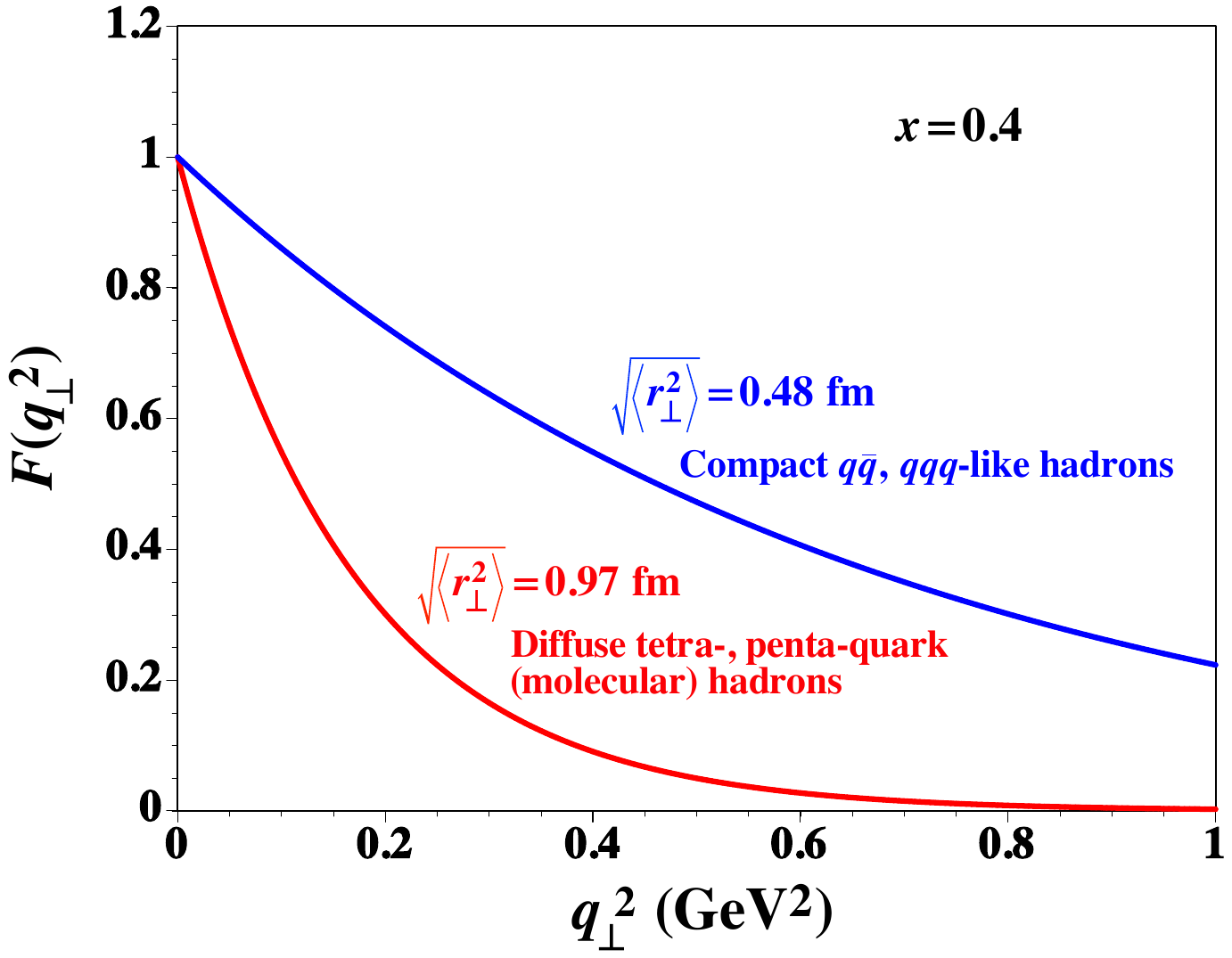}
\caption{Valence-quark distributions and 
transverse form factors of ordinary and exotic hadrons 
\cite{Kawamura:2013wfa}.}
\label{fig:exotic-gpds}
\end{figure}

The exotic nature should be reflected in both the PDF
and the transverse form factor \cite{Kawamura:2013wfa}.
For the PDFs, a simple function of $x$ 
\begin{align}
f_n (x) = C_n \, x^{\alpha_n} \, (1-x)^{\beta_n},
\label{eqn:fn}
\end{align}
is often used.
For valence-quark distributions,
the parameters $C_n$, $\alpha_n$, and $\beta_n$
are determined by the valence-quark number
$ \int_0^1 dx \, f_n (x) = n$, and the quark momentum
$ \int_0^1 dx \, x \, f_n (x) = \langle x \rangle_q$.
The parameter $\beta_n$ is determined 
the constituent counting rule in QCD
as $\beta_n = 2n -3+2\Delta S_z$ with the spin factor 
$\Delta S_z=|S_z^q-S_z^h|$.
Using these constraints together with $\langle x \rangle_q =0.47$,
we obtained the valence-quark distribution functions
of the pion, proton, tetraquark hadron, and
pentaquark hadron in Fig.\,\ref{fig:exotic-gpds}
\cite{Kawamura:2013wfa}.
The pion and proton distributions roughly agree with
the current PDF parametrizations at $Q^2=2$ GeV$^2$.
These PDFs have a clear indication of their exotic nature
because the peak of the valence-quark distributions moves
toward the smaller-$x$ region as $n$ increases.

The transverse form factor $F_n^h (t, x)$ also reflects the exotic nature.
A simple exponential function could be used for the form factor,
with the cutoff parameter $\Lambda$ for the transverse momentum,
as \cite{Guidal:2004nd}
\begin{align}
F_n^h (t,x) = e^{(1-x) t/(x \Lambda^2)} .
\end{align}
The transverse spatial density is given by its Fourier transform,
and then the root-mean-square (rms) radius is given by 
the parameter $\Lambda$  as
$ \left <  r_\perp^2 \right > = 4(1-x) / (x \Lambda^2)$.
The transverse form factor is shown in Fig.\,\ref{fig:exotic-gpds}
at $x=0.4$ by taking $\Lambda$=0.5 and 1.0 GeV as examples.
These choices mean that the rms radii are 
$\sqrt{ \langle r_\perp^2 \rangle}=0.48$ and 0.97 fm for 
$\Lambda$=1.0 and 0.5 GeV, respectively.
The ``ordinary'' hadrons with the $q\bar q$ and $qqq$ configurations
would have compact spatial distributions, which are shown
as the hard form factor ($\Lambda$=1.0 GeV, $\sqrt{\langle r_\perp^2 \rangle}=0.48$ fm),
whereas the exotic tetra- and pentaquark hadrons, 
or the molecular-type hadrons should have wide spatial distributions
and they have a soft form factor
($\Lambda$=0.5 GeV, $\sqrt{\langle r_\perp^2 \rangle}=0.97$ fm) 
in Fig.\,\ref{fig:exotic-gpds}.

It is interesting to measure the GPDs for exotic hadron candidates,
but their $t$-channel GPDs would not be measured easily because the exotic hadrons are unstable particles.
However, the generalized distribution amplitudes (GDAs), which
could be called $s$-channel GPDs, should be measured even for unstable
exotic hadrons as investigated at the KEK B factory for the pion \cite{Belle:2015oin,Kumano:2017lhr}.
Here, the GPDs obtained in the DVCS are called $t$-channel (or spacelike) GPDs, whereas the GDAs obtained in the two-photon processes could be called
$s$-channel (or timelike) GPDs because they contain timelike form factors
in them. 

\begin{figure}[t]
\centering
\includegraphics[width=0.30\textwidth]{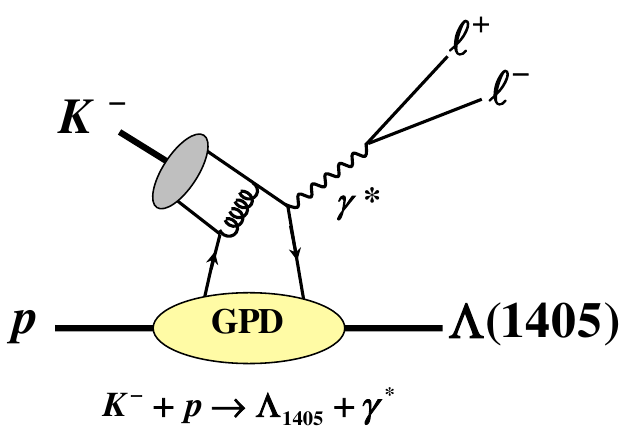}
\caption{Transition GPDs for the exotic candidate $\Lambda$(1405).}
\label{fig:tran-gpd-exotics}
\end{figure}

On the other hand, the exotic-hadron GPDs could be measured
in the spacelike region as transition GPDs.
For example, the $\Lambda$(1405) production process
$K^- p \to \ell^+ \ell^- \Lambda(1405)$ in Fig.\,\ref{fig:tran-gpd-exotics}
could be investigated at J-PARC by using the future kaon beam,
as an extension project of the pion Drell-Yan process
$\pi^- N \to  \ell^+ \ell^- N$ for measuring the nucleon GPDs 
\cite{Sawada:2016mao,J-PARC-GPD-LoI}.
The $\Lambda$(1405) cannot be described by
constituent-quark models, so that it is considered
as a $\bar K N$ molecule or a pentaquark ($uud\bar u s$) hadron.
Such transition GPDs have not been investigated in both
theoretically and experimentally, so future efforts are needed
for these kinds of studies.

There are also possibilities for studying exotic nature
in high-energy reactions by using the constituent-counting rule
in perturbative QCD \cite{Kawamura:2013iia,Chang:2015ioc}
because the constituent number should be different 
between ordinary and exotic hadrons.
This rule indicates that the cross-section scales
by the total number of constituents ($n$)
as $d\sigma /dt \sim 1/s^{n-2}$
where $s$ is the center-of-mass-energy squared.
For example, the high energy $\pi^-  p \to K^0  \Lambda(1405)$
reaction could be measured at J-PARC \cite{Kawamura:2013iia}.
There is an interesting indication that $\Lambda(1405)$
could be a pentaquark state at medium energies 
and a three-quark one at high energies by the analysis of
photo-production data on $\gamma  p \to K^+  \Lambda (1405)$ \cite{Chang:2015ioc}.


\section{Transition processes: diffractive scattering}
\label{sec:transGPDdiffractive}

\subsection{Inelastic diffraction and quantum fluctuations}
%
%
\begin{figure}[t]
\centering
\includegraphics[width = 0.65\columnwidth]{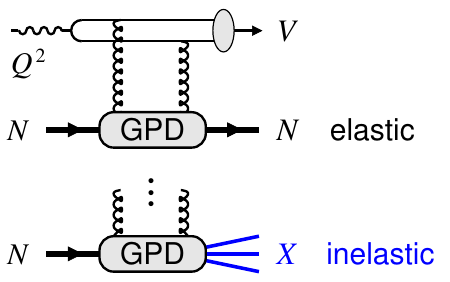}
\caption{Exclusive vector meson production in diffractive channels at small $x$
($V = \rho^0, \omega, \phi, J/\psi, \Upsilon$).
The amplitude is proportional to the gluon GPD. Upper graph: $N \rightarrow N$ transitions (elastic diffractive scattering). Lower graph: 
$N \rightarrow X(\textrm{low-mass})$ transitions
(inelastic diffractive scattering).}
\label{fig:hardexcl_fluct}
\end{figure}

In $eN/\gamma N$ scattering at high energies $W \sim 10^1-10^2$ GeV, the dominant exclusive meson production channels are vector mesons with the same quantum numbers as the (virtual) photon, 
$V = \rho^0, \omega, \phi, J/\psi, \Upsilon$.
In these channels, the production process does not exchange quantum numbers between the nucleon and the produced meson
and has the characteristics of diffractive scattering.
If the momentum transfer is large or the produced meson is heavy ($Q^2, M_{Q\bar Q}^2 \gg$ 1 GeV$^2$), the amplitude of
such high-energy exclusive processes can be factorized (see Sec.~\ref{sec:overview}) and samples the gluon GPD
of the nucleon at $x, \xi \ll 1$
(see Fig.~\ref{fig:hardexcl_fluct})
\cite{Frankfurt:2005mc}.
At such values of $x$ and $\xi$ the gluon GPD is close to diagonal and can be obtained from the gluon density ($\xi = 0$)
in a controlled approximation \cite{Frankfurt:1997ha,Shuvaev:1999ce}.

Now, in such high-energy exclusive processes, the nucleon can remain intact in the final state, $N \rightarrow N$
(elastic diffractive scattering) or dissociate into a low-mass state with nucleon quantum numbers, $N \rightarrow X$
(inelastic diffractive scattering)
(see Fig.~\ref{fig:hardexcl_fluct}).
According to the general theory of diffractive scattering, the elastic cross section is proportional to the
quantum average squared of the strength of interaction of the high-energy probe with the target configurations,
while the inelastic cross section is proportional to the quantum fluctuations, i.e., the variation of the
strength of interaction between the various configurations \cite{Good:1960ba,Miettinen:1978jb}.
Applying these concepts to the factorized processes at high $Q^2$, one can show that elastic scattering
probes the quantum average of the gluon density in the nucleon, while inelastic scattering probes its
quantum fluctuations \cite{Frankfurt:2008vi}. Specifically,
\begin{align}
\omega_g \equiv 
\frac{\langle G^2 \rangle - \langle G \rangle^2}{\langle G \rangle^2}
= 
\left[ \frac{d\sigma_{\text{inel}}}{dt} \! \right/ \! \left.
\frac{d\sigma_{\text{el}}}{dt} \right]_{t=0}^{\gamma^\ast_L N             \rightarrow VX} .
\label{fluctuations}
\end{align}
where $G$ denotes the gluon density operator and $\langle ... \rangle$ the quantum average over configurations
in the nucleon (the relation is valid at small $x$ and $\xi$, where the gluon GPD can be obtained
from the gluon density; see Ref.~\cite{Frankfurt:2008vi} for details). The relation Eq.~(\ref{fluctuations}) allows one to
extract the quantum fluctuation of the gluon density from the ratio of inelastic and elastic diffractive
scattering. Note that the relation is valid only at $t = 0$; the inelastic cross section can be
measured also at $t < 0$ but is connected with the quantum fluctuations only at $t = 0$.

The concept of fluctuations of the gluon density and their connection with inelastic diffractive scattering
is derived here in the	context	of collinear factorization of exclusive processes \cite{Frankfurt:2008vi}.
In this way, they are directly connected with the
transition GPDs	describing the individual amplitudes for $N \rightarrow X$ inelastic diffractive processes.
A similar relation between inelastic diffraction and the fluctuations of the gluon density has	been
derived in the context of the dipole model of $eN/\gamma N$ scattering
\cite{Schlichting:2014ipa,Mantysaari:2016ykx,Mantysaari:2016jaz}

\subsection{GPDs at HERA}
\label{sec:HERA}

Measurements sensitive to GPDs in lepton-proton collisions have been performed at the H1 and
ZEUS HERA collider experiments. The covered photon-proton center-of-mass energies of these
experiments ranges from 30~GeV to 300~GeV, providing access hereby to $10^{-4}\le x_B\le 10^{-2}$. Triggers from far-forward detectors and calorimeters allowed for the separation of elastic and dissociative diffractive events, and, depending on the period of data collection, the experiments were equipped with far-forward proton spectrometers able to detect, with coverage in acceptance of a few percent, directly the proton scattered in the elastic process. 
Hard diffractive production of the light mesons $\rho^0$~\cite{H1:2009cml,ZEUS:2007iet} and $\phi$~\cite{H1:2009cml, ZEUS:2005bhf}, of pion pairs~\cite{ZEUS:2011tzw} and of the quarkonia
$J/\psi$~\cite{H1:1996kyo, H1:2002voc, H1:2005dtp, H1:2010udv, H1:2013okq, ZEUS:1997wrc, ZEUS:2002src, ZEUS:2004yeh, ZEUS:2009qug, ZEUS:2012qog}, $\psi(2S)$~\cite{H1:2002yab,
ZEUS:2002src, ZEUS:2012qog}, and $\Upsilon$~\cite{ZEUS:2009asc,ZEUS:2011spj} have been studied in both the elastic and the dissociative channels, while the hard exclusive production of photons was limited to the elastic channel~\cite{H1:2005gdw, H1:2007vrx, H1:2009wnw, ZEUS:2003pwh, ZEUS:2008hcd}. 
A weak dependence on $W_{\gamma p}$ of the elastic diffractive cross section is observed in the absence of a hard scale. In contrast, the cross-section is seen to rise with increasing $W_{\gamma
p}$ in the presence of a hard scale, either through large $Q^2$ or large heavy-quark mass, where the
rise is steeper for increasing values of the hard scale. This behavior reflects the increase in gluon density with smaller $x_B$ (or larger $W_{\gamma p}$). The measurements of the elastic channel
also show a $t$ slope, $b$, evolving from large values in the absence of a hard scale to a small,
constant value once a sufficiently hard scale is reached, and this irrespective of the produced
final-state particle. In the dipole picture, this can be understood as $b$ being a reflection of the
transverse dipole size and the transverse proton size, where at large scales the dipole size
becomes negligible. For the inelastic channel, a very different $t$ dependence is observed.
Spin-density matrix elements (SDMEs), which are related to the GPDs, have been extracted for the DVMP process from the H1 and ZEUS data through the measurement of the angular distribution of the
meson decay products. It is observed that for quarkonia $s$-channel helicity is conserved and that,
apart from some slight deviations for some SDMEs, the same is true for the light vector mesons.
This is in contrast with the low-energy, fixed-target experiments, where for some SDMEs a strong
violation of s-channel helicity conservation is observed; see {\it e.g.}~\cite{HERMES:2009oim}. The
extracted SDMEs at the collider experiments also indicate that the interaction between the virtual
photon and beam proton proceeds through the exchange of particles of natural parity, which can
be at first order equated to the exchange of two gluons. Also here, differences are seen with the
fixed-target experiments, where at low energy unnatural-parity exchange also seems to
contribute~\cite{HERMES:2017qwt}. Finally, an increase in the dominance of the longitudinal cross-section over the transverse cross-section with increasing values of $Q^2$ has been observed
both for the light and the heavy mesons.
In relation to the measurement of processes involving transitions of the beam proton to another
state, the H1 experiment performed a measurement of $e+p\rightarrow \rho^0+n+\pi^+$ with
$Q^2<2$~GeV, where the $\rho^0$ meson is reconstructed in the central detector, the neutron is
tagged in a forward detector system, and the $\pi^+$ escapes along the beam line without being
detected~\cite{H1:2015bxa}. The interpretation of this interaction can be attributed to scattering
off the proton pion cloud or off the neutron core, but also to the creation of a $N^*$ resonance, which subsequently decays into $\pi^+ n$. The measurement of this process exemplifies one possibility to study transition GPDs at electron-proton colliders.


\subsection{Transition GPDs in ultraperipheral collisions}
\label{sec:LHC}

Ultra-peripheral collisions of hadrons at very high energies can be used to access GPDs.
Hard exclusive production, in particular, the photoproduction of heavy quarkonium  ($J/\psi$, $\psi(2S)$, $\Upsilon({\rm n}S)$) states, have been studied in $\gamma-p$ and $\gamma-$lead nucleus collisions at RHIC and the LHC~\cite{Klein:2020fmr}, providing unique access to the low $x_B$ region, down to $x_B=10^{-6}$. We focus here on $\gamma-$p reactions. They can be studied in proton-proton collisions and proton-nucleus collisions at these two running hadron colliders. 

For the access to GPDs through meson production, the quasi-real nature of the exchanged photon is limiting the useful studies to the production of quarkonia, while there are also proposals to study exclusive photon-meson pair production at sufficiently high invariant mass~\cite{Duplancic:2022ffo,Duplancic:2023kwe} as to introduce the needed hard scale. Exclusive production of the quarkonia $J/\psi$, $\psi(2S)$ and $\Upsilon$ has been successfully studied in proton - proton~\cite{LHCb:2013nqs,LHCb:2014acg,LHCb:2015wlx,LHCb:2018rcm}, proton lead~\cite{ALICE:2014eof,CMS:2018bbk,ALICE:2018oyo} and lead-lead collisions~\cite{ALICE:2012yye,ALICE:2013wjo,CMS:2016itn,ALICE:2019tqa,ALICE:2021gpt,ALICE:2021tyx,LHCb:2021hoq,LHCb:2022ahs}. The photon-proton cross sections for exclusive $J/\psi$ production extracted from the measurements in lepton-proton, proton-proton, and lead-proton collisions are compatible with each other in the common $W_{\gamma p}$ range, which hints at the universality of the underlying probed physics. At present, the exclusive quarkonium production cross section has been used in the study of standard PDFs~\cite{Flett:2020duk}, but not yet to constrain GPDs. 

A measurement of the DVCS process is not possible due to the small virtuality of the initial state photon that originates from the electrostatic field of the colliding ultra-relativistic ions. Time-like Compton scattering measurements are in principle possible \cite{Pire:2008ea}. A $t$-integrated cross-section result on exclusive dilepton production in $\gamma-p$ collisions already exists in a kinematic domain of interest (dilepton mass between 1 and 2.5~GeV$/c^2$)~\cite{ALICE:2023mfc}.

In addition to hard exclusive photoproduction, ultra-peripheral collisions allow the study of hard dissociative photoproduction processes at high energies, where the target proton breaks up.   This observable class is within the Good-Walker-formalism sensitive to the fluctuations of the proton structure~\cite{PhysRevD.18.1696}, whereas the exclusive production is sensitive to the average. A first measurement of dissociative $J/\psi$ photoproduction in $\gamma-$p collisions has been published recently~\cite{ALICE:2023mfc}. Future measurements at the LHC can be expected based on the interest of the community and the luminosity increase in the 2020s compared to previous data-taking campaigns, thanks to the HL-LHC upgrades~\cite{Citron:2018lsq}.

Ultra-peripheral collisions can also be studied in the fixed-target mode of the LHC ($\sqrt{s}\approx$~100~GeV). A few observables and corresponding publications are summarized in Ref.~\cite{Hadjidakis:2018ifr}. LHCb performed with the SMOG gas target originally designed for luminosity measurements via beam imaging~\cite{Barschel:2014iua}, a first set of low luminosity cross section measurements, {\it e.g.}, on charm and antiproton production~\cite{LHCb:2018jry,LHCb:2018ygc}. 

Transition GPDs can, in principle, be accessed within dissociative production provided that the final state nucleon resonance can be at least partially reconstructed. At the existing hadron colliders, this is a difficult undertaking given the existing forward instrumentation. Dedicated simulation studies are required to assess the feasibility with the current experimental set-ups or with dedicated additional instrumentation.

\section{Future experimental facilities}
\label{sec:perspectives}

The data from already existing measurements allows the first extraction of observables, sensitive to transition GPDs. However, for more detailed studies, higher statistics and also higher beam energies are favorable. In addition, hadron beams at J-PARC will allow the study of transition GPDs in hadronic reactions.
The following section will summarize upcoming opportunities with existing and planned experiments.

\subsection{CLAS12 luminosity upgrade and JLab 22 GeV}
\label{sec:JLAB24}

The $N \rightarrow N^{*}$ DVCS as well as the $N \rightarrow N^{*}$ DVMP processes will both strongly profit from a luminosity upgrade of CLAS12 \cite{CLAS12lum} as well as from an energy upgrade of JLab \cite{Accardi:2023chb}.

From the statistics point of view, the low efficiency for the detection of the multi-particle final states, in combination with the background suppression cuts, strongly limits the available statistics of the final sample. At the moment, the statistics that can be collected with CLAS12 are mainly limited by the rate capability of the detector, which is determined primarily by the forward tracking in the drift chambers, while the CEBAF accelerator could deliver around two orders of magnitude more beam current.
Therefore, a luminosity upgrade of CLAS12, which is currently in progress, will help collect data more efficiently and increase the amount of data that can be collected in a certain time. Based on this upgrade, it is expected that sufficient data can be collected in a reasonable time frame to perform a fully differential study of the $N \rightarrow N^{*}$ DVCS and DVMP processes, which will allow a partial wave decomposition of the resonance-spectrum.  

\begin{figure}[t]
	\centering
		\includegraphics[width=0.40\textwidth]{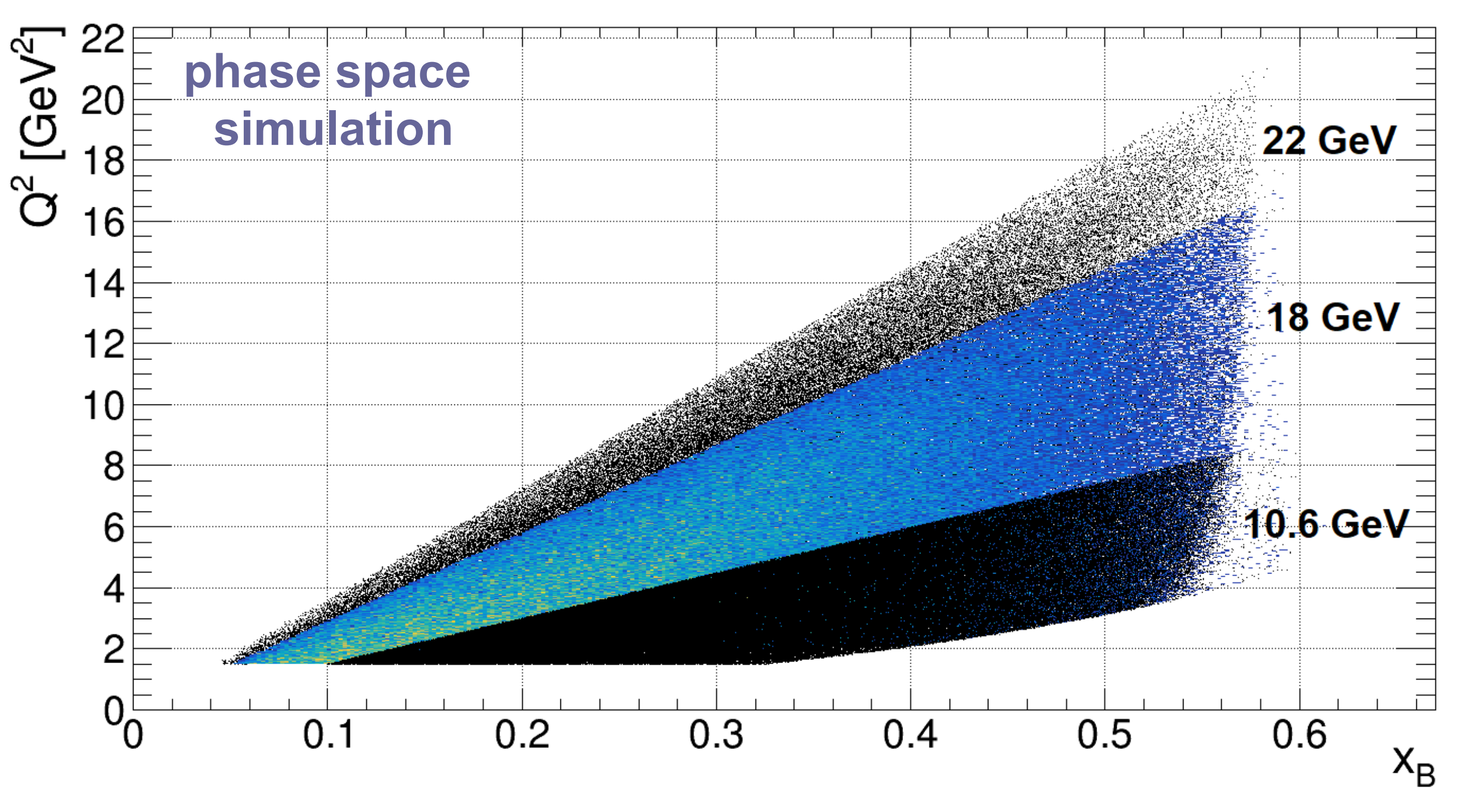}
		\includegraphics[width=0.40\textwidth]{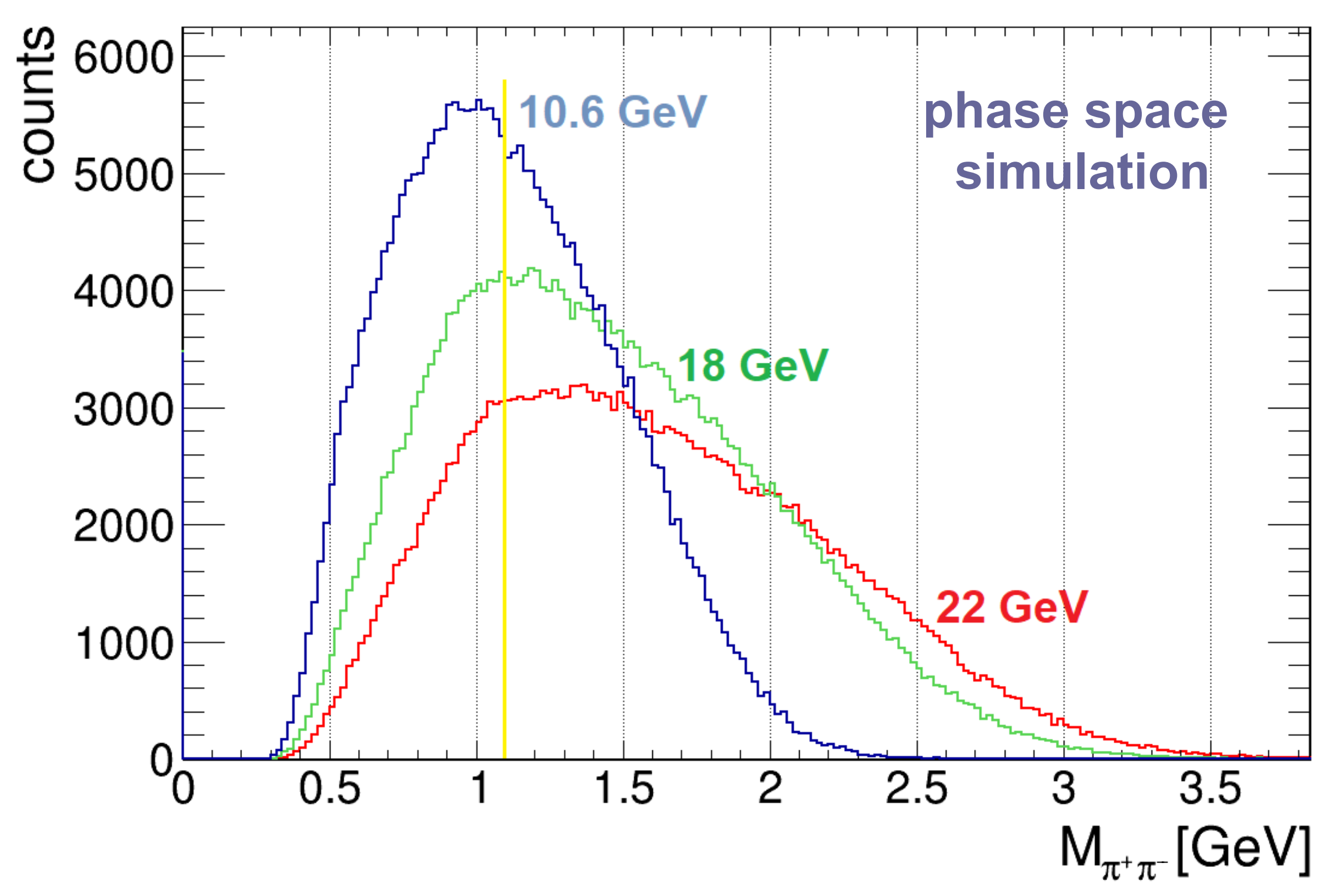}
	\caption{Comparison of the available phase space, accessible with the present CLAS12 setup, in $Q^{2}-x_{B}$ or the $\pi^{-}\Delta^{++}$ process under forward kinematics ($-t <$~1.5~GeV$^{2}$) [up] and for the $\pi^{+}\pi^{-}$ invariant mass of the same process, which is used to suppress the dominant $\rho$ production background by the cut on $M(\pi^{+}\pi^{-}) >$~1.1~GeV, indicated by the yellow line [down] for a 10.6~GeV, 18~GeV and 22~GeV electron beam.}
	\label{fig:q2_xB_coverage}
\end{figure}

From the beam energy point of view, the currently available beam energy of 10.6 GeV allows a study of the lower lying nucleon and $\Delta$ resonances in a limited $Q^{2}$ range. However, especially for higher-mass resonances, the factorization requirement $Q^{2} \gg m^{2}_{N^{*}}$ strongly limits the options based on a 10.6 GeV electron beam. Here, a 22 GeV upgrade of JLab would enable the investigation of higher-mass resonances and extend the accessible $Q^{2}$ range for the lower mass resonances. Based on this extended range, a detailed study of the scaling behavior of the different observables will become possible.  
Fig. \ref{fig:q2_xB_coverage} shows the available phase space, accessible with the present CLAS12 setup, in $Q^{2}-x_{B}$ for the $\pi^{-}\Delta^{++}$ process under forward kinematics and the $\pi^{+}\pi^{-}$ invariant mass of the same process for a 10.6~GeV, 18~GeV, and 22~GeV electron beam. The corresponding distributions of the $N \rightarrow \Delta$ DVCS process show similar characteristics.
It can be seen that a 22 GeV upgrade of JLab would provide a significantly increased $Q^{2}$ range for a fixed value of $x_{B}$. This leads to a big advantage for the study of these processes since the factorization of the $N \rightarrow N^{*}$ DVCS and DVMP processes requires high virtuality $Q^{2}$ to be above the resonance mass squared. While this condition can already be fulfilled with a 10.6 GeV electron beam for lower mass resonances, like the $\Delta(1232)$, an energy upgrade to 22 GeV will be a big advantage to ensure the factorization of the process for higher mass resonances and to study the scaling behavior of the observables.
As shown in the lower part of Fig. \ref{fig:q2_xB_coverage} the increase in phase space for the different invariant mass combinations will allow a more efficient suppression of non-resonant background from exclusive meson production and also from other (differently charged) nucleon resonance production channels, which is mostly expected at lower masses. Higher beam energies will therefore also provide a more efficient event selection and a better suppression of the non-resonant background.
While the presently available 10.6~GeV electron beam already allows a study of the 3D structure of lower-mass nucleon resonances, a 22~GeV upgrade of JLab in combination with a luminosity upgrade of CLAS12 will provide ideal conditions for the study of the 3D structure of nucleon resonances via transition GPDs.
Also, future studies with the high-resolution two-arm spectrometer in JLab hall C will strongly benefit from an energy upgrade.


\subsection{Muon and meson beams at COMPASS/AMBER}
\label{sec:COMPASSAMBERtransGPD}

So far, no physics program for the measurement of transition GPDs at COMPASS \cite{COMPASS:2007rjf} or AMBER \cite{Adams:2018pwt} exists. However, the existing COMPASS data based on a 160 GeV muon beam will provide potential access to lower $x_B$ (0.01 $< x_B <$ 0.2) and higher $Q^2$ values. It can therefore help to constrain the impact of sea quarks on the resonance properties and also provide valuable inputs for an extrapolation to the gluonic regime, accessible with the EIC.

Since the main spectrometer of COMPASS is focused at very forward angles and provides relatively high momentum thresholds, it can only be used to detect the scattered muons and the DVCS photon or DVMP pion. The decay products of the resonance have to be detected by the CAMERA detector, which consists of two regions of plastic scintillator bars, covering the region around the target. Since only charged tracks can be detected by this device, it mainly limits the channels that can be studied with COMPASS to two $N \to N^{*}$ DVMP channels:
\begin{itemize}
    \item~~$\mu^{\pm} ~p ~\to~ \mu^{\pm}~ \pi^{-}~\Delta^{++} ~~~~~\to~ \mu^{\pm}~ \pi^{-}~[p\pi^{+}]$
    \item~~$\mu^{\pm} ~p ~\to~ \mu^{\pm}~ \pi^{+}~\Delta^{0} (N^{*}) ~\to~ \mu^{\pm}~ \pi^{+}~[p\pi^{-}]$
\end{itemize}

Currently, feasibility studies for these two channels are ongoing based on existing data taken with COMPASS in 2016. With AMBER also the high-intensity pion and kaon beams with up to 190 GeV/c are expected to provide an excellent basis for the study of meson-induced transitions within the exclusive Drell-Yan process. Feasibility studies for potential future muon and meson-based measurements with AMBER, including potential detector upgrades, are planned.


\subsection{Electron-Ion Collider EIC}
\label{sec:EICtransgpd}

\begin{figure}[t]
	\centering
		\includegraphics[width=0.45\textwidth]{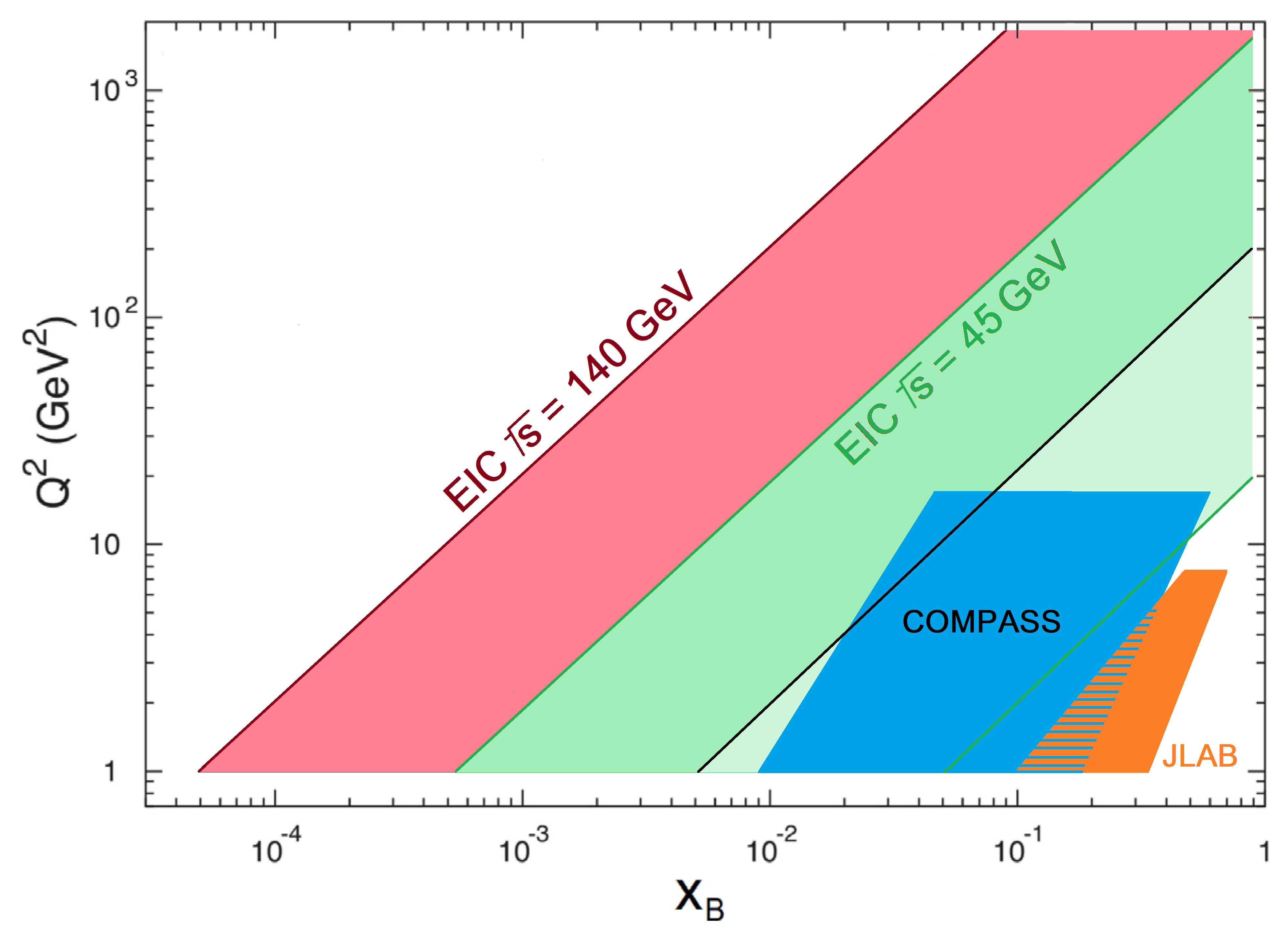}
	\caption{Comparison of the available phase space in $Q^{2}-x_{B}$, accessible with a 12 GeV electron beam at JLab (orange), a 160 GeV muon beam at COMPASS (blue), and different CM energies of the EIC (green, red).}
	\label{fig:EICq2xB}
\end{figure}

The planned electron-ion collider (EIC) \cite{Accardi:2012qut} will allow an extension of the kinematic regime to the gluonic sector at low $x_B$ and towards higher $Q^2$ values (see Fig. \ref{fig:EICq2xB}), which are especially important for the factorization of the process in relation to higher mass resonances.

The upcoming data from the EIC will provide insights into the contributions of the gluons to the excitation process and the characteristics of baryon resonances. The planned configuration of the far-forward detectors will provide excellent opportunities to study the low $-t$ region of hard exclusive processes. 
So far, no theoretical predictions exist for the low $x_{B}$ regime of the $N \to N^{*}$ DVCS and DVMP processes, but following the predictions for the ground-state DVCS and DVMP processes in this regime, a measurement is expected to be feasible from the cross-section point of view.


\subsection{Hadron beams at J-PARC and FAIR}
\label{sec:JPARC}

As discussed in section \ref{sec:transGPDhadrons}, theoretical models also allow the study of transition GPDs with hadronic beams in $2 \rightarrow 3$ processes and based on the exclusive Drell-Yan process. The high-momentum hadron beam line of J-PARC \cite{Aoki:2021cqa} provides excellent opportunities for such studies. Further opportunities with proton and antiproton beams in a similar energy region will become available at the Facility for Antiproton and Ion Research (FAIR) \cite{FAIRrep22}.

The feasibility of measuring the $2 \rightarrow 3$ process in the ongoing E16 experiment in the high-momentum beamline of J-PARC is currently being investigated. Equipping additional detectors in the forward direction will become necessary to increase the acceptance of these exclusive hard events. Also, studies of this process at FAIR may be considered and further investigated.

For the measurement of the exclusive Drell-Yan process in the E50 experiment, a letter of intent~\cite{J-PARC-GPD-LoI} was submitted to the J-PARC 2019 PAC. A full proposal with the design of the required mu-ID system and dimuon trigger setting is currently being prepared. The ``$\pi 20$'' project of delivering 10-20 GeV meson beam in the high-momentum beamline of J-PARCs hadron hall has been included in the hadron hall extension project~\cite{Aoki:2021cqa}. This six-year project has been selected as the top priority in the KEK mid-term plan (KEK-PIP2022) and is scheduled to start in 2024. After the upgrade, high-resolution pion beams with a momentum of up to 20 GeV will be available.

\subsection{LHC fixed-target and ultraperipheral collisions}

In 2024, results from the first unpolarized fixed-target set-up for high-luminosity data taking at the LHC are expected from LHCb SMOG-2~\cite{Bursche:2649878,2707819}. This setup, installed in 2022, will allow to run LHCb in fixed-target and collider mode in parallel throughout the full data-taking period and with up to two orders of magnitude larger instantaneous luminosities than SMOG, depending on the gas. An effort towards a polarized gas target in LHCb is ongoing~\cite{Aidala:2019pit}.

With the HL-LHC upgrades~\cite{Citron:2018lsq} also future measurements of dissociative $J/\psi$ photoproduction in $\gamma-p$ collisions at the LHC can be expected with significantly increased statistics.
However, the relevant luminosity increase strongly depends on the experimental setup, collision system, and operational decisions.


\section{Challenges of GPD analysis}
\label{sec:extraction}

It is instructive to compare the phenomenology of transition GPDs with that of diagonal counterparts, the latter of which has been extensively studied over the last three decades. 
Both types of objects naturally share many similar features. However, studies of transition GPDs come with a number of unique difficulties.

Both cases require the simultaneous extraction of several multidimensional objects from observables for exclusive processes, presenting a non-trivial task. Specifically, distinguishing among many contributing objects can only be achieved by measuring a variety of exclusive processes (DVCS, TCS, DVMP, double DVCS~\cite{Deja:2023ahc}, diphoton production~\cite{Pedrak:2020mfm, Grocholski:2022rqj}, {\it etc.}) and observables (cross-sections, beam/target / charge asymmetries, {\it etc.}).\\
This motivates a global experimental program conducted in laboratories covering complementary kinematic domains and diverse setups equipped with sophisticated detection systems capable of determining the exclusivity condition. In the case of diagonal GPDs, the experimental programs carried out in laboratories such as JLab, DESY, and CERN have already resulted in numerous measurements, fueling phenomenological analyses like~\cite{Kumericki:2019ddg, Moutarde:2019tqa, Guo:2023ahv}. As indicated in Sec.~\ref{sec:current_results}, so far only the first data sensitive to transition GPDs have been collected, allowing for exploratory studies that help to prove the correctness and usefulness of the formalism. We stress again that processes sensitive to both diagonal and transition GPDs can, in principle, be measured together. However, the latter typically come with smaller cross-sections and are more prone to detector effects threatening the reconstruction of all particle states.
{In addition, processes sensitive to transition GPDs typically contain at least three hadrons in the final state and therefore include an, at least, partial overlap of different resonances, as well as non-resonant background, which needs to be separated from the reaction of interest. This is a well-known issue from the field of resonance spectroscopy and can be handled based on a partial wave decomposition (see e.g. Ref. \cite{Thiel:2022xtb} for a recent review). However, this method introduces the need for sufficient statistics and detector acceptance to perform a differential measurement in terms of the relevant kinematic variables.}

Similarly to diagonal counterparts, transition GPDs also enter observables via convolutions with coefficients describing the hard part of a given process and, in the case of, for example, DVMP, also with additional non-perturbative objects describing produced hadrons. This requires the development of sophisticated deconvolution methods, which in practice for processes like DVCS and lowest-order coefficient functions turns out to be an arduous task (for a full discussion, see both Refs.~\cite{Bertone:2021yyz} and~\cite{Moffat:2023svr}). Despite this difficulty, the phenomenology of diagonal GPDs demonstrates that analyses of amplitudes (not requiring deconvolution) still allow us to learn a lot about the structure of hadrons. In particular, one can directly study the nuclear tomography at low $\xi$ thanks to certain similarities with the optical theorem~\cite{ZEUS:2008hcd, H1:2009wnw, COMPASS:2018pup}, mechanical properties via the subtraction constant appearing in the dispersion relation~\cite{Burkert:2018bqq, Kumericki:2019ddg, Dutrieux:2021nlz}, and response of hadrons on probes carrying different angular momenta thanks to the recent development of techniques based on the Froissart-Gribov projections~\cite{Semenov-Tian-Shansky:2023ysr}. The same can be done in the non-diagonal case, assuming experimental data allow for it.

Another common challenge is defining models that fulfill 
a maximum of theory-driven 
requirements. 
This proves to be a challenge, particularly due to the non-trivial task of combining polynomiality with positivity constraints. Many modeling techniques have been proposed in the diagonal case, including those involving Radyushkin’s double distribution Ansatz~\cite{Goloskokov:2005sd, Vanderhaeghen:1998uc, Mezrag:2013mya}, conformal moment expansion~\cite{Kumericki:2015lhb, Guo:2023ahv}, dual parametrization~\cite{Polyakov:2002wz,Polyakov:2009xir,Muller:2014wxa} and light front wave functions~\cite{Mezrag:2023nkp}. This selection is supplemented by non-parametric models based on machine learning techniques~\cite{Dutrieux:2021wll} used to study model dependency. All these modeling techniques can be re-used in the non-diagonal case. The issue of incorporating diagonal GPDs into models of non-diagonal objects relying on the large $N_{c}$ limit, as discussed in Ref.~\cite{Kroll:2022roq}, remains an open and interesting phenomenological issue. Additional difficulty arises from the limited knowledge of forward limits and elastic form factors, which are used as basic ingredients in the modeling procedures; see, for instance, Eqs.~(\eqref{eq:eff_unpol}) and~(\eqref{eq:eff_pol}).

It is also very important to generalize for non-diagonal hard exclusive reactions the consistent methods for accounting the higher-order kinematical effects relying on the approaches developed in Refs.~\cite{Braun:2012hq,Guo:2021gru,Braun:2022qly}.

Assuming we have experimental data and a model we would like to constrain, and we understand well the connection between these data and the model, the last non-trivial task is the creation of an extraction framework. In this framework, all elements needed for the extraction must be implemented and well-tested. Since we expect data to span a broad kinematic domain, to fully facilitate their potential, it is required to use evolution equations but also a suitable description of exclusive processes, particularly in terms of $\alpha_S$ and twist expansions. Evolution also acts as a useful tool, allowing us to distinguish between various contributions. For instance, in the analysis of the subtraction constant, evolution allows us to decompose this object to access the part related to the energy-momentum tensor form factor $C$~\cite{Dutrieux:2021nlz}. In the GPD field, two open-source platforms dedicated to diverse phenomenological studies exist: Gepard~\cite{Kumericki:2007sa} and PARTONS~\cite{Berthou:2015oaw}. Both of them can be adapted to the non-diagonal case.

The constrained models can be employed in Monte Carlo generators to support ongoing and future experimental campaigns, such as those planned at JLab~\cite{Accardi:2023chb} and EIC~\cite{AbdulKhalek:2021gbh}. The generators should incorporate radiative corrections, the understanding of which is crucial for interpreting data, and ideally, they should be capable of generating multiple exclusive processes. A potential candidate for such a generator is EpIC~\cite{Aschenauer:2022aeb}, which, due to its modular 
architecture can easily accommodate elements required to generate events related to non-diagonal GPDs.

\section{Summary and future transition GPD program} 
\label{sec:future}

Transition GPDs provide new tools for quantifying and interpreting the structure of baryon resonances in QCD.
{
Transition GPDs describe matrix elements of nonlocal partonic QCD operators between ground and excited baryon states
and contain much more information than transition form factors of vector/axial currents traditionally employed to study resonance excitation.
They allow the performance of tomographic imaging of resonance excitation 
processes 
and provide insights into the mechanical properties of hadronic medium such as angular momentum, pressure, and shear force.}
Systematic theoretical methods are available for characterizing the $N \rightarrow N^\ast$ transition GPDs
and connecting them with the $N \rightarrow N$ ground state GPDs.

Transition GPDs are sampled in hard exclusive electroproduction processes with $N \rightarrow N^\ast$ transitions,
such as DVCS $e + N \rightarrow e' + \gamma + N^\ast$ and meson production $e + N \rightarrow e' + M + N^\ast$.
QCD factorization is applied to the amplitudes along the same lines as for $N \rightarrow N$ processes.
First experimental results from JLab 12 GeV in $\pi^-\Delta^{++}$ and $\pi^0 \Delta^+$ meson production, as well
as $p \rightarrow \Delta^{+}$ DVCS, demonstrate the feasibility of the measurements and provide a first test of
the applicability of the framework. Much more extensive	data are expected from the ongoing JLab	12 GeV program.
Further measurements of transition processes in	electron/muon scattering could be performed with COMPASS,
EIC and the possible JLab 22 GeV	upgrade. Complementary information comes from hadron-induced processes such
as exclusive dilepton production $\pi + N \rightarrow (\ell^+\ell^-) + N^\ast$, which will be measured at J-PARC.
The concept of transition GPDs can also be applied to gluon-mediated diffractive scattering at	small $x$,
with measurements performed in ultraperipheral collisions at LHC and photo/electroproduction at the future EIC, and EicC.

This opens the prospect of a new scientific program exploring baryon resonance structure using transition GPDs
as a unifying concept and quantitative analysis tool. To realize this program, further efforts are needed to
develop	the theoretical framework, extend the experimental studies at existing facilities, and simulate measurements with future facilities. On the theoretical side, this includes:

\begin{itemize}

\item
Revisit and standardize the structural decomposition of the $N \to \Delta$ transition matrix elements;
establish the connections between the various definitions of transition GPDs used in the literature;
clarify the manifestation of their basic properties in the different parametrizations.
Extend the structural decomposition to other $N \to N^{*}$ transition GPDs.

\item 
Construct a general formalism for non-diagonal hard exclusive reactions based on $N \to \pi N$ 
transition GPDs in the full $\pi N$ resonance domain, using methods of amplitude analysis to separate
resonant and non-resonant contributions. To avoid complications associated with the spins of the target 
and produced hadrons, a first step might be to treat the toy example of $\pi \to \pi \pi$ transition GPDs.

\item
Develop the interpretation of the EMT transition form factors and the
mechanical properties of baryon resonances, using representations of the transition matrix elements in
terms of 2-dimensional light-front densities or 3-dimensional Breit frame densities.

\item
Apply the $1/N_c$ expansion to achieve a complete classification of the $N \rightarrow \Delta$
transition GPDs, including subleading structures and $1/N_c$ corrections to leading structures,
using algebraic methods based on the spin-flavor symmetry group. Extend the $1/N_c$ analysis
to $N \rightarrow N^\ast$ transitions with mass difference $\mathcal{O}(N_c^0)$ and orbital 
excitations.

\item Explore quantitative features of transition GPDs in dynamical models of baryons such as 
light-front quark models, the chiral quark-soliton model based on the large-$N_c$ limit, 
relativistic bound state models based on Dyson-Schwinger equations, or GPD models based on holographic QCD.

\item Classify and compute kinematic power corrections and dynamic higher-twist corrections
in the non-diagonal DVCS process

\item Develop and implement the transition GPD formalism for hard exclusive reactions with 
strangeness production $\gamma^* N \to \{\Lambda,\,\Sigma\} K$, using methods of 
SU(3) flavor symmetry. Study the feasibility of strange baryon production in the
kinematic conditions of JLab, EIC, and EicC.

\item Develop dynamical models of diffractive $N \rightarrow X$ gluon transition GPDs at small $x$
based on quantum fluctuations of the gluon fields in the nucleon, explore the sensitivity and impact of measurements of inelastic diffraction in ultraperipheral $pA$ collisions at LHC and in photo/electroproduction at EIC and EicC.

\item Explore applications of non-diagonal hard exclusive reactions with $N \to \pi N$ transitions
to resonance spectroscopy, particularly the production of exotic states enriched by large gluonic or multiquark components. 
This requires employing theoretical tools developed for conventional partial wave 
analysis, 
such as the $K$-matrix approach, $N/D$-method, dispersive methods {\it etc}.

\item Calculate and interpret polarization observables depending on the decay angles of the meson-nucleon system, as a necessary step towards hadron spectroscopy applications.  

\item {Development of methods for calculation of transition GPDs within the framework of lattice QCD. This involves specifying the description of hadron resonance states within lattice QCD and appropriately generalizing advanced calculation techniques, such as the quasi-distribution formalism and
the large momentum effective theory (LaMET) framework. }

\end{itemize}
On the experimental side, this includes:
\begin{itemize}

\item 
Perform high-statistics, fully differential measurements of the $N \to N^{*}$ DVMP and DVCS processes with CLAS12, enabling a full partial wave decomposition of the resonance spectrum. For this purpose, an upgrade of the rate capability of CLAS12 will be essential to fully exploit the luminosity that can be provided by CEBAF at JLab. Adding a calorimeter to the central detector of CLAS12 would allow for the detection of further final states for both processes.

\item 
Perform cross-section measurements of the $p \to \Delta$ transition and a potential L-T separation of the unpolarized cross section for certain $N \to N^{*}$ DVMP processes in JLab Hall C.

\item 
Explore the options for transition GPD measurements with a possible 22 GeV upgrade 
of JLab, extending the phase space and kinematic coverage for hard exclusive processes, especially in $Q^{2}$, and allowing for more efficient background separation.

\item 
Investigate the possibilities for measuring DVMP reactions with hyperon transitions at JLab 12 GeV, to probe the nucleon-to-hyperon transition GPDs.

\item 
Explore options for studies of transition GPDs with muon and meson beams at
CERN COMPASS/AMBER.

\item 
Simulate transition GPD measurements with EIC using the far-forward detectors, 
especially $N \rightarrow \Delta$ transition processes in EIC kinematics.
Far-forward $\Delta$ detection can use complementary decay channels, {\it e.g.}\ 
$\Delta^+ \rightarrow \pi^+ n$ or $\pi^0 p$ (= one charged and one neutral particle), 
$\Delta^0 \rightarrow \pi^- p$ or $\pi^0 n$ (= all charged or all neutral particles);
each of these channels presents particular challenges or opportunities.
The detection of far-forward pions is also needed for the $u$-channel exclusive processes.

\item Study opportunities for transition GPD measurements at J-PARC, which will provide access to the time-like regime and are complementary to lepton scattering experiments.

\item Simulate measurements of diffractive transition GPDs $p \rightarrow X$ 
in ultraperipheral collisions at the LHC.

\end{itemize}
 
Altogether, the available theoretical concepts and methods and the existing and planned 
experimental facilities provide excellent opportunities for measurements of transition GPDs
and next-generation studies of baryon resonances.


\section{Acknowledgements}

The workshop on ``Exploring resonance structure with transition GPDs'' has been supported by STRONG-2020 ``The strong interaction at the frontier of knowledge: fundamental research and applications'' which received funding from the European Union’s Horizon 2020 research and innovation program under grant agreement No 824093, by the European Centre for Theoretical Studies in Nuclear Physics and Related Areas (ECT*) and by the Asia Pacific Center for Theoretical Physics (APCTP).

S. Diehl was partially supported by Deutsche Forschungsgemeinschaft Project No. 508107918 and the German Bundeministerium f\"ur Bildung und Forschung (BMBF).

S. Kumano was partially supported by Japan Society for the Promotion of Science (JSPS) Grants-in-Aid for Scientific Research (KAKENHI) Grant Numbers 19K03830 and 24K07026.

K. Semenov-Tian-Shansky was supported by Basic Science Research Program through the National Research Foundation of Korea (NRF) funded by the Ministry of Education  RS-2023-00238703; and under Grants  No. NRF-2018R1A6A1A06024970 (Basic Science Research Program); and by the Foundation for the Advancement of Theoretical Physics and Mathematics ``BASIS''.

M. Vanderhaeghen was supported by the Deutsche Forschungsgemeinschaft (DFG, German Research Foundation), in part through the Cluster of Excellence [Precision Physics, Fundamental Interactions, and Structure of Matter] (PRISMA$^+$ EXC 2118/1) within the German Excellence Strategy (Project ID 39083149).

The work presented by A. Usman was supported by Natural Science and Engineering Research Council of Canada (NSERC) Grant No. SAPIN-2021-00026 and the National Science Foundation of USA (NSF) Grant No. PHY2012430 and PHY2309976

Parada T. P. Hutauruk was supported by the National Research Foundation of Korea Grant Nos. 2018R1A5A1025563, 2022R1A2C1003964, 2022K2A9A1A0609176, and 2023R1A2C1003177.

C.-H. Lee was supported by the National Research Foundation of Korea(NRF) grant funded by the Korea government(MSIT) (No. 2023R1A2C1005398). 

H.-D. Son was supported by the National Research Foundation of Korea(NRF) grant funded by the Korea government(MSIT) (RS-2023-00210298).

H.S. Jo was supported by the National Research Foundation of Korea (NRF) grant funded by the Korea government (MSIT) (RS-2023-00280845) and by Basic Science Research Program through the National Research Foundation of Korea (NRF) funded by the Ministry of Education (RS-2018-NR031074).

This material is based upon work supported by the U.S.~Department of Energy, Office of Science,
Office of Nuclear Physics under contract DE-AC05-06OR23177.

The research reported here takes place in the context of the Topical Collaboration ``3D quark-gluon
structure of hadrons: mass, spin, tomography'' (Quark-Gluon Tomography Collaboration) supported by
the U.S.~Department of Energy, Office of Science, Office of Nuclear Physics.


\bibliographystyle{elsarticle-num}

\bibliography{refs}

\end{document}